\documentclass[11pt, oneside]{article}
\usepackage{jheppub}
\usepackage{graphicx}
\usepackage{amsmath}								
\usepackage{color}
\usepackage{mathtools}
\usepackage{amssymb}

\usepackage{hyperref}

\hypersetup{
    colorlinks,
    citecolor=blue,
    filecolor=black,
    linkcolor=blue,
    urlcolor=blue
}

\usepackage{overpic}
\usepackage{array}
\usepackage{rotating}
\usepackage{datetime}

\newcommand{\pd}{\partial}
\newcommand{\LL}{\mathcal{L}}

\newcommand{\bea}{\begin{eqnarray}}
\newcommand{\eea}{\end{eqnarray}}
\newcommand{\be}{\begin{equation}}
\newcommand{\ee}{\end{equation}}

\newcommand{\dint}{\ensuremath{\: \mathrm{d}}}

\newcommand{\diff}[2]{\frac{\mathrm{d} #1}{\mathrm{d} #2}}

\title{Carroll stories}
                                          \author[a]{Jan de Boer,}
                                           \author[b] {Jelle Hartong,}
                                           \author[c,d]{Niels A. Obers,}
                                           \author[e]{Watse Sybesma,}
                                           \author[f]{and Stefan Vandoren}

 \affiliation[a]{Institute for Theoretical Physics and Delta Institute for Theoretical Physics, \\
University of Amsterdam, P.O.Box 94485 1090 GL
Amsterdam, The Netherlands}
\affiliation[b]{School of Mathematics and Maxwell Institute for Mathematical Sciences, University of Edinburgh, Peter Guthrie Tait Road, Edinburgh EH9 3FD, UK} 
\affiliation[c]{Nordita, KTH Royal Institute of Technology and Stockholm University, \\
Hannes Alfvéns väg 12, SE-106 91 Stockholm, Sweden} 
\affiliation[d]{The Niels Bohr Institute, Copenhagen University, \\  Blegdamsvej 17, DK-2100 Copenhagen {\O}, Denmark}
\affiliation[e]{Science Institute, University of Iceland, \\Dunhaga 3, 107 Reykjav\'{i}k, Iceland}
\affiliation[f]{Institute for Theoretical Physics, Utrecht University, \\ Princetonplein 5, 3584 CE Utrecht, The Netherlands}

\emailAdd{J.deBoer@uva.nl, Jelle.Hartong@ed.ac.uk, niels.obers@su.se, \\zhws2@cam.ac.uk, S.J.G.Vandoren@uu.nl}

\abstract{We study various aspects of the Carroll limit in which the speed of light is sent to zero. A large part of this paper is devoted to the quantization of Carroll field theories. We show that these exhibit infinite degeneracies in the spectrum and may suffer from non-normalizable ground states. As a consequence, partition functions of Carroll systems are ill-defined and do not lead to sensible thermodynamics. These seemingly pathological properties might actually be a virtue in the context of flat space holography.

Better defined is the Carroll \textit{regime}, in which we consider the leading order term in an expansion around vanishing speed of light without taking the strict Carroll limit. Such an expansion may lead to sensible notions of Carroll thermodynamics. An interesting example is a gas of massless particles with an imaginary chemical potential conjugate to the momentum. In the Carroll regime we show that the partition function of such a gas leads to an equation of state with $w=-1$.

As a separate story, we study aspects of Carroll gravity and couplings to Carrollian energy-momentum tensors. We discuss many examples of solutions to Carroll gravity, including wormholes, Maxwell fields, solutions with a cosmological constant, and discuss the structure of geodesics in a Carroll geometry. The coupling of matter to Carroll gravity also allows us to derive energy-momentum tensors for hypothetical Carroll fluids from expanding relativistic fluids as well as directly from hydrostatic partition functions.

}

\begin{document}
                                           \maketitle

\section{Introduction}

Carroll symmetry arises in the limit of vanishing speed of light, in which the Poincar\'e algebra is contracted to the Carroll algebra \cite{Levy1965,sen1966analogue,Bacry:1968zf}. The most notable features of this algebra are that Carroll boosts commute and the Hamiltonian is a central charge. The physics arising in this Carroll limit is full of strange phenomena and mysteries. In contrast to Galilean relativity, Carroll symmetry implies that under boosts, space is absolute and time is relative,
\begin{equation}
    t'=t-\vec b\cdot\vec x\ ,\qquad \vec{x}^{\,'}=\vec{x}\ ,
\end{equation}
where $\vec b$ is the boost parameter. The light cone closes up as $c\to 0$ so particles with timelike worldlines cannot move in the Carroll limit and the theory becomes ultralocal. On the other hand there exist other types of Carroll particles with zero energy but they cannot stand still.\footnote{In \cite{Bergshoeff:2014jla,Casalbuoni:2023bbh} it was demonstrated that non-trivial dynamics for coupled Carroll particles can be realized when introducing interactions. In \cite{Marsot:2021tvq} particles in the extended Carroll group were studied and in \cite{Figueroa:2023jpi} fractonic particles.} They can be understood as the $c\to 0$ limit of relativistic tachyons \cite{deBoer:2021jej}. As a consequence, the theory seems to have potential problems with causality \cite{Levy1965,deBoer:2021jej}. Furthermore, as we will see in this paper, Carroll quantum field theories have some pathologies, as they do not seem to have well defined partition functions and there are uncontrolled divergences in perturbation theory that are difficult to regularize. So why is it worth to continue with this research?

\subsection*{Flat space holography}

The main motivation for studying systems with Carroll symmetry comes from the expectation that conformal Carroll field theories might be dual to quantum gravity in asymptotically flat spacetime. This expectation is substantiated by the fact that the asymptotic BMS symmetry group \cite{Bondi:1962px} of flat space is the conformal extension of the Carroll group living on its null boundary \cite{Duval:2014uva,Duval:2014lpa,Hartong:2015usd,Bagchi:2016bcd}. 
There is by now more evidence that conformal Carrollian field theories play a role in the celestial holography approach to flat space holography \cite{Donnay:2022aba,Donnay:2022wvx,Saha:2023hsl,Bagchi:2022emh,Bagchi:2023fbj}.

If Carrollian field theories are dual to quantum gravity in flat space, their thermal properties should say something about black holes.\footnote{Carroll symmetries in relation to black holes have also been studied in, e.g., the context of Love numbers \cite{Penna:2018gfx} and the black hole membrane paradigm \cite{Donnay:2019jiz,Freidel:2022bai,Redondo-Yuste:2022czg}.} 
But it is well known that black holes in flat space do not have well-defined partition functions at non-zero temperature; they are never in thermal equilibrium \cite{Hawking:1974rv}. Therefore, we expect similar problems with defining partition functions for Carroll field theories, and this is what we will demonstrate in this paper as one of the main results. So the seemingly pathological properties of Carroll quantum field theory may actually be a virtue of being a consequence of flat space holography. It is interesting to
contrast this with large black holes in AdS, which can be in thermal equilibrium with the Hawking radiation.  Its partition function is well defined, but diverges in the large radius limit in which AdS becomes flat space. The small black hole in AdS is of course unstable, and faces the same problems as in flat space.

\subsection*{Cosmology and dark energy}

Carroll symmetry might also be relevant for de Sitter cosmology and inflation \cite{deBoer:2021jej}. In the Carroll limit where we keep the Hubble constant fixed, the Hubble radius $R=c/H$ goes to zero and outside it recessional velocities are naturally large compared to the speed of light. 
 As we send the speed of light to zero, essentially the entire universe becomes super-Hubble and hence Carrollian. The Hubble radius defines the causal patch of an observer, and as the Hubble radius goes to zero, the theory becomes ultralocal, one of the main characteristics of Carrollian physics. We have already given an example in \cite{deBoer:2021jej} of a scalar field that in the Carroll limit yields an equation of state corresponding to dark energy, i.e. $w=-1$ and so ${\cal{E}}+P=0$, leading to a de Sitter universe when coupled to gravity.

 In this paper, we look at another example, and start with a Boltzmann gas of relativistic massless particles with a chemical potential (with the dimensions of velocity) conjugate to the momentum. We compute from the partition function the energy and pressure and consider these quantities in the Carroll regime, i.e. to leading order in the small $c$-expansion. The strict $c=0$ Carroll limit is not a well-defined statistical system, but any small value of $c$ is. We then show that for imaginary values of the chemical potential, such gasses in the Carroll regime have an equation of state with $w=-1$. This analysis is done in section \ref{sec:Carrollmicrogas}.

\subsection*{Carroll gravity}

A related aspect of Carrollian physics is Carroll gravity. This is obtained by considering the small speed of light (i.e. ultra-local) limit of General Relativity (GR) which was first considered in \cite{Henneaux:1979vn}. The more general
small $c$  expansion \cite{Dautcourt:1997hb} can be seen as a perturbative expansion around the (singular) Carroll point, complimentary to the large $c$ expansion that gives rise to Post-Newtonian corrections. 
Recently \cite{Hansen:2021fxi}, this ultra-local expansion of GR was considered using the modern perspective of non-Lorentzian geometry, incorporating that  Carroll geometry arises from Lorentzian geometry when taking $c\rightarrow 0$. As a result one obtains the electric (time-like) Carroll gravity action at leading order from the
Einstein-Hilbert action, while the magnetic (space-like) Carroll gravity action is a truncation of the next-to-leading order term. These theories are considered from a Hamiltonian point of view in \cite{Henneaux:2021yzg}
(see also \cite{Figueroa-OFarrill:2022mcy,Campoleoni:2022ebj}). Carroll gravity appears to describe
interesting dynamics of limits of important solutions in gravity. For example, it is closely related to the Beliniski-Khalatnikov-Lifshitz limit \cite{Belinski:2017fas} describing the near-singularity dynamics of general relativity. We also show that wormholes arise as the Carroll limit of black hole solutions in section \ref{sec:Cargra}. More generally, various aspects of Carroll gravity and geometry have appeared in a wide variety of recent studies \cite{Duval:2014uoa,Nzotungicimpaye:2014wya,Hartong:2015xda,Bekaert:2015xua,Bergshoeff:2017btm,Duval:2017els,Ciambelli:2018ojf,Morand:2018tke,Penna:2018gfx,Donnay:2019jiz,Bergshoeff:2019ctr,Ravera:2019ize,Gomis:2019nih,Ciambelli:2019lap,Ballesteros:2019mxi,Bergshoeff:2020xhv,Niedermaier:2020jdy,Gomis:2020wxp,Grumiller:2020elf,Concha:2021jnn,Guerrieri:2021cdz,Henneaux:2021yzg,deBoer:2021jej,Perez:2021abf,Figueroa-OFarrill:2021sxz,Herfray:2021qmp,Baiguera:2022lsw,Perez:2022jpr,Fuentealba:2022gdx,Campoleoni:2022wmf}.

\subsection*{Carroll hydrodynamics}
Hydrodynamics is an important framework for computing quantities such as pressure and energy-density for quantum systems. 
Coupling such a system to curved spacetime allows for effective computation of such quantities, but does not guarantee thermodynamical consistency.
In this paper we study two candidates of perfect fluids, timelike and spacelike, that satisfy Carroll symmetries. 
Using complementary geometric and thermodynamical arguments we highlight that the timelike candidate cannot be a true hydrodynamical fluid.

The study of Carroll symmetry in hydrodynamics has been initiated starting from two distinct notions of Carroll symmetry. 
The main distinction is that one notion starts from a fully diffeomorphism covariant approach with local tangent space Carroll boosts as hallmark of Carroll symmetry \cite{deBoer:2017ing}, which has a Ward identity that  constrains the energy flux to vanish. 
This notion is adopted in this paper. 
The other notion of Carroll fluids considers Carroll diffeomorphisms as manifestation of Carroll symmetry, a less restrictive requirement that, e.g., translates to a less constrained energy flux compared to the former approach and was pioneered in \cite{Ciambelli:2018xat,Ciambelli:2018wre}. 
For more details on this comparison we refer to \cite{Baiguera:2022lsw}.

\subsection*{Outline}

In this paper we work out specific aspects of this counterintuitive Carroll setting and report on their non-trivial features, but also possible limitations of our (non-)relativistic intuition or, perhaps, the limitations of Carrollian physics.
The examples we study are free quantum models, various solutions to the Carroll version of general relativity, and manifestations of thermodynamics and hydrodynamics.
One of the recurring themes we find is that although we can construct models that adhere to the Carroll symmetries, their thermodynamical nature or statistical mechanical realizations seem elusive. 

In the context of quantum field theories we work out two models, the so-called electric and magnetic Carroll contractions of a free massive scalar field, and point out features that arise when quantizing such theories. This is the topic of discussion in Section \ref{sec:CCFT}. In Section \ref{sec:carrollgeometryEM} we review
how Carroll geometry and gravity follows from the ultra-local (i.e. small speed of light) expansion of GR and study the coupling to matter in the Carroll limit. We Subsequently  report in Section \ref{sec:Cargra}  on various Carroll solutions that arise for specific models, including a non-trivial solution to Carroll gravity in the presence of matter. 
Afterwards, energy-momentum tensors are studied via an expansion around zero speed of light and using the hydrostatic partition function, and we furthermore provide an analysis concerning the statistical nature of a Carroll gas. This is done in Section \ref{sec:emtcarroll}. Some appendices are added with more material and technical details.

\section{Carroll quantum field theory}\label{sec:CCFT}

In this section we explore some properties of Carroll quantum field theories to gain some insight in the structure of their physical Hilbert
space, their correlation functions and their thermodynamics\footnote{Some aspects of Carroll quantum field theories have also been discussed in the recent works \cite{Banerjee:2023jpi,Figueroa-OFarrill:2023qty}.
}. For the sake of the discussion below, we will assume that the fields in a Carroll QFT 
transform under Carroll coordinate transformations in exactly the same way as they would transform under a more general coordinate transformation. 
It is conceivable that completely other realizations of the Carroll algebra exist but in those cases
the implications would need to be examined on a case-by-case basis, in particular Ward identities would not take the usual form, 
and it is not clear that these other realizations could, e.g., be applied to a putative theory at future null infinity in flat space holography.

Carroll QFT's can roughly be divided in (i) ``electric" theories, (ii) ``magnetic" theories and (iii) a combination of these two. This nomenclature
has its origin in considering Carrollian limits of the Maxwell equations of motion \cite{Duval:2014uoa}, where in one limit only the  electric field survives, and in another only the
magnetic field survives. An off-shell formulation was introduced in \cite{Henneaux:2021yzg,deBoer:2021jej}.
Departing from free relativistic theories, heuristically, the electric and magnetic limit, respectively, correspond to considering timelike or spacelike excitations before the relativistic causal structure collapses by taking the Carroll limit.

More generally, electric theories are theories which are ultralocal in space and have non-trivial time-dependence, whereas 
magnetic theories have a very simple time-dependence and non-trivial space-dependence. Theories of type (i) and (ii) can be coupled together
to give rise to theories of type (iii). 

In what follows, we will consider a simple illustrative example of an electric and magnetic scalar field theory and consider their properties. 
We will also make some comments on more general electric and magnetic theories, and in particular point out that one can construct electric Carroll 
theories starting from any quantum mechanical system, and $d$-dimensional magnetic theories starting 
from any $(d-1)$-dimensional Euclidean field theory (as was also eluded to in \cite{Baiguera:2022lsw}). 
We postpone a study of theories of type (iii) to future work. 

It is interesting whether the electric theory could be describing the ``hard" particles and the magnetic theory the ``soft"
particles as one has in flat space holography. The magnetic theory has arbitrarily soft particles but no 
normalizable zero energy states as we will see below. It would be interesting to explore the precise connection between the magnetic theory 
and soft modes in flat space holography in more detail.

In addition to the aforementioned works, conformal Carroll symmetry realizations in field theory were studied in \cite{Bagchi:2016bcd,Bagchi:2019xfx,Bagchi:2019clu,Banerjee:2020qjj,Chen:2021xkw,Rivera-Betancour:2022lkc,Baiguera:2022lsw,Saha:2022gjw,Chen:2023pqf}, Carroll fermions in \cite{Hao:2022xhq,Banerjee:2022ocj,Bagchi:2022eui}, Carroll electrodynamics in \cite{Basu:2018dub}, Carroll Yang-Mills in \cite{Islam:2023rnc}, SUSY realizations in \cite{Barducci:2018thr}, and fractonic realizations in \cite{Bidussi:2021nmp,Figueroa-OFarrill:2023vbj}. For a recent discussion on quantum effects in Carroll field theory see \cite{Banerjee:2023jpi}.

\subsection{Electric scalar theory}

A simple example of an electric scalar theory is given by the Lagrangian 
\be\label{eq:elelagrangian}
S = \frac{1}{2}\int {\rm d}t\, {\rm d}^d \vec{x}(\dot{\phi}^2 - m^2 \phi^2)\ .
\ee
This theory is ultralocal, in the sense that there are no spatial derivatives and therefore spatial points appear to be completely independent
from each other: the theory consists of an infinite number of harmonic oscillators, one for each value of the coordinate $x$. In 
particular, if we were to put the theory on a lattice, there would be no need to add any coupling between different lattice points.

\subsubsection*{Canonical quantization}

The most general solution of the field equation is
\be
\phi= e^{imt} \int {\rm d}^d \vec{k}\, a_{\vec{k}^\dagger}\, e^{i\vec{k}\cdot \vec{x}} + c.c.\ ,
\ee
as was pointed out by \cite{Bagchi:2022emh}.
Upon quantization, the canonical commutator
\be
[\phi(t,\vec{x}),\pi(t,\vec{y})]=i\delta^{(d)} (\vec{x}-\vec{y})\ ,
\ee
implies that 
\be
[a_{\vec{k}},a_{\vec{l}}^{\dagger}]=\frac{1}{2 m (2\pi)^d} \delta^{(d)}(\vec{k}-\vec{l})\ ,
\ee
so up to some irrelevant normalization this is indeed just an infinite collection of harmonic oscillators (we assume $m>0$ for now).    
The obvious choice for a ground state is to take a state which is the ground state for each harmonic oscillator. 
To write down Ward identities, we need to make sure that the Carroll symmetry is not spontaneously broken, which follows fairly 
straightforwardly from the explicit form of the symmetries of the theory which we will now review.

\subsubsection*{Symmetries}

Consider a transformation
of the form
\be
\delta\phi = \xi^t \partial_t \phi + \xi^i \partial_i \phi + \xi \phi\ .
\ee
If we vary the action, we find the following necessary conditions for this to be a symmetry
\bea
2\xi + \partial_t \xi^t - \partial_i \xi^i & = & 0 \nonumber\ , \\
\partial_t\xi^i & = & 0 \nonumber\ , \\
m^2 \partial_t\xi^t + m^2 \partial_i \xi^i - \partial^2_t\xi - 2 m^2 \xi & = & 0 \,\,  .
\eea
This has some peculiar $m$-dependent symmetries which we will not investigate. The symmetries which
are $m$-independent are of the form
\be
\delta\phi = a(x^i) \partial_t \phi + b^i(x^i) \partial_i \phi + \frac{1}{2} \partial_i b^i \phi\ ,
\ee
which includes Carroll transformations but is in fact a much larger group of transformations, presumably due to the fact 
that the theory is ultralocal in $x$ and quadratic. Higher order interactions will restrict $b$ to be constant, but
the function $a$ remains unconstrained. 

In fact, the transformations of the form $\delta \phi = a(x^i)\partial_t\phi$ are reminiscent of supertranslations. There is no
analogue of superrotations in this toy model unless we consider the massless ``conformal" case. With the mass equal to zero the 
symmetries become 
\begin{equation} \label{j3}
\delta\phi = \Big(a(x^i)+t(\partial_ib^i-2\xi_0)\Big) \partial_t \phi + b^i(x^i) \partial_i \phi + (\xi_0+t\xi_1) \phi\ ,
\end{equation}
for any $\xi_0,\xi_1$ depending only on spatial components. This indeed contains terms linear in $t$ in $\xi^t$, as expected for 
superrotations. 
For simplicity, we continue below with the mass turned on.

The conserved currents are found to be
\bea
J^t & = & a (\dot{\phi}^2 - m^2 \phi^2) - 2 (a(x^i) \partial_t \phi + b^i(x^i) \partial_i \phi + \frac{1}{2} \partial_i b^i \phi)\dot{\phi} \nonumber\ , \\
J^i & = & b^i (\dot{\phi}^2 - m^2 \phi^2)\ ,
\eea
which for constant $a$, $b$ yield the energy-momentum tensor of the system.

The conserved charges read
\bea
Q_a & = & \int {\rm d}^d x\, a(\dot{\phi}^2 + m^2 \phi^2) \nonumber\ , \\
Q^i_b & = & \int {\rm d}^d x \,(2 b^i \partial_i\phi \dot{\phi} + \partial_i b^i \phi\dot{\phi})\ .
\label{conscharg}
\eea
The first conserved charge in terms of modes reads
\be
Q_a = 4m^2 \int {\rm d}^dx\, {\rm d}^dk\, {\rm d}^dl\, a_k^\dagger a_l \,e^{i(k-l)\cdot x} a(x)\ ,
\ee
so that in particular the Hamiltonian reads 
\be
H = 2 m^2 (2\pi)^d \int {\rm d}^dk\, a_k^\dagger a_k\ ,
\ee
We get the usual quantization rules with $a^\dagger$ creation and $a$ annihilation operator and we normal order such that the annihilation operator is put to the right. 

The Carroll boost charges are (take $a=\epsilon \cdot x$) 
\be
C_i = -4i m^2 (2\pi)^d\int {\rm d}^dk \,{\rm d}^dl\, a_k^\dagger a_l \,\partial_i \delta(k-l)\ ,
\ee
with $\partial_i$ the derivative with respect to $k^i$, while momenta read
\be 
P^i = 4m(2\pi)^d \int {\rm d}^dk\, k^i a_k^\dagger a_k\ ,
\ee
and one can check that these obey the right type of commutator, 
\begin{equation}
    [P^i,C_j]=4i\delta^i_j H\ .
\end{equation}
We can then normal order the generators of the Carroll algebra which will preserve their 
commutation relations and also guarantee that they annihilate the unique ground state.

\subsubsection*{Spectrum}

Therefore for these types of Carroll theories there is a unique normalizable Carroll invariant ground state satisfying $a_k |0\rangle =0$, and the energies come in multiples of $m$ with infinite degeneracy - all states $a^\dagger_k |0\rangle $ have energy $m$ regardless
of the choice of $k$,
\begin{equation}
    H\,a^\dagger_k |0\rangle = m \,a^\dagger_k |0\rangle\ .
\end{equation}
This is just a manifestation of the ultralocality of the theory. States of the form $a^\dagger_k  a^{\dagger}_{l}|0\rangle $ are all
degenerate with energy eigenvalue $2m$, etc.

\subsubsection{Correlation functions}

We can compute the commutator
\be
[\phi(t,\vec{x}),\phi(t',\vec{x}^{\,'})]=-\frac{i}{m}\sin [m(t-t')]\delta^d(\vec{x}-\vec{x}^{\,'})\ ,
\ee
which agrees with the $c\to 0$ limit of the commutator of a relativistic massive scalar field. The main difference is the appearance of the delta function in space, so the commutator vanishes for any two separated points in space. 

We can also compute correlators in this theory. The time-ordered correlator is computed to be
\begin{eqnarray}
    \label{prop-electric}
\langle 0 | {\rm T} (\phi(t,\vec{x}) \phi(t',\vec{x}^{\,'}) )| 0 \rangle &=& \frac{1}{2m}\Big(e^{-im(t-t')}\theta(t-t') + e^{im(t-t')}\theta(t'-t)\Big)\delta^{(d)} (\vec{x}-\vec{x}^{\,'})\ ,\nonumber\\
&=&\frac{1}{(2\pi i)^{d+1}}\int{\rm d}q \,{\rm d}^d\vec{p}\,\,\frac{e^{-iq(t-t')+i\vec{p}\cdot(\vec{x}-\vec{x}')
}}{-q^2+m^2-i\varepsilon}\ .
\end{eqnarray}
This has a form which agrees with the Carroll Ward identity which allows solutions of the form
$f(t-t') \delta^{(d)}(\vec{x}-\vec{x}^{\,'})$ \cite{deBoer:2021jej}, which is explicitly Carroll boost invariant.

Notice that the second line is the (electric) Carroll limit of the correlator of relativistic scalar field, in which the spatial momenta are suppressed in the $c\to 0$ limit. The integral over spatial momenta then trivially gives rise to the ultralocality in space, in the form of the delta function. Furthermore, notice that from the first line, it is easy to see that the two-point correlation function is again Carroll boost invariant. This may seem somewhat surprising, since it would mean that propagation is causal in any boosted frame. Causality is however not guaranteed in Carroll systems, as was already noticed in the original paper \cite{Levy1965}. This is because in general, time-ordering is not a Carroll-boost invariant notion. The reason is that under Carroll boosts, we have the transformation laws
\begin{equation}
\Delta t'=\Delta t - \vec{b}\cdot\Delta \vec{x}\ .
\end{equation}
For large enough boost parameter $|\vec{b}|$, time ordering between any two events can change sign between boosted Carroll observers, at least if $\Delta x \neq 0$. Fortunately, in the above correlator, when $\Delta x=x'-x$ is non-zero, the entire correlator vanishes and so time ordering is boost invariant and causality at the level of the propagator is satisfied because the theory is ultralocal.

It is interesting to consider the massless limit, which corresponds to a free conformal Carroll scalar. After subtracting a diverging constant, we find for the $m\to 0$ limit of \eqref{prop-electric}
\begin{equation}
    \label{prop-electric-massless}
\langle 0 | {\rm T} (\phi(t,\vec{x}) \phi(t',\vec{x}^{\,'}) )| 0 \rangle = -\frac{i}{2}|t-t'|\,\delta^{(d)} (\vec{x}-\vec{x}^{\,'})\ .
\end{equation}
The massless case is most interesting in the context of flat space holography. The Feynman propagator is then 
\begin{equation}
  G_F(x,x')=-i\langle 0 | {\rm T} (\phi(t,\vec{x}) \phi(t',\vec{x}^{\,'}) )| 0 \rangle= -\frac{1}{2}|t-t'|\,\delta^{(d)} (\vec{x}-\vec{x}^{\,'})
  \ ,   
\end{equation}
whereas the retarded propagator is 
\begin{equation}
    G_{\rm {ret}}(x,x')=i\langle 0 | [\phi(t,\vec{x}), \phi(t',\vec{x}^{\,'})] | 0 \rangle \theta(t-t') = (t-t')\theta(t-t')\delta^d(\vec {x}-\vec{x}^{\,'})\ .
    \end{equation}

\subsubsection{Scale invariance}

We can replace the mass term in \eqref{eq:elelagrangian} by a general potential $V(\phi)$. This will generically lead to an equation
for symmetries which looks like
\be
(\partial_t\xi^t +  \partial_i \xi^i)V(\phi) - \phi V'(\phi) \xi=0\ ,
\ee
which forces $\xi=0$ unless the potential is proportional to a single power of $\phi$. So for generic potentials
we still have all the symmetries as above, but with $\partial_i b^i=0$ (so these do not fix $b$ to be constant in
higher dimension), and if $V$ is a pure power we get more symmetry, in particular a dilation symmetry. For example,
for $\dot{\phi}^2 - \phi^p$ there is a symmetry under which $t\rightarrow \lambda^z t$, $x\rightarrow \lambda x$ 
and $\phi \rightarrow \lambda^{\xi} \phi$ with
\be z=\left(1-\frac{p}{2}\right)\xi ,\qquad   d
= -\left(1+\frac{p}{2}\right) \xi\ ,
\ee
and the corresponding generator will extend the Carroll algebra as
\be\label{eq:lifshitz-carroll-algebra}
[D,H]=-z H,\qquad [D,P_i] = -P_i, \qquad [D,C_i]= (1-z)C_i .
\ee

Conformal Carroll symmetry has $z=1$, obtained from a limit of relativistic scale invariance where space and time scale the same way. The boosts then commute with dilations, and we furthermore have
\begin{equation}
p=2\,\frac{d+1}{d-1}\ .
\end{equation}
This constrains the values of $p$ for a given dimension, for instance we have $p=4$ for three spatial dimensions and $p=6$ for $d=2$. The case $d=1$, so two spacetime dimensions, is special as there is no solution for $p$ when $z=1$; only the free scalar with only a kinetic term is scale invariant.

\subsubsection{General electric theories}

A much larger class of electric theories is obtained as follows. We take a $d$-dimensional spatial manifold $M$ and 
some quantum mechanical system with Hilbert space ${\cal H}$ and Hamiltonian $H$, and associate one copy 
of the quantum mechanical system to each point in $M$. If the quantum mechanical system has a Lagrangian description, then we can 
simply let all the fields depend on the spatial coordinates $x$ and write ${\cal L}= \int_M {\rm d}x L_{QM}[\phi(t,x)]$ for the full 
Lagrangian of the system. Here ${\rm d}x$ represents the measure on $M$, in coordinates 
it would read ${\rm d}^{d}x \sqrt{g}$ with $g$ the metric on $M$. 

These theories have conserved charges similar to those in (\ref{conscharg}) where 
\be
Q_a = \int_M {\rm d}x\, a(x) H_x\ ,
\ee
with $H_x$ the Hamiltonian of the quantum mechanical system at the point $x$. Since $[H_x,H_y]=0$ these charges all commute with each other
for all space-dependent functions $a(x)$. There are also charges $Q_b$ associated to the isometries of $M$ that are a bit less 
straightforward to write down, but they simply translate the system along $M$. In particular, when exponentiated these will act on $Q_a$ as
\be
Q_a \rightarrow Q_{a'} \equiv \int_M {\rm d}x\, a(x-b) H_x\ .
\ee
If $M$ is simply ${\mathbb R}^{d}$ then these theories are in particular invariant under the standard Carroll algebra.

The finite energy eigenstates of the system consist of the somewhat singular states which are a tensor product of the ground state for almost 
all $x$ times a finite number of finite energy states at a finite number of points on $x$. The energy eigenvalue is the finite sum of the individual non-zero energies and excited states are infinitely degenerate. The ground state is the tensor product of the ground states for all $x$. It is clearly annihilated 
by both $Q_a$ and $Q_b$. The first excited state is the tensor product of ground states times the first excited state of the
quantum mechanical system at one point $x\in M$. These states are degenerate and there is an $L^2(M)$ worth of such states. We saw a momentum space version of these states in our scalar example earlier. 

Correlation functions of operators with vanishing one-point functions in the ground state $\langle 0| O_{x_1} \ldots O_{x_n}|0\rangle$
with the operators $O_{x_i}$ part of the quantum mechanical theory associated to the point $x_i$ will vanish unless for each $x_i$ 
there is at least another $x_j$ with $x_i=x_j$. In other words, these correlators will contains products of delta-functions in the spatial coordinates. This is therefore a general feature of these electric theories. One can also get derivatives of delta-functions if
one considers explicit realizations in terms of fields and if one constructs operators involving spatial derivatives but the theory remains ultra-local. Since these theories are by assumption built from well-defined quantum mechanical building blocks, they require no further regularization or renormalization.

\subsection{Magnetic scalar theory}

We now consider the magnetic theory
\be\label{eq:magnetic-action}
S= \int {\rm d}t \,{\rm d}^d x \, \left( \chi \dot{\phi} + {\cal L}(\phi) \right)\ ,
\ee
where ${\cal L}$ can be any Lagrangian depending on $\phi$. One could consider different versions of this theory, e.g. one where the
leading term is $\dot{\phi}_1 \dot{\phi}_0$ (as one obtains in Taylor series expansions around $c=0$ \cite{deBoer:2021jej}), but this theory has
a phase space which is twice as large as the original theory and the theory with a Lagrange multiplier does not have 
this feature. 

\subsubsection*{Canonical quantization}

The field equations are $\dot{\phi}=0$ and $\dot{\chi}= {\cal L}'(\phi)$ where $\int \delta {\cal L}\equiv 
\int \delta \phi {\cal L}'(\phi)$ so ${\cal L}'$ is just shorthand for the equations of motion obtained from ${\cal L}$.
The general solution of the field equations is 
\bea\label{eq:gensolele}
\phi & =& \phi(x)\ , \nonumber \\
\chi & = & \chi(x) + t {\cal L}'(\phi(x))\ .
\eea
The canonical momenta are 
\bea
\pi_{\chi} & = & 0 \ ,\nonumber \\
\pi_{\phi} & = & \chi + \pi_{\cal L}(\phi)  \ ,
\eea
which show that this is a system with constraints. As the solutions of the field equation show, the complete phase space is 
spanned by $\phi(x)$ and $\chi(x)$ so it is sufficient to find their Poisson bracket and commutator. To make life a little bit simpler, we will assume that with a suitable shift of $\chi$ we can always absorb all terms
with time derivatives in ${\cal L}$, so that ${\cal L}$ no longer contains time derivatives and $\pi_{\cal L}=\frac{\partial\mathcal{L}}{\partial\dot\phi}=0$. Then the definition of the canonical momenta take the form of two second-class constraints and using a standard Dirac bracket for systems with constraints we obtain
\be
[\chi(x),\phi(y)]=-i\delta^{(d)}(x-y)\ .
\ee
If we write 
\be \label{magmodes}
\phi = \int {\rm d}^d k\, e^{ikx}\, a_k\ ,\qquad  \chi=\int {\rm d}^d k\, e^{ikx}\, b_k\ .
\ee
Then
\be
[a_k,b_l]=\frac{i}{(2\pi)^d}\delta^{(d)}(k+l) 
\ee
with $a_k^{\dagger}=a_{-k}$ and similar for $b_k$.

\subsubsection*{Symmetries}

We next examine the global symmetries of the magnetic scalar theory using an ansatz similar to the electric case
\be
\delta\phi = \xi^t \partial_t \phi + \xi^i \partial_i \phi + \xi \phi\ .
\ee
The variation of the
action becomes
\be
\delta S = \int {\rm d}t{\rm d}^dx \left(\delta \chi \dot{\phi} -\dot{\chi}(\xi^t \partial_t \phi + \xi^i \partial_i \phi + \xi \phi)
+ \xi^t \partial_t \phi {\cal L}'(\phi) + \xi^i \partial_i \phi {\cal L}'(\phi)+ \xi \phi {\cal L}'(\phi) \right).
\ee

For generic ${\cal L}$ the last term cannot be canceled against anything and we must choose $\xi=0$. For
special ${\cal L}$ we may be able to choose a $\xi$ so that the theory has an additional scale symmetry but we
will ignore that possibility for now. The term $ \xi^i \partial_i \phi{\cal L}'$ can also not be canceled by 
anything, and to cancel this term $\xi^i$ must be a symmetry of ${\cal L}$. The term $\xi^t \partial_t \phi {\cal L}'(\phi)$ can be rewritten as\footnote{The term $\xi^t \partial_t \phi {\cal L}'(\phi)$ could have been canceled by a suitable transformation of $\chi$. The reason we have to rewrite this term is because otherwise $\xi^t$ corresponds to a trivial gauge transformation parameter due to an equation of motion symmetry of the action whereby $\chi$ transforms into the equation of motion of $\phi$, and $\phi$ into the equation of motion of $\chi$.}
\begin{equation}\label{eq:rewriting}
    \xi^t \partial_t \phi {\cal L}'(\phi) =-\mathcal{L}\partial_t\xi^t+\dot\phi\partial_i\xi^t\frac{\partial\mathcal{L}}{\partial\partial_i\phi}\,,
\end{equation}
up to total derivative terms. The term $\mathcal{L}\partial_t\xi^t$ cannot be canceled and so we must set $\partial_t\xi^t=0$. The second term in \eqref{eq:rewriting} can be cancelled by assigning an appropriate transformation law to the $\chi$ field. Finally, the term $\dot{\chi} \xi^i \partial_i \phi$ can also
not be canceled unless $\partial_t \xi^i=0$. There are no further restrictions so we get the following
symmetries (with $\xi^i$ a symmetry of ${\cal L}$)
\be
\delta\phi = \xi^t \partial_t \phi + \xi^i \partial_i \phi\ ,\quad
\delta\chi = \xi^t \partial_t \chi +  \partial_i (\chi \xi^i)  - \partial_i \xi^t 
\frac{\partial {\cal L}}{\partial(\partial_i\phi)}      
\ ,\quad
 \partial_t \xi^i = \partial_t \xi^t =0\ .
\ee
Interestingly, we once more get many more symmetries than just Carroll, we seem
to get supertranslations, just like what we got in the electric case.

The conserved currents are found to be (assuming again $\pi_{\cal L}=0$)
\bea
J^t & = & \xi^t {\cal L} - \chi \xi^i \partial_i \phi\ , \nonumber \\
J^k & = & \chi \xi^k \dot{\phi} - \xi^t \dot{\phi} \frac{\partial {\cal L}}{\partial(\partial_k\phi)} +
\Lambda^k[\xi^i] - \xi^i \partial_i \phi  \frac{\partial {\cal L}}{\partial(\partial_k\phi)}\ ,
\eea
where $\Lambda^k[\xi^i]$ is the total derivative obtained by varying ${\cal L}$ with respect to $\xi^i$. 
Recall that we assumed that $\xi^i$ is a symmetry, so $\delta_{\xi^i}  {\cal L} =  \partial_k \Lambda^k[\xi^i]$
by assumption. One can explicitly check, using the field equations, that this current is conserved. 

The conserved charges are therefore given by 
\be \label{j5}
Q[\xi]=\int {\rm d}^d x \left( \xi^t {\cal L} - \chi \xi^i \partial_i \phi  \right).
\ee
We could go ahead and express the charges in terms of the modes (\ref{magmodes}) of $\phi$ and $\chi$ but this is not particularly
instructive. It is more insightful to write the quantum charges in a Schr\"odinger representation as
\be
Q[\xi]=\int {\rm d}^d x \left( \xi^t {\cal L} + i  \xi^i \partial_i \phi \frac{\delta}{\delta \phi} \right)
\ee
which indeed have the same commutation relations as the vector fields $\xi^\mu\partial_\mu$. We could also have
put the functional derivative to the left of $\partial_i \phi$ with a similar result. 

In particular, the Hamiltonian of the system is 
\be
H=-\int {\rm d}^d x \, {\cal L}\ ,
\ee
where we introduced the correct sign to be in agreement with convention chosen in the starting Lagrangian (\ref{eq:magnetic-action}).
It is easy to see with the bracket given above that this indeed generates time translations of the
solutions of the field equations.

\subsubsection*{Spectrum}

To analyze the spectrum of the theory we will first consider a special case where the volume is finite (so momenta are discrete) and
where  ${\cal L}$ is quadratic 
and contains a term proportional to $\phi^2$. Then in the zero mode sector we have a structure of the form
$[a_0,b_0]=i$ (we dropped an irrelevant normalization factor) and $H\sim a_0^2$. 
We see that $a_0$ and $b_0$
are like coordinates and momenta. If we take $b_0$ to be like momentum and $a_0$ like position we can consider
position eigenstates $|a_0\rangle$ which are delta-function normalizable. In this notation, the energy eigenstates
for $E\geq 0$ are $|\sqrt{E}\rangle$ and $|-\sqrt{E}\rangle$. We therefore get a continuous spectrum with 
delta-function normalizable eigenstates. In particlar, there is no normalizable ground state \footnote{Note that in the relativistic case, 
there is an extra term in the action proportional to $c^2\chi^2$, which makes the ground state normalizable.}.

For non-zero modes we pick some momentum $k$ and consider the modes with momentum $\pm k$. The Hamiltonian
will be proportional to $a_k a_{-k}$. There might be a $k$-dependent prefactor but will ignore that prefactor
as it is just the overall normalization. So the structure that we have is (dropping irrelevant factors)
\be
[b,a^{\dagger}] = [b^{\dagger},a]=-i, \qquad H=aa^{\dagger}.
\ee
We might be able to directly study this in coordinate space but it is instructive to do it in terms
of harmonic oscillators as well. We can redefine
\be
b= - \frac{i}{2} c + \frac{i}{2} d^{\dagger},\quad
b^{\dagger} = \frac{i}{2} c^{\dagger} - \frac{i}{2} d,
\quad
a=c+d^{\dagger}, \quad a^{\dagger} = c^{\dagger} + d, 
\ee
which results in standard harmonic oscillators $[c,c^{\dagger}]=[d,d^{\dagger}]=1$ and a Hamiltonian
$H=(c+d^{\dagger})(c^{\dagger}+d)$. It is an amusing exercise to find the spectrum of this system whose details
we defer to Appendix \ref{qmtoymodel}. The result of this computation is that the spectrum is
infinitely degenerate and continuous and does not possess a normalizable ground state. 
 
A more general analysis could proceed as follows using the Schr\"odinger representation of 
states as wave functionals $\Psi[\phi(x)]$. 
Consider the classical solution space to the equation $-\int {\rm d}^dx{\cal L}(\phi)=E$. Any Schr\"odinger wave functional on the space of $\phi(x)$ with support on this subspace will
be an energy eigenstate. Since the support is on a submanifold of function space (because the Hamiltonian does not depend on the momenta) we expect these states to 
be at best delta-function normalizable (certainly with respect to other energy eigenstates). In particular,
there is no normalizable ground state (unless we have the trivial case ${\cal L}=0$). If the equation ${\cal L}(\phi)=0$ has $\phi=0$ as a solution, then because ${\cal L}=0$ and $\partial_i \phi=0$ the 
wave functional with delta-functional support at $\phi=0$ appears to be invariant under the Carroll symmetries, but this
is a purely formal statement due to the non-normalizability of this ground state. 

\subsubsection{Correlation functions}

Consider the simplest case of a magnetic scalar theory, given by
\be\label{eq:magnetic-action2}
S= \int {\rm d}t \,{\rm d}^d x \, \chi \dot{\phi} \ .
\ee
Under Carroll boosts, the two scalars transform as
\begin{equation}
    \delta \chi=\vec{b}\cdot \vec{x}\, \dot{\chi}\ ,\qquad \delta \phi=\vec{b}\cdot \vec{x}\, \dot{\phi}\ ,
\end{equation}
and the Lagrangian transforms into a total time derivative.

The Green's function is now
\begin{equation}
    G_{\chi\phi}(t,\vec{x}\,; t',\vec{x}^{\,'})=\frac{i}{2}{\rm{sgn}}(t-t')\,\delta^d(\vec{x}-\vec{x}^{\,'})\ ,
\end{equation}
and satisfies $\partial_tG(t,t')=i\delta(t-t')\delta^d(\vec{x}-\vec{x}^{\,'})$. The result is basically the Fourier transform of $1/\omega$, and the pole at $\omega=0$ produces the discontinuity at $t=t'$ appearing in the sign function. This pole, which appears generically in magnetic field theories, could be interpreted as the dual of a bulk soft mode propagating to the Carroll boundary. 

One can also consider more general Lagrangians for the magnetic theory, e.g. by adding spatial derivatives to ${\cal L}$, but we will not discuss these theories 
further here. We will make some more comments about the structure of the correlation functions when we discuss general magnetic theories below.
The form of the correlation functions for the free electric and magnetic scalar theories have also been discussed in \cite{Chen:2023pqf}.

\subsubsection{Scale symmetry}

Just as in the electric case, we can consider theories with additional scale symmetries. 
Assume for example that under $x\rightarrow \lambda x$ and $\phi\rightarrow \lambda^{\xi}\phi$ the scaling of ${\cal L}$ reads
$\cal L \rightarrow {\lambda}^{\alpha}L$, then the action is scale invariant under the additional assignment $t\rightarrow \lambda^{-\alpha-d} t$
and $\chi \rightarrow \lambda^{-\xi-d} \chi$. A simple theory with ${\cal L}\sim (\partial_i\phi)^2$  would have a scale 
invariance with $\alpha=2\xi-2$. It is therefore easy to construct scale invariant magnetic Carroll theories starting from a suitable
scale covariant ${\cal L}$.

\subsubsection{General magnetic theories}

A general construction of magnetic theories starts from any $d$-dimensional Euclidean field theory $-{\cal L}(\phi_a)$ which depends on
fields $\phi_a(x)$. We can then write down the following magnetic theory
\be
S=\int {\rm d}t {\rm d}^d x\, (\chi^a \dot{\phi}_a + {\cal L}(\phi_a) )\ .
\ee
If ${\cal L}$ has a more complicated field content we similarly need to double the field content and add first order couplings which
force the fields in ${\cal L}$ to be time-independent, and which force the additional fields to be linear in time on-shell, and such that the new fields serve as canonical momenta for the fields in ${\cal L}$. 
For example, if $-{\cal L}$ is Euclidean Maxwell theory, the theory will take the form \cite{deBoer:2021jej}
\be
S=\int {\rm d}t {\rm d}^d x\, (\chi_i E_i  - \frac{1}{4} F_{ij}^2 )\ ,
\quad
E_{i}=\partial_{i}A_{t}-\partial_{t}A_{i}
\ ,
\quad
F_{ij}=\partial_{i}A_{j}-\partial_{j}A_{i}
\ ,
\ee
where $A$ is a 1-form.
In the remainder we will restrict attention to a Euclidean field theory with a single scalar field for simplicity. We already discussed
various aspects of such theories above, including the symmetries and the spectrum of the theory. They are most easily understood in
a Schr\"odinger wave functional formalism, where states are wave functionals $\Psi[\phi(x)]$ with inner product given by the path 
integral
\be
\langle \Psi | \Psi \rangle = \int {\cal D}\phi\, |\Psi[\phi(x)]|^2\ .
\ee
Since the Hamiltonian is given by $H=-\int {\rm d}^dx{\cal L}(\phi(x))$, the equation for energy eigenstates reads
\be
-\int {\rm d}^dx{\cal L}(\phi(x)) \Psi[\phi(x)] = E \Psi[\phi(x)]\,,
\ee
which means that the wave function must have support on the space of solutions of the equation $-\int {\rm d}^dx{\cal L}(\phi(x))=E$ only. This shows that the spectrum will be generically continuous and infinitely degenerate. 

Correlation functions of $\chi \equiv -i \frac{\delta}{\delta \phi}+ t{\cal L}'(\phi)$ (see \eqref{eq:gensolele}) and $\phi$ take the form
\be
\langle \Psi | F(\chi, \phi) |\Psi \rangle \rightarrow \int {\cal D}\phi\, \Psi^{\ast}[\phi(x)] F( -i \frac{\delta}{\delta \phi}
+ t{\cal L}'(\phi),\phi)
\Psi[\phi(x)]\ .
\ee
From this we observe a few general properties. 
\begin{itemize}
    \item Correlation functions involving only $\phi$ are time-independent in any state and therefore invariant under time-translations and Carroll boosts. 
    \item Correlation functions involving a finite number of $\chi$-fields will
be polynomial in $t$, up to possible theta-functions associated to a choice of time-ordering. 
\item Correlation functions of $\phi$'s in states $\psi_E$ which are invariant under translations and rotations will obey all Carroll Ward identities even if $\psi_E$ is not annihilated by all Carroll generators. 
\item For wave-functionals of the form
$\psi_E = N\exp(-S_E[\phi]/2)$ with some auxiliary Euclidean ``action" $S_E$ and normalization factor $N$, correlation functions will be given by correlation functions in an auxiliary Euclidean QFT with action $S_E$. 
\end{itemize}

Finally we remark that, complimentary to the general construction of magnetic theories described above,
there is another method. This is based on a map that uses as input a (magnetic) Galilean and an electric Carroll action
and generates a corresponding magnetic Carroll action. This is described in Appendix \ref{app:map} (see in particular equation \eqref{newL}). According to this method the action \eqref{eq:magnetic-action} with $\mathcal{L}$ depending on $\phi$ and its spatial derivatives $\partial_i\phi$ is a magnetic Galilean theory to which we add a suitable constraint and appendix \ref{app:map} explains why the resulting theory is Carroll invariant. 

\subsection{Thermodynamics}\label{thermo}

In section~\ref{sec:emtcarroll} we will discuss energy-momentum tensors for Carrollian fluids. 
Such energy-momentum tensors have made frequent appearance in the literature, 
but in order for these energy-momentum tensors to be the actual energy-momentum tensor 
of a well-defined microscopic quantum system, we should find Carrollian quantum systems with well-defined thermodynamics and a well-defined equation of state. As we will see, it is very problematic to find such systems in the strict $c\to 0$ limit. These problems also manifest themselves in 
our discussion in section~\ref{sec:Carrollmicrogas} when we consider the $c\rightarrow 0$ limit of partition functions of ideal gases.

\subsubsection{Representation theory and partition function}

The first place where we see a potential problem in defining thermodynamics for Carroll systems is in the representation theory which we discussed in \cite{deBoer:2021jej}. The Carroll algebra contains commutators of the type $[C,P]=H$ with $C$ the Carroll boost, $P$ a momentum generator, and $H$ the Hamiltonian. The representations of the subalgebra spanned by $C,P,H$ are very simple. $H$ is a central element so we can take it to be a given number. For $H \neq 0$ this commutator is like the commutator for a single coordinate and momentum in quantum mechanics, and the relevant Hilbert space is therefore $L^2(\mathbb R)$. For $H=0$ we simply fix $C$ and $P$ to a particular value since now all three generators commute. 

We therefore see that energy-eigenstates with $E\neq0$ are necessarily infinitely degenerate. Theories with only $E=0$ states have a partition function which is temperature-independent and equal to the dimension of the Hilbert space, which will be infinite in a local QFT, and we will not consider this pathological case in what follows. To have an infinitely degenerate spectrum is not necessarily problematic if we are in flat space, because thermodynamics is only supposed to be finite in finite volume and to become extensive in the large volume limit where everything becomes proportional to volume. One can therefore ask whether the infinite degenerate finite energy eigenstates can be resolved with the help of an IR regulator. Here, there are two possibilities. If the IR regulator does preserve the Carroll algebra, it will typically make the spectrum of $P$ discrete rather than continuous. However, since the energy does not depend on $P$ at all, energy levels remain 
infinitely degenerate, leading to a divergent partition function even in finite volume. Notice that for standard quantum systems the 
energy will typically always depend non-trivially on $P$ and this pathology does therefore not arise. 

It is also possible that the IR regulator breaks the Carroll algebra. If we denote the IR regulator by some length scale $L$, energies can depend non-trivially on $L$ as $E=E_0 + L^{-\alpha}f(P) + \ldots$ with $\alpha>0$ so that we recover the infinitely degenerate 
spectrum in the limit $L\rightarrow \infty$ where we remove the IR regulator. One also expects that the IR regulator makes the 
momenta discrete in units of $1/L$ so we will write $P=n/L$. If there are $d$ momenta this yields a contribution to the partition function which heuristically looks like
\be
Z \ni L^d \sum_{n\in {\mathbb Z}^d} e^{-\beta ( E_0 + L^{-\alpha}f(n/L) + \ldots) }  \,   .
\ee
The prefactor is what in the usual case gives rise to the extensive behavior of the free energy
In order for the sum over $n$ to be regulated by $f(n/L)$ we would need to introduce a new temperature 
$\beta^{\ast} =  L^{-\alpha} \beta$ and keep this fixed as $L\rightarrow \infty$. This would however make the 
partition function vanish due to the factor $e^{-\beta E_0}$. Even if we would ignore this fact, the resulting theory
would no longer be extensive due to the extra $L$ scaling in the temperature. This extra $L$ scaling would effectively drive the theory to infinite temperature in the $L\rightarrow 0$ limit. 

Let us exemplify this starting with the free massive relativistic particle with Hamiltonian $H={\sqrt{c^2{\vec p}^{\,2}+m^2c^4}}$. In the Carroll limit $c\to 0$ with $E_0=mc^2$ fixed -- one can call this the electric limit since the one particle spectrum has non-zero energy -- we can make the expansion 
\begin{equation}
    H=E_0+\frac{1}{2}\frac{c^2\vec{p}^{\,2}}{E_0}+\cdots \ .
    \end{equation}
    The second term in the Hamiltonian breaks Carroll symmetry, but it vanishes in the $c\to 0$ limit. It can be used as a regulator and
    we can now compute the partition function quite easily. There is no real need to discretize the momenta and we find
    \begin{equation}\label{regulatedZrel_particle}
Z=\frac{V}{h^d}e^{-\beta E_0}\Big(\frac{2\pi E_0}{\beta c^2}\Big)^{d/2}\ .
    \end{equation}
    The result is diverging as $1/c^d$ in the Carroll limit, as expected~\footnote{If one would rescale the temperature as $\beta^*=\beta c^2$,  then $e^{-\beta E_0}=e^{-\beta^*E_0/c^2}\to 0$, which would make the partition function vanish for any finite $\beta^*$, and so no good thermodynamics.}. In terms of the dimensionless quantities
    \begin{equation}\label{defxy}
        x\equiv \frac{\beta h c}{R}\ ,\qquad y\equiv \beta E_0\ ,
    \end{equation}
   with length scale $R^d\equiv V$, we can write the partition function as 
\begin{equation}\label{regulatedZrel_particle_xy}
Z=x^{-d}e^{-y}(2\pi y)^{d/2}\ .
    \end{equation}
    The Carroll limit can now be taken on the dimensionless quantity $x \to 0$, and so it has a pole of order $d$. More details on the Carroll limit of relativistic particles are given in \ref{sec:Carrollmicrogas}.

While the above argument is admittedly rather sketchy, the general structure of the electric and magnetic theories that we described above implies that in both cases energy levels remain infinitely degenerate even in finite volume leading to pathological thermodynamics. As far as a possible relation to flat space holography goes, however, this may be a feature rather than a bug
because there is no finite temperature of flat space either. 

One can also ask whether there are other ways to regulate the infinities in the finite temperature partition functions. For the general electric theories it is not clear how to do that, but for a general magnetic theory one can do this as follows. Since the Hamiltonian is $H=-\int {\rm d}^d x\, {\cal L}(\phi)$, the canonical partition function can be expressed as (notice that in our conventions ${\cal L}$ is negative definite)
\be \label{magnpart}
Z={\rm Tr}(e^{-\beta H}) = \int D\Phi(x) \langle \phi(x) | e^{\beta \int {\rm d}^d x\, {\cal L}(\phi)} | \phi(x) \rangle = 
\int D\Phi(x) e^{\beta \int {\rm d}^d x\, {\cal L}(\phi)} \, .
\ee
In other words, the partition function of the magnetic theory is equal to the Euclidean partition function of theory $-{\cal L}$ seen as a Euclidean theory, seen as a theory in its own right, with a prefactor $\beta$. It is interesting that the temperature shows up
in the prefactor and not as the periodicity of an imaginary time direction. In fact, this description is somewhat reminiscent of stochastic quantization, where Euclidean theories are viewed as finite temperature statistical systems in one-dimension higher, and with Carrollian theories being a concrete realization of the higher-dimensional theory. While the regularized Euclidean partition function can potentially be computed, it is not clear whether this regularization spoils the microscopic thermodynamics interpretation, nor is there any a priori reason why the thermodynamics obtained from the regularized partition function should be compatible with Carroll symmetry. We will explore this issue in a simple scalar example in some more detail below.

We have certainly not exhausted all possibilities in the above. For theories with scale symmetry (and more generally for Carroll theories which have more symmetries beyond the Carroll algebra) one could consider partition functions that are not based on the
Hamiltonian but on the generator of scale symmetry. This is precisely what we do when we consider standard CFT's on the plane, where the dilatation generator maps to the time translation on the cylinder under a conformal transformation. For Carroll theories there
are several issues with this perspective: it is not clear the generator of scale symmetry has a discrete spectrum, it is not clear whether we
can map the plane to the cylinder in such theories, and the Hamiltonian of Carrollian theories on the cylinder is still infinitely degenerate. An object that might have a better chance of being well-defined is to write a function which counts the number of 
independent local operators with a given scaling dimension, $Z\sim \sum_{\cal O} N_{\Delta_{\cal O}} e^{-\beta \Delta_{\cal O} }$. Without a suitable operator-state correspondence, where scaling dimensions are somehow related to time translations in a possibly different geometry, it is not clear whether this function has a thermodynamics interpretation, but it appears to be well-defined, and it would be interesting to study it further. 

\subsubsection{Scalar example}

To illustrate some of the issues in finding microscopic Carroll thermodynamics, we will consider the explicit example of a free
scalar field in two dimensions with an action of the form 
\begin{equation}
    S=\int {\rm d}^2 x ( a^2 \dot{\phi}^2 -b^2 (\partial_x\phi)^2 - m^2 \phi^2)\ ,
\end{equation}
where we take the periodicity of $x$ to be $2\pi R$. 
We have introduced two parameters $a$ and $b$. For the relativistic scalar, we have $a=1/c$ and $b=1$. The electric theory can be obtained in the limit $b\rightarrow 0$, the magnetic theory is obtained in
the limit $a\rightarrow \infty$ which can be seen easily from rewriting the Lagrangian with an auxiliary field $\chi$ with $L_\chi=\chi \dot\phi-\frac{1}{4a^2}\chi^2$. The magnetic limit is a bit subtle as we will see below.

Quantization of the theory is straightforward. The partition function of the theory is
\begin{equation} \label{j1}
    Z = e^{-\beta E_C} \prod_{k\in\mathbb{Z}} \frac{1}{1-e^{-\beta E_k}}\ ,
\end{equation}
where the energy of each harmonic oscillator is
\begin{equation}
    E_k = \sqrt{\frac{ b^2 k^2/R^2 + m^2}{a^2}}\ ,
\end{equation}
and $E_C$ is the vacuum Casimir energy of the theory on the cylinder, which is equal to a suitably regulated sum of zero-point energies, $E_C=\frac{1}{2} \sum_{k} E_k$. It is not clear how important this zero point energy is, but interestingly it is UV divergent for nonzero masses. 
One way to see this is to expand $E_k$ in powers of $m^2$, so we can write
\begin{equation}
    E_C = -\frac{b}{12 Ra} +\frac{m}{2a} +  \frac{R m^2}{2 a b} \sum_{k>0} \frac{1}{k} -\frac{R^3 m^4}{8 a b^3 } \zeta(3) + \ldots
\end{equation}
where we used $\sum_{k>0} k =-1/12$. The sum $\sum_{k>0} \frac{1}{k}$ is UV divergent and can for example also not be zeta function regularized. So 
\begin{equation}
    E_C = -\frac{b}{12 Ra} +\frac{m}{2a} +  \frac{R m^2}{2 a b} \log(R\Lambda/b) -\frac{R^3 m^4}{8 a b^3 } \zeta(3) + \ldots
\end{equation}
where $\Lambda$ is some energy UV cutoff. We will ignore the Casimir energy for the time being. 

Before taking any limit, we notice that the partition function only depends on the dimensionless quantities $x=b\beta/Ra$ and 
$y=\beta m/a$. 

\begin{itemize}
\item Electric case: In the electric case with $b\rightarrow 0$, the partition function must end up being a function of $y$ alone. It can therefore not depend on $R$ and one can therefore not obtain extensive thermodynamics. 
\item Magnetic case: In the magnetic case with $a\rightarrow\infty$, the ration $x/y$ remains finite so the partition function
can only be a function of this ratio. But this ratio does not depend on temperature, and we would therefore end up with a 
temperature-independent partition function. We will confirm that this indeed is what happens if we define the partition function 
through the Euclidean path integral (\ref{magnpart}).
\item  Conformal case: As an aside, in the conformal limit with $m\rightarrow 0$ the partition function can only be a function of $x$. In order for it to be extensive, $\log Z$ must be linear in $R$ and therefore proportional to $1/x$. This is indeed the correct answer for a 2d CFT, where for large $R$ and/or high temperature we get the Cardy answer $\log Z \sim 1/x$.
\end{itemize}

One can try to define modified electric and magnetic limits in which one does not only send $a\rightarrow \infty$ or $b\rightarrow 0$ but at the same time scales some other parameters in the theory as well. If both $x$ and $y$ remain finite in this limit one is not really taking a limit but merely redefining the units of the theory, so that case is not very interesting. If we also demand that the theory is extensive (so $\log Z$ is linear in $R$) the logarithm of the partition function must be proportional to $1/x$.

To explore the constraint of extensivity, we first consider the standard $R\rightarrow \infty$ limit of the theory in which we write
\begin{equation}
    \log Z = \sum_{k,n>0} \frac{1}{n} e^{-n\beta E_k}\ ,
\end{equation}
and approximate the sum by an integral to extract the term linear in $R$. This yields
\begin{equation} \label{extpart}
    \log Z \simeq \sum_{n>0} \frac{Rm}{nb} f(\frac{\beta n m}{a}) = \sum_{n>0} \frac{y}{nx} f(ny)
\end{equation}
where 
\begin{equation}
    f(\xi) = \int dz e^{-\xi\sqrt{z^2+1}} =\frac{2}{\xi} + \ldots
\end{equation}
and where we also included the leading term in the small $\xi$ expansion. Therefore, the partition function is approximately equal to 
\begin{equation}
    \log Z \simeq \frac{Ra\pi^2}{3b\beta} = \frac{\pi^2}{3x}
\end{equation}
which is valid in the limit where $ny =\beta n m/a$ becomes small and just the standard Cardy answer for a $c=1$ theory. 
We clearly see that the partition function in the CFT regime diverges in the magnetic limit $a\rightarrow \infty$ and also in the electric limit $b\rightarrow 0$.

Turning back to (\ref{extpart}), one can verify that $\sum_{n>0} \frac{y}{n} f(ny)$ is constant for $y\rightarrow 0$ and decays
exponentially for large $y$. There is no other scaling regime where its behavior is different and well-behaved
which is what would be required for non-trivial extensive Carrollian thermodynamics. 

We therefore find no evidence for the existence of any limit, even a rather contrived one, of the partition function which 
yields extensive and Carroll invariant thermodynamics, and both the standard electric and magnetic limits seem to give rise to somewhat pathological answers. 

We conclude this section by comparing the magnetic limit of the partition function to the Euclidean path integral representation of the
magnetic partition function in (\ref{magnpart}). When $a\rightarrow \infty$ the expression in (\ref{j1}) becomes
\be \label{j2}
Z\sim \prod_k \frac{1}{\beta E_k} = \frac{1}{y} \left( \prod_{k>0} \frac{1}{x^2 k^2} \right)
\left( \prod_{k>0} \frac{1}{1+\frac{y^2}{k^2 x^2} } \right) = \frac{2\pi}{x}  \left( \prod_{k>0} \frac{1}{x^2 k^2} \right)
\frac{1}{2\sinh \pi y /x} 
\ee
where in the last step we use the infinite product representation of the $\sinh$ function. In the $a\rightarrow 0$ limit $x\rightarrow 0$ so the $x$-dependent formal prefactor in the partition function is badly divergent but it does not depend on the variable $y/x$ which we keep fixed and we could decide to remove this prefactor. This would leave us with a finite partition function
\be Z \sim \frac{1}{2\sinh \pi y /x} 
\ee
which also happens to be the partition function of an ordinary harmonic oscillator. This harmonic oscillator result is precisely what one would obtain from the expression (\ref{magnpart}), as we recognize that in the magnetic case this 1d Euclidean theory is simply the Euclidean theory of a quantum mechanical harmonic oscillator. If we translate variables more precisely we get
 $b^2 \beta=m_{ho}$, $2\pi R=\beta_{ho}$ and $\beta m^2 = m_{ho} \omega_{ho}^2$, with $\beta_{ho}, m_{ho}$ and $\omega_{ho}$ standard harmonic oscillator variables. 
The partition function is then equal that of an ordinary harmonic oscillator so that
\begin{equation}
    Z = \frac{1}{2 \sinh \beta_{ho} \omega_{ho} /2} =  \frac{1}{2 \sinh \pi Rm/b} = \frac{1}{2\sinh \pi y /x} 
\end{equation}
which agrees with (\ref{j2}) and indeed does not depend on the temperature $\beta$ as this just appears as a prefactor in the Euclidean action in agreement with our general scaling symmetry analysis. While one could argue that this is the ``correct" partition function for the magnetic theory, it does not give rise to Carrollian thermodynamics, which may be due to the implicit regularization which has been employed and which apparently breaks the Carroll symmetry of the problem.

\subsubsection{Stress tensor of the scalar example}

For completeness, we point out a few other peculiar features of the scalar example. As we pointed out in \cite{deBoer:2017ing} the stress-tensor of a general non-boost invariant fluid (we assumed the existence of a consistent thermodynamic description to obtain this form) reads
\be \label{nbstress}
T^t{}_t=-{\cal E}, \quad T^i{}_t = -({\cal E}+P)v^i, \quad  T^t{}_j ={\cal P}_j,\quad T^i{}_j = P\delta^i_j + v^i {\cal P}_j \ .
\ee
For $v^i\neq 0$ Carroll symmetry implies ${\cal E}+P=0$ as an additional constraint. If we also assume that the momentum density ${\cal P}_j$ is proportional to the velocity then the statement ${\cal E}+P=0$ also holds when $v^i=0$ \cite{deBoer:2021jej}. If we do not make that assumption and put $v^i=0$ the requirement ${\cal E}+P=0$ no longer applies. 

The stress tensor of the electric theory was given in (4.19) in \cite{deBoer:2021jej} and that of the magnetic theory in (4.29) in that same paper. One sees that the electric theory is an example of a stress tensor of the form (\ref{nbstress}) with $v^i=0$ but with ${\cal E}+P\neq 0$. Carroll symmetry therefore does not impose any additional constraints on the partition function in this case.
The stress tensor of the magnetic theory is also of the form (\ref{nbstress}) but with non-zero velocity proportional to the gradient of the scalar field and indeed ${\cal E}+P=0$.

Some of these observations may sound contradictory, however since neither the electric nor the magnetic theory has a well-defined thermodynamic description, the assumptions that were used to derive (\ref{nbstress}) do not apply anyway. 

\subsubsection{Two theories with BMS${}_3$ symmetry}

We briefly consider the conformal electric and magnetic theories in two dimensions. The symmetries for the massless electric theory were given in (\ref{j3}). The subset of transformation of the form
\be\label{j4}
\delta \phi = a(x) \partial_t \phi + t b'(x) \partial_t \phi + b \partial_x \phi\ ,
\ee
form a BMS${}_3$ algebra. The modes of $a$ are usually denoted by $M_m$ and those of $b$ by $L_m$. They form the algebra
\begin{eqnarray}
[  L_n,L_m ] & = & (n-m)L_{m+n} + c_L (n^3-n)\delta_{m+n,0} \nonumber \\\mbox{}
[  L_n,M_m  ] & = & (n-m)M_{m+n} + c_M (n^3-n)\delta_{m+n,0} \nonumber \\\mbox{}
[   M_n,M_m  ] & = & 0\ ,
\end{eqnarray}
with $c_M=0$ and $c_L=2$ \cite{Hao:2021urq}.

The $2d$ massless magnetic theory with
\be
S=\int {\rm d}t {\rm d}x \, (\chi \partial_t \phi - \frac{1}{2}( \partial_x\phi)^2)\ ,
\ee
also has a BMS${}_3$ symmetry given by the same transformation (\ref{j4}) for $\phi$ together with 
\be
\delta \chi = (a(x)  + t b'(x))\partial_t \chi +\partial_x (b(x) \chi) + (a'(x)  + t b''(x))\partial_x \phi .
\ee
The relevant conserved charges are given in (\ref{j5}). In particular, the charge at $t=0$ for the $b$-transformations is 
$Q\sim \int {\rm d}x\, b \chi \partial_x \phi$ so that the stress tensor which generates the Virasoro transformations is $T\sim \chi \partial_x \phi$. We can think of $\chi$ and $\phi$ as a bosonic beta-gamma system where $\chi$ has dimension one and $\phi$
has dimension zero. The central charge of this beta-gamma system is $c=2$. The generators $M$ are the modes of the spin two operator
$\partial_x \phi \partial_x \phi$, and interpreting this also in terms of a beta-gamma system we see that there is no central term 
between the stress-tensor and this spin-two current. We conclude that $c_L=2$ and $c_M=0$ just like in the electric case.

It is interesting to consider the decomposition of the Hilbert spaces of the electric and magnetic theory in terms of representations of the BMS${}_3$ algebra. Our theories are unitary, and since the BMS${}_3$ does not admit unitary highest-weight representations, the electric and magnetic theory will not contain such representations. For completeness, we briefly review the argument why BMS${}_3$ does not have unitary highest weight representations. 

Consider a highest weight state with $L_0=\Delta$ and $M_0=\xi$. At the first excited level 
there are two states obtained by acting with $L_{-1}$ and $M_{-1}$. The matrix of inner products is (see (2.14) in
\cite{Bagchi:2019unf}) is
\be 
\left( \begin{array}{cc} 2\Delta & 2 \xi \nonumber \\ 2 \xi & 0 \end{array} \right)\ .
\ee
This matrix has one positive and one negative eigenvalue for $\xi\neq 0$ because the determinant is $-4\xi^2<0$. Therefore 
the inner product is not positive definite unless $\xi=0$. If $\xi=0$, there is a null vector obtained by acting with
$M_{-1}$. Similarly, at higher levels, all states which involve at least one $M$ raising operator are null states when $\xi=0$.
So we either have negative norm states, or we have a unitary highest weight representation of Virasoro where all $M_n$ map
all states to zero. The latter case is a bit pathological but it would be an example where all states have zero energy.

There is a different way to see that there cannot be unitary representations with $\xi\neq 0$. If we look at the action of $M_0$
on a basis of states of a given level we find a triangular matrix with $\xi$ on the diagonal and only non-trivial upper triangular
matrix elements. See (3.11) and below in \cite{Bagchi:2019unf}. Such a matrix has only one proper eigenvector (similar to matrices in Jordan normal form), 
but if we had a positive definite inner product with respect to which $M_0$ would be hermitian, we should be able to 
find a complete basis of eigenvectors. Therefore, there cannot exist a positive definite inner product\footnote{Though maybe not obvious, triangular matrices can be self-adjoint with respect to mixed signature inner products.}.

In the electric theory, the conserved currents whose modes correspond to $L_m$ and $M_m$ are $M(x)\sim \dot{\phi}^2$ and 
$L(x)\sim \dot{\phi} \partial_x (1-t\partial_t) \phi$ which are both time-independent on-shell. On-shell we can write
\be
\phi=\gamma(x) + t\beta(x), \quad M(x)\sim \beta(x) \beta(x), \quad L(x)\sim \beta \partial_x \gamma\,.
\ee
Just like in the magnetic case, the structure is reminiscent of a 
beta-gamma system with a field of dimension one and a field of dimension zero. The difference with the magnetic case is that
here $M$ is expressed in terms of the weight-one degree of freedom, whereas in the magnetic case it was expressed in terms of the weight-zero degree of freedom. 

To write an explicit basis, we are going to decompose the fields in Fourier modes along the spatial $S^1$. The non-zero modes come in pairs with opposite momenta along the circle, and we will decompose each pair in a radial and an angular variable. Moreover, we will 
Fourier transform wavefunctionals of the modes with respect to the angular variables of the non-zero modes in order to isolate the eigenvalues under rotations which corresponds to the $L_0$ generator. For each pair of non-zero spatial Fourier modes, this will yield sets of states of the form $|p,m\rangle$ which 
represents the wave function $\delta(|z|-p)e^{im\phi}$ on the complex plane, with inner product $\langle p,m | p',m'\rangle=2 \pi p \delta(p-p')\delta_{m,m'}$. The reason why we use these somewhat peculiar basis states is that the electric and magnetic theory resemble a free particle rather than a harmonic oscillator. The most natural representation of the Hilbert space therefore uses states with continuous coordinate or momentum labels rather than raising and creation operators. We now summarize the result that one obtains for the spectrum of $L_0$ and $M_0$. The Hilbert
space will be given by states of the form
\be 
{\cal H} = |p_0\rangle \otimes \otimes_{k>0} |p_k,m_k\rangle
\ee
using basis states as described above, with $p_0\in \mathbb{R}$, $p_k\geq 0$, and $m_k\in \mathbb{Z}$. For the magnetic theory, the numbers will refer to the modes of the field itself, and for the electric theory the modes will refer to momentum conjugate to the scalar, but we will keep the same notation for either case.

We then find that 
\be L_0^{\rm electric} = L_0^{\rm magnetic} = \sum_{k>0} k m_k\ ,
\ee
and
\begin{eqnarray}
    M^{\rm electric}_0 & = & p_0^2 + \sum_{k>0} p_k^2\ , \nonumber \\
    M^{\rm magnetic}_0 & = & \sum_{k>0} k^2 p_k^2\ .
\end{eqnarray}
One could consider the contribution of the momentum sector to a partition function ${\rm tr} (e^{-\beta M_0 + i \theta L_0})$ 
which keeps track of both quantum numbers. This yields (up to some numerical factors)
\begin{eqnarray}
    Z_{\rm electric} & = & \frac{1}{\sqrt{\beta}} \prod_{k>0} \frac{\delta(k \theta)}{\beta} \nonumber\ , \\
    Z_{\rm magnetic} & = & \int dp_0 \prod_{k>0} \frac{\delta(k \theta)}{k^2 \beta} \ .
\end{eqnarray}
Both partition functions are rather pathological in agreement with our earlier observations that Carroll partition functions tend to not be well-behaved. 

If we focus only on the eigenvalues of $L_0$ we can try to formally separate $m_k>0$ and $m_k<0$ by introducing $q=e^{i\theta}$ and
$\tilde{q}=e^{-i\theta}$ and by writing
\be \label{weirdchar}
{\rm Tr}   (e^{i\theta L_0}) = \prod_{k>0} \frac{1-q^k \tilde{q}^k}{(1-q^k)(1-\tilde{q}^k)}\ .
\ee
We could subsequently consider regulating this expression by taking $|q|<1$ and $|\tilde{q}|<1$. It is however unclear from the present perspective what the physical meaning of this procedure is. With $L_0$ generating a compact $U(1)$, a generalized character would naturally be a distribution on the group (as our delta-functions above) and the introduction of $q$ and $\tilde{q}$ seems somewhat arbitrary and not in line with generalized group characters. 

Expressions similar to (\ref{weirdchar}) which are reminiscent of standard free boson/Virasoro characters appear when computing characters for the BMS algebra in non-unitary highest-weight representations \cite{Bagchi:2019unf,Hao:2021urq,Yu:2022bcp} and also when considering characters for induced representations \cite{Oblak:2015sea} and in the partition function for thermal flat space \cite{Barnich:2015mui}. In the latter two cases the characters are formally infinite and require a regularization similar to the one above. The lack of suitable Carroll thermodynamics in our examples suggests that any regularization which makes the partition function well-behaved will also automatically break the Carroll symmetry. It therefore remains unclear what the precise physical meaning of these regulated characters and corresponding partition functions is.

Finally, we notice that the electric and magnetic theory do not appear to enjoy a form of modular invariance due to the asymmetric treatment of space and time. If anything, the electric and magnetic theory could be related to each other under a modular transformation, but such a relation is not manifest in our ill-defined product formulas for the partition function. 

\section{Carroll geometry and energy momentum tensor from small $c$ expansion  \label{sec:carrollgeometryEM}}

In this section we first review how Carroll geometry arises from expanding Lorentzian geometry around $c=0$. For a primer on non-Lorentzian gravity theories we refer to \cite{Bergshoeff:2022eog}.
We then discuss the dynamics of Carroll gravity and its coupling to generic Carrollian field theories, including 
the notion of a Carrollian energy momentum tensor. For the gravitational part we will use the results of \cite{Hansen:2021fxi} 
for the action and equations of motion of electric/magnetic Carroll gravity obtained from the ultra-local expansion of General Relativity. See also \cite{Henneaux:2021yzg} for work on this in the Hamiltonian formalism. 

\subsection{Ultra-local expansion of General Relativity}\label{sec:carrollgeometry}

Following \cite{Hansen:2021fxi}, we start by briefly reviewing  the geometry and dynamics obtained from expanding Lorentzian geometry around $c=0$, yielding Carroll geometry to leading order. This parallels the non-relativistic expansion around $c=\infty$  \cite{VandenBleeken:2017rij,Hansen:2018ofj,Hansen:2020pqs}
(see \cite{Hartong:2022lsy} for a review). 

Consider the expansion of a Lorentzian metric $g_{\mu\nu}$ around $c=0$, which is called the Carrollian or ultra-local expansion. The starting point is to split time and space in a covariant way by writing 
\begin{equation}
\label{eq:gexp} 
    g_{\mu\nu}=-c^{2}T_{\mu}T_{\nu}+\Pi_{\mu\nu} \ , 
\end{equation}
 where $T_\mu$ is the time-like vielbein
 and $\Pi_{\mu \nu}=\delta_{ab}E^{a}_{\mu}E^{b}_{\nu}$ the spatial part of the metric expressed in terms of
 the spatial vielbeins $E^{a}_{\mu}$, with Latin indices running over the tangent space spatial directions only. Local (boost) Lorentz transformations correspond in this form to 
 \begin{equation}\label{eq:LLtrafo}
     \delta T_\mu =  c^{-2}{\Lambda}_a E^a_\mu\,,\qquad \delta \Pi_{\mu\nu} =  \Lambda_a T_\mu E^a_\nu+ \Lambda_a T_\nu E^a_\mu\,. 
 \end{equation}
  For use below we note that 
  $\sqrt{-g}=cE$ with $E = {\rm det} (T_\mu, E^a_\mu)$, which is invariant
  under local Lorentz boosts. The inverse metric can be written as 
  \begin{equation}\label{eq:inversemetric}
    g^{\mu\nu}=-c^{-2}T^{\mu}T^{\nu}+\Pi^{\mu\nu} \ , 
\end{equation}
where 
\begin{equation}
    T^\mu T_\mu=-1\,,\qquad T^\mu\Pi_{\mu\nu}=0\,,\qquad \Pi^{\mu\nu}T_\nu=0\,,\qquad\Pi^{\mu\rho}\Pi_{\rho\nu}=\delta^\mu_\nu+T^\mu T_\nu\,.
\end{equation}

 By assumption the fields introduced above start at order $c^0$ in a Taylor expansion around $c=0$, so that the vielbeins and their inverses may be expanded as\footnote{As in
\cite{Hansen:2018ofj,Hansen:2020pqs} we make the self-consistent choice of only even powers in $c$.}\footnote{We note that the field $M^\mu$ was already introduced in Ref.~\cite{Hartong:2015xda} 
 using the relation between Carrollian geometry and null hypersurfaces  (in order to construct appropriate Carroll boost invariants). Here we see that it also arises naturally from the small $c$ expansion of a Lorentzian metric \cite{Hansen:2021fxi}.}
\begin{equation}
    T_{\mu}
    =
    \tau_{\mu}
    +
    \mathcal{O}(c^{2})
    \,,
    \quad
    E^{a}_{\mu}
    =
    e^{a}_{\mu}
    +
    c^{2}\pi^{a}_{\mu}
     +
    \mathcal{O}(c^{4})  
    \,,
\end{equation}
\begin{equation}
    T^{\mu}
    =
    v^{\mu} + c^{2} M^{\mu} 
    +
    \mathcal{O}(c^{4})
    \,,
    \quad
    E_{a}^{\mu}
    =
    e_{a}^{\mu}
    +
    \mathcal{O}(c^{2})  
    \,.
\end{equation}
Expanding the local Lorentz transformations \eqref{eq:LLtrafo} leads to the local Carroll boost transformations (arising from $\Lambda_a=O(c^2)$ which follows from the requirement that the form of the $c=0$ expansion is preserved by the transformation \eqref{eq:LLtrafo})
\begin{eqnarray}
&&\delta\tau_{\mu}=\lambda^{a}e^{a}_{\mu}\,,\qquad
\delta  e^{a}_{\mu} =0\,,\qquad \delta\pi^{a}_{\mu}=\lambda^{a}\tau_{\mu}\,,\\
&&\delta v^{\mu}= 0\,,\qquad \delta M^\mu = \lambda^a e^\mu_{a}\,, \qquad\delta e_{a}^{\mu} = \lambda_a v^\mu\,.
\end{eqnarray}
The corresponding expansion for the metric and its inverse is then
\begin{equation}\label{exp-metric}
    g_{\mu\nu}
    =
    h_{\mu\nu}
    +
    c^{2}
    (
        \Phi_{\mu\nu}
        -
        \tau_{\mu}\tau_{\nu}
    )
    +
    \mathcal{O}(c^{4})
    \,,
\end{equation}
\begin{equation}\label{exp-invmetric}
    g^{\mu\nu}
    =
    -
    \frac{1}{c^{2}}v^{\mu}v^{\nu}
    +
    h^{\mu\nu}
    -
    2v^{(\mu}M^{\nu)}
    +
    \mathcal{O}(c^{2})
    \,,
\end{equation}
where we have defined
\begin{equation}
h_{\mu\nu}:=\delta_{ab}e^{a}_{\mu}e^{b}_{\nu}\,,\qquad
h^{\mu\nu}:=\delta^{ab}e_{a}^{\mu}e_{b}^{\nu}\,,\qquad 
\Phi_{\mu\nu}:=2e^{a}_{(\mu}\pi^{a}_{\nu)}\,.
\end{equation}
Note that $\Phi_{\mu \nu}$ obeys the property that $v^{\mu}v^\nu\Phi_{\mu\nu}=0$. 

As a direct consequence of local Lorentz invariance, one may check that the terms appearing at a given order in $c^2$ in  both the metric and inverse metric given above are invariant under Carroll boosts.
For use below we record here that while
$h_{\mu \nu}$ is Carroll invariant we have that 
\begin{equation}
    \delta h^{\mu\nu}=2v^{(\mu} e^{\nu)}_a\lambda^a\,,
\end{equation}
which is consistent with the 
completeness relation  $-v^\mu\tau_\nu+h^{\mu\rho}h_{\rho\nu}=\delta^\mu_\nu$.
We also introduce the following Carroll boost invariant combinations 
\begin{eqnarray}
    \bar h^{\mu\nu} & = & h^{\mu\nu}-M^\mu v^\nu-M^\nu v^\mu\,,\\
    \hat\tau_\mu & = & \tau_\mu-h_{\mu\nu}M^\nu\,,\\
    \bar\Phi_{\mu\nu} & = & \Phi_{\mu\nu}-\tau_\mu\tau_\nu\,.
\end{eqnarray}

Next we introduce a Carroll metric-compatible `connection' $\tilde\nabla_\mu$
satisfying
\begin{equation}
  \label{eq:tangent-carroll-compatibility}
  \tilde\nabla_\mu v^\nu = 0,
  \qquad
  \tilde\nabla_\rho h_{\mu\nu} = 0.
\end{equation}
We will assume that the torsion is purely intrinsic, i.e. expressed in terms of $K_{\mu\nu}$ which is defined as
\begin{eqnarray}
K_{\mu\nu}=-\frac{1}{2}\mathcal{L}_v h_{\mu\nu}\,.
\end{eqnarray}
This tensor is also purely spatial, since it satisfies $v^\mu K_{\mu\nu}=0$. We put the word connection in quotation marks because the ones we work with are not Carroll boost invariant. A convenient choice is \cite{Hartong:2015xda,Bekaert:2015xua,Hansen:2021fxi} 
\footnote{This connection can be obtained  \cite{Hansen:2021fxi} from the small speed of light expansion of the Levi-Civita connection. It also appeared in \cite{Bekaert:2015xua} and is a special case of the general class of Carroll connections satisfying the compatibility requirements~\eqref{eq:tangent-carroll-compatibility}, which was determined in~\cite{Hartong:2015xda,Bekaert:2015xua}.}
\begin{align}
  \label{eq:minimal-carroll-tangent-connection}
  \tilde{\Gamma}^\rho_{\mu\nu}
  &= - v^\rho \pd_{(\mu} \tau_{\nu)} - v^\rho \tau_{(\mu} \LL_v \tau_{\nu)}
  \\
  &{}\qquad\nonumber
  + \frac{1}{2} h^{\rho\lambda} \left[
    \pd_\mu h_{\nu\lambda} + \pd_\nu h_{\lambda\mu} - \pd_\lambda h_{\mu\nu}
  \right]
  - h^{\rho\lambda} \tau_\nu K_{\mu\lambda}.
\end{align}
Note that this connection is constructed to have only intrinsic torsion 
\begin{equation}
  2 \tilde{\Gamma}^\rho_{[\mu\nu]}
  = 2 h^{\rho\lambda} \tau_{[\mu} K_{\nu]\lambda}.
\end{equation}
This reflects the result that the intrinsic torsion of a Carroll metric-compatible connection is  determined by the extrinsic curvature $K_{\mu\nu}$~\cite{Figueroa-OFarrill:2020gpr}.

Using the expansion of the metric, one can obtain the expansion of the
Riemann curvature tensor in GR and subsequently expand the Einstein-Hilbert action 
(see \cite{Hansen:2021fxi} for details). 

From the expansion of the Levi-Civita connection we derive the leading order behavior of the Ricci tensor $R_{\mu\rho}=R_{\mu\sigma\rho}{}^\sigma$ where the Riemann tensor is
\begin{equation}
    R_{\mu\nu\rho}{}^\sigma=-\partial_\mu\Gamma^\sigma_{\nu\rho}+\partial_\nu\Gamma^\sigma_{\mu\rho}-\Gamma^\sigma_{\mu\lambda}\Gamma^\lambda_{\nu\rho}+\Gamma^\sigma_{\nu\lambda}\Gamma^\lambda_{\mu\rho}\,.
\end{equation}
A straightforward calculation then gives
\begin{equation}
    R_{\mu\rho}=O(c^{-2})\,,\qquad v^\mu R_{\mu\rho}=O(1)\,.
\end{equation}
This implies that the Ricci scalar is at most of order $c^{-2}$ so that for the Einstein tensor we obtain the same behaviour as for the Ricci tensor, namely
\begin{equation}\label{eq:Einsteintensorc}
    G_{\mu\rho}=O(c^{-2})\,,\qquad v^\mu G_{\mu\rho}=O(1)\,.
\end{equation}
This means that the leading order behavior of $G_{\mu\nu}$ is a $c^{-2}$ term that is orthogonal to $v^\mu$, i.e. that is purely spatial.
In the next subsection  we will use these results together with the Einstein equations to infer what the behaviour of the energy-momentum tensor should be as $c\rightarrow 0$.

Using the expansions reviewed in this section, one can compute the corresponding small $c$ expansion of the Einstein-Hilbert action. This was done in \cite{Hansen:2021fxi} and in Section \ref{sec:Cargra1} we will collect the results that are needed for the purposes of this paper.

\subsection{Expanding the energy-momentum tensor around $c=0$}\label{sec:emtfromcarrolllimit}

Now that we have decomposed the metric tensor around $c=0$ in Section \ref{sec:carrollgeometry}, we are able to consider the expansion of a general relativistic energy-momentum tensor around $c=0$ to obtain the leading order non-trivial Carroll energy-momentum tensor. 
This will be relevant in Section \ref{sec:Cargra} when we consider some examples of solutions of Carrollian gravity coupled 
to matter. Moreover, it will be the starting point in Section \ref{sec:emtcarroll} when we consider Carrollian perfect fluid stress tensors as obtained from the small $c$ expansion of relativistic perfect fluids. 

Consider the Lagrangian
\begin{eqnarray}
\mathcal{L}=\frac{c^3}{16\pi G_N}\sqrt{-g} R+\mathcal{L}_{\text{mat}}\,.
\end{eqnarray}
Note that we use $c^3$ (as opposed to $c^4$) because we have a put a factor of $c^2$ into $g_{\mu\nu}$ which amounts to rescaling $\sqrt{-g}$ by a factor of $c$. The Einstein equation then gives $G_{\mu\nu}=\frac{8\pi G_N}{c^4}T_{\mu\nu}$ where 
\begin{equation}\label{eq:EMTvarg}
\delta\mathcal{L}_{\text{mat}}=\frac{1}{2}c^{-1}\sqrt{-g}T^{\mu\nu}\delta g_{\mu\nu}\,.
\end{equation}

In the previous section we concluded that the Ricci scalar is $O(c^{-2})$. This means that the Einstein--Hilbert Lagrangian is $O(c^2/G_N)$. 
We allow ourselves the possibility that Newton's constant can scale with powers of $c$. 

Consider the Einstein equation with the indices up, i.e. $G^{\mu\nu}=\frac{8\pi G_N}{c^4}T^{\mu\nu}$. We can use \eqref{eq:Einsteintensorc} to write down the general structure of $G^{\mu\nu}$ to leading order in $c$ for an arbitrary geometry. By using the expansion of the inverse metric \eqref{eq:inversemetric} and \eqref{eq:Einsteintensorc} we learn that
\begin{equation}
    G^{\mu\nu}=c^{-4}v^\mu v^\nu G_{(-4)}+c^{-2}G_{(-2)}^{\mu\nu}+c^{-2}v^\mu v^\nu G_{(-2)}+O(1)\,,
\end{equation}
where $G_{(-2)}^{\mu\nu}$ has no $v^\mu v^\nu$ component as we explicitly split that part off and called it $G_{(-2)}v^\mu v^\nu$. The reason for this is that $G_{(-2)}v^\mu v^\nu$ is a subleading correction whereas $G_{(-2)}^{\mu\nu}$ is a leading order term. Since we will only care about leading order terms we will ignore the $G_{(-2)}v^\mu v^\nu$ part. Using the Einstein equations we can now infer what the expansion of $T^{\mu\nu}$ should be around $c=0$ for an arbitrary geometry. We thus conclude that we must have
\begin{equation}\label{eq:EMTc}
T^{\mu\nu}=c^{-N}\left(-\mathcal{T}v^\mu v^\nu+c^2\hat{\mathcal{T}}^{\mu\nu}+O(c^4)\right)\,,
\end{equation}
where $\hat{\mathcal{T}}^{\mu\nu}$ is defined up to the addition of a term proportional to $v^\mu v^\mu$ and where furthermore
\begin{equation}\label{rescaledG}
    G_Nc^{-N}\qquad \text{is independent of $c$ for some integer $N$.}
\end{equation}

We can interpret the objects $\mathcal{T}$ and $\hat{\mathcal{T}}^{\mu\nu}$ appearing in the expansion of the energy-momentum tensor as arising from variation with respect to the Carroll objects $\hat\tau_\mu$ and $h_{\mu\nu}$.
To see this we substitute \eqref{eq:EMTc} into the right hand side of \eqref{eq:EMTvarg} and expand $\delta g_{\mu\nu}$ as
\begin{equation}
    \delta g_{\mu\nu}=\delta h_{\mu\nu}+c^2\delta\bar\Phi_{\mu\nu}+O(c^4)\,.
\end{equation}
This leads to 
\begin{eqnarray}
\delta\mathcal{L}_{\text{mat}} & = & e c^{2-N}\left(-\mathcal{T}v^\mu\delta\hat\tau_\mu+\frac{1}{2}\hat{\mathcal{T}}^{\mu\nu}\delta h_{\mu\nu}+O(c^2)\right)\nonumber\\
& = & e c^{2-N}\left(-\mathcal{T}v^\mu\delta\tau_\mu+\frac{1}{2}\left(\hat{\mathcal{T}}^{\mu\nu}+\mathcal{T}\left[v^\mu M^\nu+v^\nu M^\mu\right]\right)\delta h_{\mu\nu}+O(c^2)\right)\,,\label{eq:derivingemt}
\end{eqnarray}
where $-e^2=\text{det}\,(-\tau_\mu\tau_\nu+h_{\mu\nu})$.
In deriving the above we used that $v^\mu\bar\Phi_{\mu\nu}=\hat\tau_\nu$ so that $\frac{1}{2} v^\mu v^\nu\delta\bar\Phi_{\mu\nu}=v^\mu\delta\hat\tau_\mu$. 

Note that $v^\mu v^\nu\delta h_{\mu\nu}=0$, which follows because $v^\mu h_{\mu\nu}=0$, and hence the response to varying $h_{\mu\nu}$ gives a symmetric $(0,2)$ tensor that is defined up to a part proportional to $v^\mu v^\nu$, which is in agreement with the $c=0$ expansion of the Einstein tensor results above.

The Carroll energy-momentum tensor is then
\begin{equation}\label{eq:CarEMT}
    \left(T_{\text{Car}}\right)^{\mu}{_\nu}=-\mathcal{T}v^\mu\hat\tau_\nu+\hat{\mathcal{T}}^{\mu\rho}h_{\rho\nu}=-\mathcal{T}v^\mu\tau_\nu+\mathcal{T}^{\mu\rho}h_{\rho\nu}\,,
\end{equation}
where
\begin{equation}
    \mathcal{T}^{\mu\rho}=\hat{\mathcal{T}}^{\mu\rho}+\mathcal{T}\left[v^\mu M^\rho+v^\rho M^\mu\right]\,.
\end{equation}
This does not depend on the undetermined $v^\mu v^\nu$ term in the response to varying $h_{\mu\nu}$.
The response to varying $\tau_\mu$ is the energy current which in the case of a Carrollian field theory is of the form $-\mathcal{T}v^\mu$ where $\mathcal{T}$ is the energy density. We see that this current has no components in the spatial vielbein directions. This is a consequence of local Carroll boost invariance. If we would write $-\mathcal{T}^\mu$ for the variation with respect to $\tau_\mu$ (keeping $h_{\mu\nu}$ fixed) then demanding invariance under Carroll boosts $\delta\tau_\mu=\lambda^a e^a_\mu$ forces the spatial projections of $\mathcal{T}^\mu$ to vanish. Hence the most general energy current $\mathcal{T}^\mu$ is of the form $-\mathcal{T}v^\mu$. The Carroll boost Ward identity is
\begin{equation}
    v^\nu h_{\mu\rho}\left(T_{\text{Car}}\right)^{\mu}{_\nu}=0\,.
\end{equation}
i.e. the condition that the energy flux is zero, \cite{deBoer:2017ing,deBoer:2021jej}.
From \eqref{eq:EMTc}, by lowering one index with $g_{\mu\nu}$ and expanding, we can conclude that
\begin{equation}
    T^\mu{}_\nu=c^{2-N}\left(\left(T_{\text{Car}}\right)^{\mu}{_\nu}+O(c^2)\right)\,.
\end{equation}
Hence, the Carroll energy-momentum tensor is simply the leading order term of $T^\mu{}_\nu$.

\section{Carrollian gravity, solutions and geodesics \label{sec:Cargra} }  

In this section we consider Carrollian gravity and its coupling to matter.
We focus mostly on electric (timelike) gravity theory  but also comment on the magnetic (spacelike) case.  
As a specific example we consider the coupling to electric Carrollian electrodynamics, and in particular describe the resulting equations of motion.    
We subsequently discuss various solutions: vacuum, non-zero cosmological constant and novel solutions arising from the coupling to Carrollian electrodynamics. Finally, we  discuss the properties of geodesics in a Carrollian spacetime. 

\subsection{Electric (timelike) Carroll gravity coupled to Carrollian matter \label{sec:Cargra1} } 

We follow here  \cite{Hansen:2021fxi}. 
The electric Carroll gravity (ECG) action is  
\begin{equation}
  \label{eq:LO_action}
S_{\text{ECG}} 
  = \frac{c^2}{16\pi G_N} \int_M d^{d+1}x e \left[
    K^{\mu\nu} K_{\mu\nu} - K^2
  \right]\,,
\end{equation}
where $K_{\mu\nu} = - \frac{1}{2} \LL_v h_{\mu\nu}$ is the extrinsic curvature, which is spatial, since it satisfies $v^\mu K_{\mu\nu}=0$ and where $K^{\mu\nu}=h^{\mu\rho}h^{\nu\sigma}K_{\rho\sigma}$. Varying the action with respect to $v^\mu$ and $h^{\mu\nu}$ we have
\begin{equation}
  \label{eq:LO-action-variation}
  \delta S_{\text{ECG}} 
= \frac{c^2}{8\pi G_N} \int_M d^{d+1}x e \left[
    {\cal{G}}^v_\mu \delta v^\mu
    + \frac{1}{2} {\cal{G}}^h_{\mu\nu} \delta h^{\mu\nu}
  \right]\,,
\end{equation}
This leads to the equations of motion ${\cal{G}}^v_\mu = 0$ and ${\cal{G}}^h_{\mu\nu}=0$, where 
\begin{subequations}
  \label{eq:LO-eom}
  \begin{align}
    {\cal{G}}^v_\mu
    &= - \frac{1}{2} \tau_\mu \left(K^{\rho\sigma} K_{\rho\sigma} - K^2\right)
    + \left( \tilde\nabla_\rho-\mathcal{L}_v\tau_\rho\right) h^{\rho\nu}\left( K_{\mu\nu} - K h_{\mu\nu} \right),
    \\
    {\cal{G}}^h_{\mu\nu}
    &= - \frac{1}{2} h_{\mu\nu} \left(K^{\rho\sigma} K_{\rho\sigma} - K^2\right)
    + K \left(K_{\mu\nu} - K h_{\mu\nu}\right)
    - v^\rho \tilde\nabla_\rho \left(K_{\mu\nu} - K h_{\mu\nu}\right)+A\tau_\mu\tau_\nu\,,
  \end{align}
\end{subequations}
where $A$ is an undetermined scalar. This is because $\tau_\mu\tau_\nu\delta h^{\mu\nu}=0$.
Note that here the covariant derivative is taken with respect to the Carroll metric-compatible connection \eqref{eq:minimal-carroll-tangent-connection}.

Projecting out the time and space components of each equation using $v^\mu$ and $h^{\mu\nu}$, 
we see that the time-space component of ${\cal{G}}^h_{\mu\nu}$ vanishes
($v^\mu h^{\nu\rho}{\cal{G}}^h_{\mu\nu} =0$ is a consequence of local Carroll boost invariance). 
Ignoring $v^\mu v^\nu {\cal{G}}^h_{\mu\nu}$ (which is not an equation of motion and thus plays no role whatsoever), the equations of motion can be written as
\begin{subequations}
  \label{eq:LO_EOM_alternate}
  \begin{gather}
    K^{\mu\nu}K_{\mu\nu}-K^2=0\,,\label{eq:LO_ham}\\
    \left(\tilde\nabla_\rho-\mathcal{L}_v\tau_\rho\right)h^{\rho\sigma}(K_{\sigma \mu}-K h_{\sigma\mu}) = 0\,,\label{eq:LO_mom}\\
    -\mathcal L_v K_{\mu\nu}  -2 K_\mu{}^\rho K_{\rho\nu}+KK_{\mu\nu}=0\label{eq:LO_evol}\,.
  \end{gather}
\end{subequations}
These have the form of constraint and evolution equations. The derivation of \eqref{eq:LO_evol} will be detailed further below.

\subsubsection*{Adding Carroll matter}

The leading order Carroll gravity action is order $c^2/G_N$ where we allow the possibility that $G_N$ scales with $c$ in a nontrivial way when expanding around $c=0$. In order to couple this to Carroll matter we need to make sure that the Carroll matter Lagrangian is of the same order in $c$ as the LO gravity action. Suppose that the Carroll matter Lagrangian is order $c^M$ for some $M\in\mathbb{Z}$ then the two theories couple provided that $G_N\sim c^{2-M}$.

In general, a Carroll invariant matter action is a functional of matter fields $\phi$ with Carroll metric sources, i.e. $S_{\rm M} [\phi;v^\mu,h^{\mu \nu}]$.
Varying the action gives rise to two currents
\begin{equation}\label{eq:varymatter}
\delta S_{\rm M} 
  =  c^M\int_M d^{d+1}x e \left[
    -T^v_\mu \delta v^\mu
    - \frac{1}{2} T^h_{\mu\nu} \delta h^{\mu\nu}
  \right]\,,
\end{equation} 
where we are agnostic about $M$ and where the currents $T^v_\mu$ and $T^h_{\mu\nu}$ are $c$ independent.

These currents can be combined into the Carroll energy-momentum tensor
\begin{equation}\label{eq:CarEMT2}
T^{\mu}{}_\nu =  v^\mu T^v_\nu + h^{\mu \rho} T^h_{\rho \nu}\,.
\end{equation}
We can alternatively define the responses to varying $\tau_\mu$ and $h_{\mu\nu}$. If we define
\begin{equation}
\delta S_{\rm M} 
  =  c^M\int_M d^{d+1}x e \left[
    -\mathcal{T}^\mu \delta \tau_\mu
    + \frac{1}{2}\mathcal{T}^{\mu\nu} \delta h_{\mu\nu}
  \right]\,,
\end{equation} 
then the Carroll energy-momentum tensor is given by
\begin{equation}
T^{\mu}{}_\nu = - \mathcal{T}^\mu \tau_\nu + \mathcal{T}^{\mu \rho} h_{\rho \nu}\,,
\end{equation}
where we have the relation between the currents
\begin{eqnarray}
    T^v_\rho & = & \tau_\rho\tau_\mu\mathcal{T}^\mu-h_{\rho\nu}\tau_\mu\mathcal{T}^{\mu\nu}\,,\\
    T^h_{\mu\nu} & = & h_{\mu\rho}h_{\nu\sigma}\mathcal{T}^{\rho\sigma}-\tau_\rho h_{\sigma\mu}\mathcal{T}^\mu-\tau_\sigma h_{\rho\mu}\mathcal{T}^\mu\,.
\end{eqnarray}

The sourced Carroll gravity equations of motion are thus
\begin{equation}\label{eq:sourcedECgrav}
 {\cal{G}}^v_\mu = 8\pi G_N c^{M-2}  T^v_\mu
 \quad , \quad  {\cal{G}}^h_{\mu\nu}=  8\pi G_N c^{M-2} T^h_{\mu \nu} \,.
\end{equation} 
Contracting the first of these two sourced equations with $v^\mu$ and $h^{\mu\nu}$ we obtain
\begin{subequations}
  \begin{gather}
    K^{\mu\nu}K_{\mu\nu}-K^2=16\pi G_N c^{M-2} v^\mu  T^v_\mu\,,\label{KEOM1} \\
    h^{\mu\nu}\left(\tilde\nabla_\rho-\mathcal{L}_v\tau_\rho\right)h^{\rho\sigma}(K_{\sigma \mu}-K h_{\sigma\mu}) = 8\pi G_N c^{M-2}h^{\mu\nu}T^v_\mu\,.\label{KEOM2}
  \end{gather}
\end{subequations}
The second equation in \eqref{eq:sourcedECgrav} can be written as
\begin{equation}\label{eq:intermediate}
    h_{\mu\nu}\left(\mathcal{L}_v K-K^2\right)-\frac{1}{2}h_{\mu\nu}\left(K_{\rho\sigma}K^{\rho\sigma}-K^2\right)-\mathcal{L}_v K_{\mu\nu}+KK_{\mu\nu}-2h^{\rho\sigma}K_{\mu\rho}K_{\nu\sigma}+A\tau_\mu\tau_\nu=8\pi G_N c^{M-2}T_{\mu\nu}^h\,,
\end{equation}
where we used the identity
\begin{equation}
    v^\rho\tilde\nabla_{\rho}K_{\mu\nu}=\mathcal{L}_v K_{\mu\nu}+2h^{\rho\sigma}K_{\mu\rho}K_{\nu\sigma}\,.
\end{equation}
Taking the trace of \eqref{eq:intermediate} with respect to $h^{\mu\nu}$ we find
\begin{equation}\label{eq:trace}
    \mathcal{L}_v K-K^2=8\pi G_N c^{2-M}\left(\frac{d}{d-1} v^\mu T_\mu^v+\frac{1}{d-1} h^{\mu\nu}T_{\mu\nu}^h\right)\,,
\end{equation}
where $d$ is the number of spatial dimensions and where we used the identity
\begin{equation}
    K_{\mu\nu}\mathcal{L}_v h^{\mu\nu}=2K^{\mu\nu}K_{\mu\nu}\,.
\end{equation}

We note that $T^h_{\mu\nu}$ is only determined up to a term proportional to $\tau_\mu\tau_\nu$. Furthermore, Carroll boost symmetry tells us that the $v^\mu h^{\nu\lambda}$ projection of \eqref{eq:intermediate} vanishes. Without loss of generality we can contract \eqref{eq:intermediate} with $h^{\mu\sigma}$. Using \eqref{KEOM1} and \eqref{eq:trace} we thus obtain 
\begin{equation}\label{KEOM3}
    h^{\mu\sigma}\left[-\mathcal{L}_v K_{\sigma\nu}+KK_{\sigma\nu}-2K_{\sigma}{}^{\rho}K_{\nu\rho}\right]=8\pi G_N c^{M-2}h^{\mu\sigma}\left[T_{\sigma\nu}^h-\frac{1}{d-1}h_{\sigma\nu}\left( v^\rho T_\rho^v+h^{\kappa\lambda}T_{\kappa\lambda}^h\right)\right]\,.
\end{equation}
If we take the trace of this equation we recover \eqref{eq:trace}. In the absence of sources we recover \eqref{eq:LO_evol}.

\subsection{Coupling to electric Carrollian electrodynamics}\label{sec:CarrollE&M}

We start by obtaining the action of the electric Carroll Maxwell action coupled to curved
(Carrrollian) spacetime (see \cite{Duval:2014uoa,deBoer:2021jej} for Carrollian electrodynamics on
flat space). 
One starts with the Maxwell action coupled to GR
\begin{equation}
S_{\rm Maxwell} = - \frac{1}{4 c \mu_0} \int d^{d+1}x\sqrt{-g} g^{\mu \rho} g^{\nu \sigma}  {\cal F}_{\mu \nu} {\cal F}_{\rho \sigma} \,,
\end{equation} 
and expands ${\cal F}_{\mu \nu} = F_{\mu \nu}  + {\cal{O}} (c^2)$. The constant $\mu_0$ is the vacuum magnetic permeability.
The leading order action is the electric Carroll Maxwell theory (ECM)
\begin{equation}
S_{\text{ECM}} =  \frac{1}{2 c^2 \mu_0} \int d^{d+1}x e  v^{\mu} v^{\rho }  h^{\nu \sigma} F_{\mu \nu} F_{\rho \sigma}\,,
\end{equation}
where we used equations \eqref{exp-metric} and \eqref{exp-invmetric}.
As expected this is essentially the electric field squared. 
We can compute the relevant components of the energy momentum tensor. Using \eqref{eq:varymatter} we obtain $M=-2$ and 
\begin{equation}
T^v_\mu =   -\frac{1}{2 \mu_0} \left( 2  h^{\nu \sigma} v^\rho F_{\mu \nu} F_{\rho \sigma} + \tau_\mu [v^\lambda v^\rho h^{\nu \sigma} F_{\lambda \nu}
F_{\rho \sigma} ] \right)\,,
\end{equation} 
\begin{equation}
T^h_{\mu \nu} = - \frac{1}{\mu_0} \left(  v^\rho v^\sigma  F_{\rho \mu} F_{\sigma \nu} - \frac{1}{2} h_{\mu \nu} 
[v^\lambda v^\rho h^{\alpha \sigma} F_{\lambda \alpha} F_{\rho \sigma} ] \right) +A\tau_\mu\tau_\nu\,,
\end{equation} 
where $A$ is undetermined.
Note that the Ward identity $v^\mu h^{\nu\rho} T^h_{\mu \nu } =0$ is correctly satisfied as well as the fact that the energy-momentum tensor \eqref{eq:CarEMT} is traceless for $d=3$. The equations of motion \eqref{KEOM1}, \eqref{KEOM2} and \eqref{KEOM3} become in this case
\begin{eqnarray}
 K^{\mu\nu}K_{\mu\nu}-K^2 & = &   -\frac{8\pi G_N}{c^4\mu_0}    E^2 \,,\label{eq:EGEM1}\\
 h^{\mu\nu}\left(\tilde\nabla_\rho-\mathcal{L}_v\tau_\rho\right)h^{\rho\sigma}\left(K_{\sigma \mu}-K h_{\sigma\mu}\right) & = & \frac{8\pi G_N}{c^4\mu_0}  h^{\mu\nu}h^{\lambda \sigma}  F_{\mu \lambda} E_\sigma \,,\\
    h^{\mu\sigma}\left[-\mathcal{L}_v K_{\sigma\nu}+KK_{\sigma\nu}-2K_{\sigma}{}^{\rho}K_{\nu\rho}\right] & = & \frac{8\pi G_N}{c^4\mu_0} h^{\mu\sigma}\left[-E_\sigma E_\nu+\frac{1}{d-1}h_{\sigma\nu}E^2\right]\,,\label{eq:EGEM3}
\end{eqnarray}
where we defined the electric field $E_\mu=-v^\rho F_{\rho\mu}$ and $E^2=h^{\mu\nu}E_\mu E_\nu$.

\subsection{Magnetic Carroll gravity and magnetic Carroll Maxwell} 

Both for the gravity side as well as the Maxwell side one can also consider the magnetic limit.
Together with the electric theories, this then gives rise to four different sets of sourced equations of motions as one can couple
electric/magnetic Carroll gravity to electric/magnetic Carroll Maxwell.  We will not spell out in detail
the equations of motion, but give below the corresponding actions. 

\subsubsection*{Magnetic Carroll gravity}

The LO action in the $c=0$ expansion of GR is at order $c^2$ and defines the electric theory. The magnetic theory is obtained by going to the NLO action at order $c^4$ and adding a Lagrange multiplier that kills the LO action. The details can be found in \cite{Hansen:2021fxi}. The resulting theory is the magnetic Carroll gravity theory (MCG):
\begin{equation}
\label{MCG} 
S_{\text{MCG}}=\frac{c^4}{16\pi G_N}\int {\rm d}^{d+1}x\, e\left(\phi^{\mu\nu}K_{\mu\nu}+h^{\mu\nu}\tilde R_{\mu\nu}\right)\,,    
\end{equation}
where the Lagrange multiplier is $\phi^{\mu\nu}$. Here $\tilde R_{\mu\nu}$ denotes the Ricci tensor associated with the connection $\tilde\Gamma^\rho_{\mu\nu}$.

\subsubsection*{Magnetic Carroll Maxwell}

The action for magnetic Carroll Maxwell (MCM) coupled to
Carroll gravity is given by
\begin{equation}
\label{MCM} 
S_{\text{MCM}} =  \frac{1}{2  \mu_0} \int {\rm d}^{d+1}x e \left[ \chi_{\mu} h^{\mu \nu}  v^{\sigma}F_{\nu \sigma } 
- \frac{1}{4}   h^{\mu \rho}h^{\nu \sigma}  F_{\mu \nu} F_{\rho \sigma} \right]\,,
\end{equation}
where $\chi_\mu$ is a Lagrange multiplier setting the electric field to zero. In this case we have $M=0$ (cf. equation \eqref{eq:varymatter}). This theory can also be coupled to the electric Carroll gravity theory, but we refrain from working out the details. 

\subsection{Carroll spacetimes: examples}

\label{examples}

In this subection, we study some examples
\footnote{Some of these examples were discussed earlier in \cite{deBoer:2021jej,Hansen:2021fxi,Perez:2022jpr} and are included here with further details.}
and consider Carrollian limits of  Schwarzschild, Reissner-Nordstr\"om, de Sitter, and anti de Sitter metrics. We also discuss the generic structure of the geodesic equation in Carrollian geometries.

\subsubsection{Schwarzschild black holes}

Here we consider an example of a Carroll spacetime whose connection has torsion. It arises in the Carroll limit of the Schwarzschild metric
\begin{align}
    g=-c^2\left(1-\frac{R}{r}\right){\rm d} t^2+\frac{1}{1-\frac{R}{r}}{\rm d} r^2+r^2{\rm d}\Omega^2\ ,
\end{align}
where the Schwarzschild radius in terms of the mass is given by
\begin{align}\label{SSradius}
    R=\frac{2MG_N}{c^2}\ ,
\end{align}
and ${\rm d}\Omega^2={\rm d}\theta^2+\sin^2\theta {\rm d}\phi^2$.

\subsubsection*{Electric limit and the Kasner spacetime}

In this limit one keeps fixed the black hole energy, or more precisely the combination $MG_N$. This is similar to \cite{deBoer:2021jej} if we write $MG_N=EG_C^{(el)}$, with
\begin{equation}
    E=Mc^2 \ ,\qquad G_C^{(el)}\equiv\frac{G_N}{c^2}\ ,
\end{equation} 
which can both be kept fixed in the Carroll limit. Effectively, this limit describes the region inside the black hole where gravity is strong. Perhaps the most proper way of doing this, is to use Kruskal coordinates, but it is instructive to proceed with Schwarzschild coordinates in the region $r<R$. The first step in the Carroll limit then amounts to taking $R/r\gg 1$ and one finds the metric
\begin{equation}
    {\rm d}s^2=\frac{2MG_N}{r}{\rm d}t^2-\frac{r}{2MG_N}c^2{\rm d}r^2+r^2{\rm d}\Omega^2\ .
\end{equation}
Notice that effectively, this limit is also produced by the expansion around the singularity at $r=0$, where the coordinates $r$ and $t$ reverse their role of space and time coordinates. It is known that the region close to the singularity is described by a Kasner-like metric \cite{Kasner:1921zz}. Indeed, if we redefine
\begin{equation}
    \tau=\frac{1}{{\sqrt{2MG_N}}}r^{3/2}\ ,
\end{equation}
we find
\begin{equation}
    {\rm d}s^2=-\frac{4}{9}c^2{\rm d}\tau^2+\frac{1}{(H\tau)^{2/3}}{\rm d}t^2+\frac{\tau^{4/3}}{H^{2/3}}{\rm d}\Omega^2\ ,
\end{equation}
where $H\equiv \frac{1}{2MG_{N}}$, and the dimensions of the quantities are $[\tau]=[t]=s$ and $[H]=s^2m^{-3}$.
This is a Kasner-like metric with Kasner exponents
\begin{equation}
    p_1=-\frac{1}{3}\ ,\qquad p_2=p_3=\frac{2}{3}\ .
\end{equation}
When taking the Carroll limit, we keep $H$ fixed and obtain the following quantities:
\begin{equation}
    v^\tau=-\frac{3}{2}\ ,\qquad
    h=\frac{1}{({H}\tau)^{2/3}}{\rm d}t^2+\frac{\tau^{4/3}}{{H}^{2/3}}{\rm d}\Omega^2\ ,\qquad 
    K=-\frac{1}{2{H}^{2/3}\tau^{5/3}}{\rm d}t^2+\frac{\tau^{1/3}}{{H}^{2/3}}{\rm d}\Omega^2\ .
\end{equation}
Notice that, in contrast to the magnetic limit, the extrinsic curvature $K=K_{\mu\nu}{\rm d}x^\mu{\rm d}x^\nu$ is nonzero. We checked that this solution solves equations \eqref{eq:LO_EOM_alternate}. More generally, this solution falls into the class of general
vacuum solutions to the electric theory given in \cite{Hansen:2021fxi}.\footnote{We also note that this case of Kasner spacetime solutions was independently observed in
\cite{Oling:2021} and we thank Marc Henneaux for a discussion on  the appearance of Kasner geometry in the electric Carroll limit.}

\subsubsection*{Magnetic limit and Carroll wormholes}

There is another limit we can take, by not keeping $MG_N$ fixed in the Carroll limit, but instead the 
 Schwarzschild radius $R$. In terms of the mass and Newton's constant, such a limit can be taken by keeping the quantities
\begin{equation}
    E=Mc^2\ ,\qquad G_{C}^{(m)}\equiv \frac{G_N}{c^4}\ ,
\end{equation}
fixed. We call this the magnetic limit and we read off the following quantities
\begin{equation}
    v=v^t\partial_t\ ,\qquad v^t=-\sqrt{\frac{r}{r-R}}\ ,\qquad
    K_{\mu\nu}=0\ .
\end{equation}
The extrinsic curvature vanishes as a consequence of the fact that the Carroll metric
\begin{align}
    h=\frac{1}{1-\frac{R}{r}}{\rm d} r^2+r^2{\rm d} \Omega^2\ .
\end{align}
is static, so the Lie-derivative along $v$ is zero. This metric is that of a constant time-slice of the Schwarzschild black hole. 

We can now transform metric to isotropic coordinates, defined by
\begin{align}
    \rho\equiv\frac{1+\sqrt{1-\frac{R}{r}}}{1-\sqrt{1-\frac{R}{r}}}\ ,\qquad 
    r=\frac{R}{4}\left(\sqrt{\rho}+\frac{1}{\sqrt{\rho}}\right)^2\ ,
\end{align}
It was shown in \cite{Hansen:2021fxi} that this is indeed a solution of
the magnetic Carroll gravity theory. 
In the patch on the outside of the black hole horizon, we have $\rho\in\  (1,\infty)$, but we extend it to $\rho\in\ (0,\infty)$. This defines an extension of the Carrollian Schwarzschild geometry. 
Then the resulting Carroll metric becomes conformally flat
\begin{align}
    h=\left(\frac{(\rho+1)^2R}{4\rho^2}\right)^2\left({\rm d}\rho^2+\rho^2 {\rm d}\Omega^2\right)\ .
    \label{wormholemetric}
\end{align}
The vector field $v$ in these coordinates becomes
\begin{equation}
    v=-\frac{\rho+1}{\rho-1}\partial_t\,.
\end{equation}
Notice that $h$ is invariant under $\rho\mapsto 1/\rho$ and that $v$ changes sign under this map. This is the two-fold $\mathbb{Z}_2$-symmetry that gives us the familiar diagram of the Einstein-Rosen bridge \cite{Einstein:1935tc}. Inspired by this, we refer to \eqref{wormholemetric} as the Carroll wormhole.

It is easy to recover the original black hole entropy from the Carroll wormhole. One looks at the point where the neck of the wormhole is smallest, which is at the $\mathbb{Z}_2$ fixed point $\rho=1$, corresponding to the original location of the horizon, $r=R$. The area of the sphere at $\rho=1$ then gives the entropy of the original black hole via $S=\frac{A}{4}$, as $h(\rho=1)=R^2{\rm d}\Omega^2$~\footnote{The entropy and temperature of the Schwarzschild black hole in units of magnetic Carroll gravity, i.e. in terms of $R$ and $G_C^{(m)}$, are given by $S=\frac{k_B}{c\hbar}\frac{\pi R^2}{G_C^{(m)}}$ and $T=\frac{c\hbar}{k_B}\frac{1}{4\pi R}$, where $k_B$ is Boltzmann's constant and $\hbar$ is Planck's constant. We refer to \cite{Grumilleretal} for more details about the thermodynamics and definition of Carrroll black hole geometries such as the one described here.}.

\subsubsection{Solutions with cosmological constant} 

\subsubsection*{de Sitter}
The de Sitter spacetime is one that has multiple widely used coordinate systems. We will consider comoving and static coordinates. The reason that we consider both coordinate systems is because the comoving coordinates make use of the Hubble constant $H$, while the static coordinates include the Hubble radius
\begin{align}
    R_H=\frac{c}{H}\ .
\end{align}
Similar to the Schwarzschild black hole, we can take two limits, depending on whether we take $H$ or $R_H$ fixed in the Carroll limit.

The coordinate transformation between these two coordinate systems depends on $c$, which becomes singular in the Carroll limit. Therefore, this results in two inequivalent (by diffeomorphisms and local Carroll boosts) Carroll limits, dubbed the electric and magnetic limits.

\subsubsection*{Electric limit - Inflationary/Comoving coordinates}

 The de Sitter metric in comoving coordinates is
\begin{align}
    g=-c^2{\rm d} t^2+\mathrm{e}^{2Ht}\left({\rm d} x^2+{\rm d} y^2+{\rm d} z^2\right)\ .
\end{align}
In the Carroll limit we keep $H$ fixed and obtain the following quantities:
\begin{equation}
    v^t=-1\ ,\qquad
h=\mathrm{e}^{2Ht}\left({\rm d} x^2+{\rm d} y^2+{\rm d} z^2\right)\ ,\qquad
K_{ij}=H\mathrm{e}^{2Ht}\delta_{ij}\ ,
\end{equation}
Note that the extrinsic curvature is non-zero and $i=1,2,3$ runs over spatial indices only. 
The Carroll metric is again conformally flat. If we add a cosmological constant to the LO action \eqref{eq:LO_action} then the above solution solves the corresponding equations of motion. 

The electric limit of de Sitter was used in \cite{deBoer:2021jej}. In this limit, the cosmological constant was kept fixed, as well as the rescaled Newton constant $G_C^{(el)}=G_N/c^2$, similar as in the electric limit of the Schwarzschild black hole. The precise relation between the relevant quantities is
\begin{equation}
    H^2=\frac{8\pi G_C^{(el)}}{3}\,\Lambda\ ,
\end{equation}
and $\Lambda$ has dimensions of energy. Differently from the black hole, is that the Schwarzschild radius goes to infinity, whereas the de Sitter radius $R_H=c/H$ goes to zero.

\subsubsection*{Electric and magnetic limit - Static coordinates}

To get the de Sitter metric in static coordinates, we start with the metric in comoving coordinates. The first step is to convert to spherical coordinates on the spatial part of the metric, and substitute $H=\frac{c}{R_H}$:
\begin{align}
    {\rm d}s^2=-c^2{\rm d} t^2+\mathrm{e}^{2ct/R_H}\left({\rm d} r^2+r^2{\rm d}\Omega^2\right)\ .
\end{align}
Now we perform the coordinate transformation:
\begin{align}
    \rho:=r\mathrm{e}^{ct/R_H},\qquad
    \tau:=t-\frac{R_H}{2c}\log\left(-1+\frac{r^2}{R_H^2}\mathrm{e}^{2ct/R_H}\right)\ ,\label{coordtrans}
\end{align}
to get the de Sitter metric in static coordinates:
\begin{align}\label{dSstatic}
    {\rm d}s^2=-\left(1-\frac{\rho^2}{R_H^2}\right)c^2{\rm d}\tau^2+\frac{{\rm d}\rho^2}{1-\frac{\rho^2}{R_H^2}}+\rho^2{\rm d}\Omega^2\ .
\end{align}
We note that these coordinates are only valid for $0<\rho<R_H$. Note also that the expressions of the static coordinates in terms of the comoving coordinates depend on $c$, and are not well defined in the $c\to 0$ limit. 
For the metric \eqref{dSstatic}, we have the following quantities:
\begin{align}
    \tau_\tau=\sqrt{1-\frac{\rho^2}{R_H^2}}\ ,\qquad
    v^\tau=-\sqrt{\frac{R_H^2}{R_H^2-\rho^2}}\ ,\qquad 
    K_{\mu\nu}=0\ .
\end{align}
Note that now the extrinsic curvature of de Sitter spacetime is equal to $0$. This is caused by the fact that we have taken a different Carroll limit, namely one in which $R_H$ instead of $H$ is kept fixed. In this limit, the Carroll metric is
\begin{align}\label{rhfixed}
     h=\frac{{\rm d}\rho^2}{1-\frac{\rho^2}{R_H^2}}+\rho^2 {\rm d}\Omega^2\ .
\end{align}
The relation between the Hubble radius and the cosmological constant is
\begin{equation}
    R_{H}^2=\frac{3}{8\pi G_C^{(m)}\Lambda}\ ,\qquad G_C^{(m)}=\frac{G_N}{c^4}\ .
\end{equation}
Notice that, because now we kept $G_C^{(m)}$ fixed, we can also keep the positive energy density $\Lambda$ fixed in the Carroll limit.

One can also take the electric limit of the static dS patch and this gives yet another space (not diffeomorphic to the electric limit of the FLRW form of the dS metric). This would give $R_H=cH^{-1}$ and 
\begin{equation}
    v=-H\rho\partial_\rho\,,\qquad h=H^2{\rm d}\tau^2+\rho^2 {\rm d}\Omega^2\,.
\end{equation}
If we define $R=\tau$ and $T=H^{-1}\log\rho$, then we have
\begin{equation}
    v=-\partial_T\,,\qquad h=e^{2HT}\left(H^2{\rm d}R^2 +{\rm d}\Omega^2\right)\,.
\end{equation}
This Carroll geometry is not diffeomorphic to the electric limit of the dS metric in comoving coordinates. This is a consequence of the fact that the coordinate transformation \eqref{coordtrans} is not analytic in $c$.

\subsubsection*{Anti-de Sitter}

Let us consider the case of anti-de Sitter (AdS) spacetimes. Global coordinates for AdS can be obtained by considering \eqref{dSstatic} and taking $\Lambda$ negative such that $0<\rho<\infty$ and
\begin{align}\label{AdSstatic}
    {\rm d}s^2
    =
    -\left(1+\frac{\rho^2}{R_{AdS}^2}\right)c^2{\rm d}\tau^2
    +
    \frac{{\rm d}\rho^2}{1+\frac{\rho^2}{R_{AdS}^2}}+\rho^2{\rm d}\Omega^2\ \,,
\end{align}
where now
\begin{equation}
    R_{AdS}^2
    =
    -
    \frac{3c^{4}}{8\pi G_N\Lambda}\,,
\end{equation}
and where $\Lambda$ is the negative energy density of the space, which we will treat as independent of $c$. 

In the \textbf{electric limit} $G^{(el)}_{C}=G_{N}/c^{2}$ is kept fixed as $c\to0$, which results in $\tilde H:=c/R_{AdS}$ being constant. As a result we find at leading order
\begin{equation}
    v=-\tilde H\rho\partial_\rho
    \,,
    \quad
    h=-\tilde H^2\rho^2{\rm d}\tau^2+\rho^2{\rm d}\Omega^2\,,
\end{equation}
which is the Poincar\'{e} patch with spherical slicing.
Note however that $h$ now has signature $(0,-1,1,1)$ (as opposed to $(0,1,1,1)$). These spacetimes are also known as pseudo-Carrollian\footnote{In this case the $c=0$ expansion of the Lorentzian metric reads
\begin{equation}
    g_{\mu\nu}=h_{\mu\nu}+c^2\left(\tau_\mu\tau_\nu -\Phi_{\mu\nu}\right)+\mathcal{O}(c^4)\,,
\end{equation}
which should be contrasted with \eqref{exp-metric}.
The plus sign in front of $\tau_\mu\tau_\nu$ makes it that now $h_{\mu\nu}$ has signature $(0,-1,1,1)$. We still have $v^\mu h_{\mu\nu}=0$.}.

For the \textbf{magnetic limit} we keep $G_C^{(m)}=\frac{G_N}{c^4}$ fixed as $c\to0$, which means that $R_{AdS}$ will remain constant. In this limit we find
\begin{equation}
    v=-\left(1+\frac{\rho^2}{R_{AdS}^2}\right)^{-1/2}\partial_\tau
    \,,
    \quad
    h=
     \frac{{\rm d}\rho^2}{1+\frac{\rho^2}{R_{AdS}^2}}+\rho^2{\rm d}\Omega^2
     \,.
\end{equation}

\subsubsection{Reissner-Nordstr\"{o}m}\label{sec:RN}

Considering the Carroll limit in the context of the Schwarzschild metric enables one to study a resulting metric outside the Schwarzschild horizon (magnetic limit) and a metric inside the Schwarzschild horizon (electric limit). Here we point out that for a Reissner-Nordstr\"{o}m black hole, which classically has an inner and an outer horizon, the electric limit applies to the geometry between the inner and the outer horizon. This last metric yields a charged deformation of the Kasner metric.

The Reissner-Nordstr\"{o}m metric is given by
\begin{equation}
    {\rm d}s^{2}
    =
    -
    \left(
        1
        -
        \frac{R_{S}}{r}
        +
        \frac{R^{2}_{Q,P}}{r^{2}}
    \right)
    c^{2}{\rm d}t^{2}
    +
    \left(
        1
        -
        \frac{R_{S}}{r}
        +
        \frac{R^{2}_{Q,P}}{r^{2}}
    \right)^{-1}
    {\rm d}r^{2}
    +
    r^{2}{\rm d}\Omega^{2}
    \,,
\end{equation}
where
\begin{equation}
    R_{S}
    =
    \frac{2G_NM}{c^2}
    \ ,
    \quad
    R_{Q,P}^{2}
    =
    \frac{1}{4\pi}\Big(\frac{Q^{2}}{\epsilon_{0}}+P^2\mu_0\Big)\frac{G_{N}}{c^{4}}
    \,.
\end{equation}
In here $\epsilon_{0}$ and $\mu_0$ are the electric and magnetic constants and $Q$ and $P$ the electric and magnetic charges respectively, and the units for $Q$ is Coulomb ($C$), and for $P$ it is $C\,m\,s^{-1}$. The geometry is supported by a gauge field that is given by
\begin{equation}\label{eq:RNgaugefield}
    A=-\frac{1}{4\pi\epsilon_0}\frac{Q}{r} {\rm d}t-\frac{\mu_0}{4\pi}P\cos\theta {\rm d}\phi\,.
\end{equation}
The speed of light is given by $c=1/{\sqrt{\epsilon_{0}\mu_{0}}}$ and so the Carroll limit $c\to 0$ can be reached by taking either $\epsilon_0\to \infty$ or $\mu_0\to\infty$, keeping the magnetic or electric constant fixed respectively, and in such a way that the quantization condition for the charges is preserved. These two limits are called magnetic and electric Carroll limits, which we discuss now.

In section \ref{sec:CarrollE&M} we looked at electric and magnetic limits of Maxwell. We expanded the gauge field such that it is $O(1)$ plus corrections. The LO action only sees the $O(1)$ part of the electric part of the gauge field, whereas the NLO action sees the $O(1)$ part of the magnetic part of the gauge field, essentially the spatial part of $A$. The coupling constant of the LO electric theory is $c^2\mu_0=\epsilon_0^{-1}$ and the coupling constant of the magnetic theory is $\mu_0$. 

Comparing this with the results of the current section we see that the gauge field in \eqref{eq:RNgaugefield} needs to be order $O(1)$.
This can be achieved either by keeping $\epsilon_0 c^2$ fixed which implies that $\epsilon_0\rightarrow\infty$ and $\mu_0$ is fixed. This is the magnetic limit. Or, alternatively, we keep $\epsilon_0$ fixed and take $P= c^2\tilde{P}$, so that $c\rightarrow 0$ implies $\mu_0\rightarrow\infty $. This is the electric limit.

\subsubsection*{Electric limit}

In this case we take $Q^{2}/\epsilon_{0}$ to be constant when taking the Carroll limit and consider $\mu_0\to\infty$. This seems to only make sense when the magnetic charge is zero, so henceforth we set $P=0$. As before, 
when taking the electric limit, we introduce $E=Mc^{2}$ and $G_{C}^{(el)}=G_{N}/c^{2}$ which we keep fixed. Then we find
\begin{eqnarray}
    v & = & -\left[\frac{2G_C E}{r}\left(1-\frac{Q^2}{8\pi\epsilon_0 E}\frac{1}{r}\right)\right]^{1/2}\frac{\partial}{\partial r}\,,\\    
    h
    & = &
    \frac{2EG_{C}}{r}
    \left(
        1
        -
        \frac{Q^{2}}{8\pi\epsilon_{0} E}\frac{1}{r}
    \right){\rm d}t^{2}
    +
    r^{2}{\rm d}\Omega^{2}
    \,.
\end{eqnarray}
The Carroll data $v$ and $h$ are not defined at $r=b:=Q^2/(8\pi\epsilon_{0}E)$ and we need to restrict $r\in(b,\infty)$ in order that $h$ is positive semi-definite. This limit describes the region between the inner and outer horizon of the RN black hole with the outer horizon sent to infinity. If we define $a = 2EG_{C}$ then the inner and outer horizons of the RN metric are at $r_\pm$ given by
\begin{equation}
    r_\pm=\frac{a}{2c^2}\left(1\pm\sqrt{1-4c^2b/a}\right)\,.
\end{equation}
Expanding this around $c=0$ we see that indeed $r_+$ goes to infinity while $r_-$ goes to $b$.

Let us perform the following coordinate transformation, $(t,r)\mapsto (\rho, T)$, defined by
\begin{equation}
    \rho =t \,,\qquad \frac{\partial r}{\partial T}=a^{1/2}\frac{\sqrt{r-b}}{r}\,,
\end{equation}
then we find for $v$ and $h$,
\begin{eqnarray}\label{eq:EClimitRN}
    v & = & -\frac{\partial}{\partial T}\,,\\    
    h
    & = &
    \frac{a}{r}
    \left(
        1
        -
        \frac{b}{r}
    \right){\rm d}\rho^{2}
    +
    r^{2}{\rm d}\Omega^{2}
    \,,
\end{eqnarray}
where $r=r(T)$. The (electric) Carroll gauge field becomes
\begin{equation}\label{eq:EClimitRN2}
A = -\frac{1}{4\pi \epsilon_{0}}\frac{Q}{r(T)}d\rho\,.
\end{equation} 
This geometry is a kind of ``charge deformation'' of the Kasner geometry. It has a non-trivial electric field since the vector potential has a radial component which depends on time. It can be checked that equations \eqref{eq:EClimitRN2} and \eqref{eq:EClimitRN2}
satisfy equations \eqref{eq:EGEM1}--\eqref{eq:EGEM3}.

\subsubsection*{Magnetic limit}

In the magnetic limit we keep both $R_{S}$ fixed and send $\epsilon_0\to\infty$ keeping $\mu_0$ and the charges $Q$ and $P$ fixed. Keeping $R_S$ fixed can be achieved by keeping $E=Mc^2$ and $G_C^{(m)}=G_N/c^4$ fixed as before. In this limit, we get
\begin{equation}
R^2_{Q,P}\to R_P^2=\frac{\mu_0}{4\pi}P^2\, G_{C}^{(m)}
    \ .
\end{equation}
Notice that the electric charge has dropped out and only the magnetic charge survives, as expected from taking a magnetic limit. The Carroll metric obtained in this limit then is
\begin{eqnarray}
v & = & -\left(1
        -
        \frac{R_{S}}{r}
        +
        \frac{R^{2}_{P}}{r^{2}}\right)^{-1/2}\frac{\partial}{\partial t}\,,\\
    h
    & = &
    \left(
        1
        -
        \frac{R_{S}}{r}
        +
        \frac{R^{2}_{P}}{r^{2}}
    \right)^{-1}
    {\rm d}r^{2}
    +
    r^{2}{\rm d}\Omega^{2}
    \,,
\end{eqnarray}
so we find a case similar to the Schwarzschild wormhole. Assuming that $1
        -
        \frac{R_{S}}{r}
        +
        \frac{R^{2}_{P}}{r^{2}}$ has two distinct real roots (which will be the case provided $R_S^2>4R_P^2$), we see that $h$ is positive semi-definite and $v$ is real for $r>r_+$ and $r<r_-$, where $r_+$ is the outer and $r_-$ the inner horizon.

        The extrinsic curvature vanishes again. The geometry is then supported by a magnetic field only, given by $B=\frac{\mu_0P}{4\pi}\frac{1}{r^2}dr$, and survives in the Carroll limit where $\mu_0$ and $P$ are fixed.

We define a new radial coordinate $\rho$ such that 
\begin{equation}\label{eq:coordtrafo}
    \frac{{\rm d}\rho}{\rho}=\frac{{\rm d}r}{\sqrt{(r-r_-)(r-r_+)}}\,.
\end{equation}
This allows us to write 
\begin{equation}
    h=\frac{r^2(\rho)}{\rho^2}\left({\rm d}\rho^2+\rho^2{\rm d}\Omega^2\right)\,,
\end{equation}
where $r$ is now a function of $\rho$. This function can be found by integrating \eqref{eq:coordtrafo} and inverting which leads to 
\begin{eqnarray}
r & = & \frac{1}{2}\left(r_+(x+1)-r_-(x-1)\right)\,,\\
    x & = & \frac{1}{2}\left(\rho+\frac{1}{\rho}\right)\,.
\end{eqnarray}
In terms of $\rho$ we can write $v$ as
\begin{equation}
    v=-\frac{2\left(r_+(x+1)-r_-(x-1)\right)}{(r_+-r_-)\left(\rho-\frac{1}{\rho}\right)}\partial_t\,.
\end{equation}
The region outside the outer horizon corresponds to $r>r_+$. This is equivalent to demanding $x>1$. If we take $\rho>1$ then this captures the region $r>r_+$. Since $x$ is symmetric under $\rho\leftrightarrow 1/\rho$. We see that we can extend the $\rho$ coordinate to also cover $0<\rho<1$. This gives us again the wormhole geometry. We see that $h$ is invariant under $\rho\leftrightarrow 1/\rho$ whereas $v$ changes sign. 

We finally note that the magnetically charged wormhole solution described above is expected to 
be a solution of MCG  coupled to MCM. It would be an interesting
exercise to explicitly check this using the actions  \eqref{MCG} and \eqref{MCM}. 

\subsection{Geodesics}

To discuss the properties of the geodesics, we start with the action of a massive relativistic
particle coupled to a pseudo-Riemannian metric
\begin{align}
    S=\mp |m|c\int\sqrt{\mp g_{\mu\nu}\diff{x^\mu}{\sigma}\diff{x^\nu}{\sigma}}\dint \sigma =\int L\, {\rm d}\sigma \ ,
\end{align}
where $\sigma$ is a worldline parameter and there is worldline diffeomorphism invariance of the action.
The upper signs will be chosen for timelike geodesics, and the lower signs for spacelike geodesics. The momenta $p_\mu\equiv \partial L/\partial {\dot x}^\mu$ (the dot denotes the derivative with respect to $\sigma$) satisfy
\begin{equation}
    p^\mu p_\mu \pm |m|^2 c^2=0\ .
\end{equation}
One can write this in the more familiar way, $p^\mu p_\mu + m^2 c^2=0$, if for spacelike geodesics we let the particle be tachyonic with $m^2=-|m|^2<0$.

The equations of motion are
\begin{align}\label{EOMgeo}
    g_{\alpha\nu}\diff{^2x^\nu}{\tau^2}
    +\frac{1}{2}\Big[\partial_\mu g_{\alpha\nu}
    +\partial_\nu g_{\alpha\mu}
    -\partial_\alpha g_{\mu\nu}\Big]\diff{x^\mu}{\tau}\diff{x^\nu}{\tau}=0\ ,
\end{align}
where, up to rescaling and translation, $\tau$ is the proper time for a timelike geodesic or the proper length $s$ for a spacelike geodesic (${\rm d}s^2=-c^2{\rm d}\tau^2$). For the proper time/length, one has
\begin{equation}\label{eps}
    \varepsilon\equiv -g_{\mu\nu}\frac{{\rm d}x^\mu}{{\rm d}\tau}\frac{{\rm d}x^\nu}{{\rm d}\tau}=c^2,0,-1\ ,
\end{equation}
for timelike, null, or spacelike geodesics, respectively. Notice the factor of $c^2$ for timelike geodesics which will vanish in the Carroll limit. This implies that timelike geodesics become null in the Carroll limit, and since $c\to 0$, light cones close up and a timelike Carroll particle in this class can no longer move.

We can therefore concentrate on the spacelike geodesics, for which one should read \eqref{eps} with $\tau$ replaced by $s$. The units therefore are different and $\varepsilon$ is dimensionless and equal to unity. From here on, we therefore use $s$ whenever we talk about spacelike geodesics.

We could contract the equations of motion \eqref{EOMgeo} with the inverse metric to get the more familiar geodesic equation
\begin{align}
    \frac{{\rm d}^2x^\rho}{{\rm d}s^2}+\Gamma^\rho_{\mu\nu}\diff{x^\mu}{s}\diff{x^\nu}{s}=0\ ,
\end{align}
but we don't do so below in order to make the $c$-expansion easier. This way, we do not need to refer to the connection and any of its properties, but instead expand the metric again as in \eqref{exp-metric}, 
\begin{equation}
    g_{\mu\nu}=h_{\mu\nu}+c^2\overline{\Phi}_{\mu\nu}+O(c^4)\ .
\end{equation}
Furthermore, we assume that, given a solution $x^\rho(s)$, that $\diff{x^\rho}{s}$ has a Taylor expansion around $c=0$,
\begin{align}\label{aannameLaurent}
    \diff{x^\rho}{s}=\sum_{i=0}^\infty c^{i}\left.\diff{x^\rho}{s}\right|_{(i)}\ ,
\end{align}
where $(i)$ indicates the expansion order.
For spacelike geodesics an expression like the one above should always be possible whenever a Carroll limit exists. Now we can derive new equations order by order in powers of $c$. In the Carroll limit, we are interested in the leading term. To lowest order, we find

\begin{align}
    h_{\alpha\nu}\left.\diff{^2x^\nu}{s^2}\right|_{(0)}
    +\frac{1}{2}\left(\partial_\mu h_{\alpha\nu}+\partial_\nu h_{\alpha\mu}-\partial_\alpha h_{\mu\nu}\right)\left.\diff{x^\mu}{s}\right|_{(0)}\left.\diff{x^\nu}{s}\right|_{(0)}=0\ .
    \label{geodvglC-S}
\end{align}
This equation follows from an action given by (from now on we will drop the subindex ${(0)}$)
\begin{align}
S_\text{Carroll}=p_0\int\sqrt{h_{\mu\nu}\diff{x^\mu}{\sigma}\diff{x^\nu}{\sigma}}\dint \sigma\ ,\label{carrollaction}
\end{align}
where $p_0$ is a constant with the dimensions of a momentum, independent of $c$, and which we take to be positive.

In contrast to the Lorentzian case, the action is positive definite, since the metric $h_{\mu\nu}$ is of rank $D-1$ and positive definite along the spatial directions. The minima of the action then correspond to the case where the action vanishes. Those are the particles at rest. In appropriate coordinates $x^\mu=\{t,x^i\}$, the particles at rest satisfy
\begin{equation}
    \frac{{\rm d}\vec x}{{\rm d}s}=0\ ,
\end{equation}
and ${\rm d}t/{\rm d}s$ is undetermined because $h$ is of rank $D-1$. These Carroll geodesics correspond to the limit of timelike geodesics.

The other set of solutions is less trivial, and corresponds to non-trivial solutions of \eqref{geodvglC-S}. By contracting this equation with $v^\alpha$, one easily finds other identities, such as 
\begin{align}
    K_{\mu\nu}\diff{x^\mu}{s}\diff{x^\nu}{s}=0\ .\label{Kxx}
\end{align}
In terms of the Carroll connection \eqref{eq:minimal-carroll-tangent-connection}, it is an easy exercise to rewrite \eqref{geodvglC-S} as follows,
\begin{align}
    h_{\sigma\rho}\left[\diff{^2x^\rho}{s^2}
    +\tilde{\Gamma}^\rho_{\mu\nu}\diff{x^\mu}{s}\diff{x^\nu}{s}\right]=-K_{\sigma\mu}\tau_\nu\diff{x^\mu}{s}\diff{x^\nu}{s}\ ,
    \label{hgeovgl}
\end{align}
where we made use of \eqref{Kxx}. 

Notice that for the magnetic Carroll limit, $K_{\mu\nu}=0$ by definition and the geodesic equation in \eqref{hgeovgl} takes a more familiar form. 

Notice furthermore that $t(s)$ is again undetermined when $K_{\mu\nu}=0$, as a consequence of $h$ having only spatial components. In fact, for this magnetic case, $t$ does not appear in the action.  

The momenta for the action \eqref{carrollaction} satisfy 
\begin{equation}
    p_\mu h^{\mu\nu} p_\nu=p_0^2\ ,
\end{equation}
as one can easily check. In the adapted coordinates $x^\mu=\{t,x^i\}$, it implies $\vec{p}^{\,2}=p_ih^{ij}p_j=p_0^2$ together with the constraint $E=0$ as the Hamiltonian vanishes on-shell. These correspond precisely to the type of particles found in \cite{deBoer:2021jej}, obtained from the Carroll limit of relativistic tachyonic particles, but now generalized to arbitrary Carroll geometries.

Examples of geodesics were worked out in  \cite{ArjanvD}. It includes particles traveling through the Carroll wormhole. 

\section{The energy-momentum tensor of a putative Carroll fluid}\label{sec:emtcarroll}

Despite the fact that, as far as we know, there currently does not exist a bona fide microscopic quantum system with well-defined Carrollian thermodynamcs, the notion of Carrollian fluids and energy-momentum tensors that allegedly describe such fluids frequently appears in the literature. 

To facilitate comparison with the literature we will consider some aspects of energy-momentum tensors for would-be Carrollian fluids. Despite their purely hypothetical nature, we will in this section (and only in this section) use the term ``Carrollian fluid" to refer to this possibly empty set of quantum systems. Energy-momentum tensors that take a perfect Carrollian fluid\footnote{Transport was studied in for example \cite{Poovuttikul:2019ckt,Petkou:2022bmz,Freidel:2022vjq}.} form can also appear as expectation values of the energy-momentum tensor in particular states in well-defined quantum theories and/or curved backgrounds. These energy-momentum tensors are perfectly fine but one should not interpret quantities such as energy density and pressure as actual thermodynamical quantities. 

In this section we construct two distinct types of Carroll perfect fluid energy-momentum tensors using two different methods that give coinciding results: an expansion around $c=0$ starting from the relativistic fluid energy-momentum tensor and by employing the hydrostatic partition function. 

\subsection{Carroll expansion of perfect fluid energy-momentum tensors}

To illustrate these statements let us take a look at the $c=0$ expansion of the energy-momentum tensor of a relativistic perfect fluid on an arbitrary curved background. Consider the energy-momentum tensor of relativistic perfect fluid
\begin{equation}\label{eq:Einsteinfluid}
    T^\mu{}_\nu=\mp\frac{\tilde{\mathcal{E}}+P}{c^2}U^\mu U_\nu+P\delta^\mu_\nu\,,
\end{equation}
where $U^{2}=\pm c^{2}$ and $\tilde{\mathcal{E}}$ and $P$ are the energy and pressure associated to the fluid, respectively.
The upper (lower) sign corresponds to a fluid with spacelike (timelike) velocities. We parameterize the velocity $U^{\mu}$ in the following explicit manner
\begin{equation}
    U^{\mu}
    =
    \frac{u^{\mu}}{\sqrt{\mp\left(1-\frac{u^{2}}{c^{2}}\right)}}
    \,,
\end{equation}
where $u^2=\Pi_{\mu\nu}u^\mu u^\nu$, $T_{\mu}u^{\mu}=1$ and $u^{\mu}\Pi_{\mu\nu}$ is the `three'-velocity.\footnote{
As a consequence of the $U^{2}=\pm c^{2}$ normalization choice, $U$ is real for either sign. Alternatively one can always choose the normalization $U^{2}=- c^{2}$, with the consequence that $U^{\mu}$ is imaginary for the spacelike choice.
}

We can consider two distinct expansions for the relativistic fluid vector $U^{\mu}$, depending on whether we use the upper or lower sign. For the upper sign (spacelike case) we expand 
\begin{equation}
U^\mu=c\frac{u^\mu}{u}+O(c^3)\,,
\end{equation}
where $u=\sqrt{u^{2}}$.
For the lower sign (timelike case) we need to take into account that $u<c$, even when we take $c\to0$. As such we assume the leading order expansion around zero of $u=\sqrt{h_{\mu\nu}u^\mu u^\nu}$ to be of order $c^{2}$. We find
\begin{equation}
U^\mu=-v^\mu+c^2 U_{(2)}^\mu+O(c^4)\,,
\end{equation}
where $U_{(2)}^{\mu}$ is some subleading term and $v^{\mu}\tau_{\mu}=-1$ and $v^{\mu}h_{\mu\nu}=0$. 

In both cases (both signs) we take $\tilde{\mathcal{E}}$ and $P$ to be of order $c^{2-N}$ and we will denote the leading order terms in the expansion of the energy density and the pressure by the same symbols. Here $N$ is defined in \eqref{eq:EMTc}. We then obtain for the upper sign the following Carroll energy-momentum tensor
\begin{equation}\label{eq:EMTcaseI}
  {\rm spacelike:}\qquad   \left(T_{\text{Car}}\right)^{\mu}{_\nu}=-\frac{\tilde{\mathcal{E}}+P}{u^2}u^\mu h_{\nu\rho}u^\rho+P\delta^\mu_\nu\,.
\end{equation}
This agrees with a result found in \cite{deBoer:2017ing}.
For the lower sign we obtain
\begin{equation}\label{eq:EMTcaseII2}
   {\rm timelike:} \qquad \left(T_{\text{Car}}\right)^{\mu}{_\nu}=(\tilde{\mathcal{E}}+P)v^\mu \left(\hat\tau_\nu-h_{\nu\rho}U_{(2)}^\rho\right)+P\delta^\mu_\nu\,.
\end{equation}
The combination $\hat\tau_\nu-h_{\nu\rho}U_{(2)}^\rho$ is the leading order term in the expansion of $U_\nu$ for which we have
\begin{equation}\label{eq:redeffl   uid}
    U_\nu=c^2 U_{(2)\nu}+O(c^4)\,,
\end{equation}
with $U_{(2)\nu}=-\left(\hat\tau_\nu-h_{\nu\rho}U_{(2)}^\rho\right)$\,.

\subsection{Hydrostatic partition function}\label{sec:relarecap}

The hydrostatic partition function \cite{Banerjee:2012iz,Jensen:2012jh} is a thermal partition function evaluated on a weakly curved and stationary background geometry. Due to the stationarity of the background, one can construct the hydrostatic partition function explicitly by relating thermodynamical quantities to the background geometry and a corresponding Killing vector. This method has been applied to non-Lorentzian setups in, e.g., \cite{Jensen:2014ama,deBoer:2020xlc}.

It is instructive to derive the two distinct Carroll invariant perfect fluid energy-momentum tensors given in \eqref{eq:EMTcaseI} and \eqref{eq:EMTcaseII2}, respectively, directly using the hydrostatic partition function in a Carroll geometry and taking an expansion around $c\to0$ of the relativistic hydrostatic partition function. Let us first review the relativistic setup.

\subsubsection*{Relativistic hydrostatic partition function}

Let $\beta^\mu$ be a Killing vector of the relativistic geometry
\begin{equation}
    \mathcal{L}_{\beta}g_{\mu\nu}=0
    \,.
\end{equation}
The choice of $\beta^{\mu}$ enables one to make a choice of local frame invariant temperature $\tilde{T}$, where the hydrostatic partition function takes the form $\mathcal{L}=e P(\tilde{T})$ where $-e^{2}=\det g_{\mu\nu}$ and 
\begin{equation}
    g_{\mu\nu}\beta^{\mu}\beta^{\nu}
    =
    \pm\frac{c^{2}}{\tilde{T}^{2}}
    \,,
\end{equation}
where we choose $\beta^{\mu}=U^{\mu}/\tilde{T}$ to be spacelike or timelike oriented corresponding to the four-velocity normalization $U^{2}=\pm c^{2}$. To arrive at the energy-momentum tensor corresponding to the hydrostatic setup, we vary with respect to the metric keeping $\beta^\mu$ fixed. On top of that we impose the thermodynamic relations $\frac{\partial P}{\partial \tilde{T}}=\tilde{s}$ and $\tilde{s}\tilde{T}=\tilde{\mathcal{E}}+P$. Defining the energy-momentum tensor through $\delta \mathcal{L}=\frac{1}{2}e T^{\mu\nu}\delta g_{\mu\nu}$ we find the spacelike (upper sign) or timelike (lower sign) relativistic perfect fluid energy-momentum tensor:
\begin{equation}
    T^{\mu\nu}
    =
    \mp\frac{\tilde{\mathcal{E}}+P}{c^{2}}U^{\mu}U^{\nu}
    +
    P g^{\mu\nu}
    \,,
\end{equation}
which was presented in \eqref{eq:Einsteinfluid}.

\subsubsection*{Spacelike Carroll case}

To consider Carroll perfect fluids we require $\beta^\mu$ to be a Killing vector of the Carroll geometry, i.e. 
\begin{equation}
    \mathcal{L}_\beta h_{\mu\nu}=0\,,\qquad\mathcal{L}_\beta v^\mu=0\,.
\end{equation}
The hydrostatic partition function has a Lagrangian of the form $\mathcal{L}=e P(\tilde T)$ where $-e^2=\text{det}\,(-\tau_\mu\tau_\nu+h_{\mu\nu})$ and where $\tilde T$ is related to $\beta^\mu$. There are two options, either $\beta^{\mu}$ spacelike oriented
\begin{equation}\label{eq:option1}
    h_{\mu\nu}\beta^\mu\beta^\nu=\frac{c^{2}}{\tilde T^2}
    \,,
\end{equation}
or we say that $\beta^\mu$ is proportional to $v^\mu$, i.e. timelike oriented, and 
\begin{equation}\label{eq:option2}
    \tau_{\mu}\beta^\mu=\frac{1}{\tilde T}\,.
\end{equation}
This condition is Carroll boost invariant because $h_{\mu\nu}\beta^\nu=0$.

For the spacelike case we take $\beta^\mu=c\beta^\mu_{(1)}+O(c^3)$
where $\beta^\mu_{(1)}=u^\mu/(\sqrt{u^{2}}\tilde{T})$ and subsequently drop the $(1)$ subscript. The condition $T_{\mu}u^{\mu}=1$ implies $\tau_{\mu}u^{\mu}=1$. This leads to the Carrollian hydrostatic partition function
\begin{equation}\label{eq:hydro1}
    \mathcal{L}=e P(\tilde T)\,,
\end{equation}
where $\tilde T$ is defined as in \eqref{eq:option1}. 

The variation of the hydrostatic Lagrangian with respect to
the Carroll geometry (keeping $\beta^\mu$ fixed), 
as derived in \eqref{eq:derivingemt}, can be written as 
\begin{equation}
\label{eq:dL} 
\delta\mathcal{L}=e\left(-\mathcal{T}^\mu\delta\tau_\mu+\frac{1}{2}\mathcal{T}^{\mu\nu}\delta h_{\mu\nu}\right)\,.
\end{equation}
For \eqref{eq:option1} and \eqref{eq:hydro1} we find
\begin{equation}\label{eq:T}
\mathcal{T}^\mu=Pv^\mu\,,\qquad \mathcal{T}^{\mu\nu}=Ph^{\mu\nu}-\frac{\tilde{\mathcal{E}}+P}{u^2}u^\mu u^\nu\,,
\end{equation}
where we used that $\frac{\partial P}{\partial\tilde T}=\tilde s$ and that $\tilde s\tilde T=\tilde{\mathcal{E}}+P$. The Carroll energy-momentum tensor thus is
\begin{equation}\label{eq:EMTspace}
(T_{\text{Car}})^\mu{}_\nu
=
-\mathcal{T}^{\mu}\tau_{\nu}+\mathcal{T}^{\mu\rho}h_{\rho\nu}
=
P\delta^\mu{_\nu}-\frac{\tilde{\mathcal{E}}+P}{u^2}u^\mu u^\rho h_{\rho\nu}
\,.
\end{equation}
This reproduces the general form of the energy-momentum tensor obtained in \eqref{eq:EMTcaseI} from the $c=0$ expansion of the spacelike fluid. 

\subsubsection*{Timelike Carroll case}

If we view the timelike case as an expansion from the relativistic case, we take $\beta^\mu=\beta^\mu_{(0)}+c^2\beta^\mu_{(2)}+O(c^4)$. In this case we have
\begin{equation}\label{eq:expansionbeta}
    g_{\mu\nu}\beta^\mu\beta^\nu=h_{\mu\nu}\beta^\mu_{(0)}\beta^\nu_{(0)}+c^2\bar\Phi_{\mu\nu}\beta^\mu_{(0)}\beta^\nu_{(0)}+2c^2h_{\mu\nu}\beta^\mu_{(0)}\beta^\nu_{(2)}+O(c^4)=-c^2\tilde T^{-2}\,.
\end{equation}
This requires the constraint $h_{\mu\nu}\beta^\mu_{(0)}\beta^\nu_{(0)}=0$ so that $\beta^\mu_{(0)}$ is proportional to $v^\mu$. Using $\bar\Phi_{\mu\nu}=\Phi_{\mu\nu}-\tau_\mu\tau_\nu$ with $v^\mu v^\nu\Phi_{\mu\nu}=0$ we then find $\tau_\mu\beta^\mu_{(0)}=\tilde T^{-1}$ where we have taken the positive root. This leads to \eqref{eq:option2} after dropping the subscript $(0)$. Hence, in order to do the timelike expansion we need to supplement the hydrostatic partition function with a Lagrange multiplier term that enforces $h_{\mu\nu}\beta^\nu=0$. In the timelike case we thus end up with the Carrollian `hydrostatic' partition function 
\begin{equation}\label{eq:hydro2}
    \mathcal{L}=e P(\tilde T)
    +
    e\chi^{\mu}h_{\mu\nu}u^{\nu}\,,
\end{equation}
where $\chi^\mu$ is a Lagrange multiplier field and where $\tilde T$ is defined as in \eqref{eq:option2} and $\beta^{\mu}\tilde{T}=u^{\mu}$. In the previous sentence, we wrote hydrostatic partition function in quotation marks since as we will see shortly it does not actually define a perfect fluid energy-momentum tensor.

The term involving the Lagrange multiplier field can also be interpreted as required in the context of the relativistic expansion. Namely, if we use \eqref{eq:expansionbeta} for the temperature $\tilde{T}$, there is the risk that the leading order term becomes positive. The introduced constraint makes sure that this is circumvented.

Using a similar approach for \eqref{eq:option2} and \eqref{eq:hydro2} we find, supplied with $h_{\mu\nu}\beta^{\nu}=0$ coming from varying $\chi$, that 
\begin{equation}\label{eq:T2}
\mathcal{T}^\mu=-\tilde{\mathcal{E}}v^\mu\,,\qquad \mathcal{T}^{\mu\nu}=Ph^{\mu\nu}
+
2\chi^{(\mu}v^{\nu)}
,
\end{equation}
where we used that $\frac{\partial P}{\partial\tilde T}=\tilde s$, $\tilde s\tilde T=\tilde{\mathcal{E}}+P$ and $\beta^{\mu}\tilde{T}=u^{\mu}=-v^{\mu}$. Combining these currents, the Carroll energy-momentum tensor thus is
\begin{equation}\label{eq:EMTtime}
(T_{\text{Car}})^\mu{}_\nu
=
-(\tilde{\mathcal{E}}+P)v^{\mu}U_{(2)\nu}
+
P\delta^\mu{_\nu}
\,,
\end{equation}
where we introduced $U_{(2)\nu}=-(\tau_{\nu}+\frac{\chi^{\rho}h_{\rho\nu}}{\tilde{\mathcal{E}}+P})$.
This reproduces the general form of the energy-momentum tensor obtained in \eqref{eq:EMTcaseII2} from the $c=0$ expansion of the timelike fluid. From the fact that $U_{(2)\nu}$ is in fact a Lagrange multiplier, an additional hydrodynamic quantity that is not reflected in the thermodynamics, it is clear that this energy-momentum tensor falls outside the scope of conventional fluids.

\subsection{On microscopic Carroll gasses}\label{sec:Carrollmicrogas}

Before we consider Carroll gasses in this subsection, let us briefly review the description of a classical Boltzmann gas of free relativistic particles. We express the relativistic velocity $U^{\mu}$ and relativistic momentum $P_{\mu}$ to spatial momentum $p_{i}$ and energy $E$ via
\begin{equation}
	U^{\mu}
	=
	\gamma(1,v^{i})
	\,,
	\quad
	P_{\mu}
	=
	(-E,p_{i})
	\,,
	\quad
	P^{2}=-m^2c^2
	\,,
	\quad
	\Rightarrow 
	\quad
	E^{2}=c^{2}\vec{p}^{\,2}+m^2c^4\,,
\end{equation}
where $\gamma$ is the Lorentz factor and $m$ the mass of the particles. The single-particle partition function\footnote{This agrees with the ideal gas model used in \cite{deBoer:2017ing} in which $z=1$ and $\lambda=c$.} is defined as
\begin{equation}\begin{aligned}\label{eq:Z1}
	Z_{1}(T,V,v^{i})
	=
	\frac{1}{h^{d}}\int {\rm d}^{d}x\int {\rm d}^{d}p\,e^{\tilde\beta U^{\mu}P_{\mu}}
	=
	\frac{V}{h^{d}}\int {\rm d}^{d}p\,e^{-{\beta} H_{1}(p)+{\beta} v^{i}p_{i}}	
	\,,
\end{aligned}\end{equation}
where $H_{1}(p)$ is the single-particle Hamiltonian, $h$ represents Planck's constant and $V$ the volume of space. 
The chemical potential $v^{i}$ conjugated to the momentum $p_i$ can be understood as the average total velocity, see e.g. \cite{deBoer:2017ing}.
The introduced inverse temperature
\begin{equation}
	\beta=\frac{1}{k_B T}=\gamma \tilde{\beta}=\frac{\gamma}{k_B\tilde T}
	\,,
\end{equation}
is constructed in such a way that $\tilde{\beta}$ is the boost invariant rest-frame inverse temperature (see e.g. \cite{deBoer:2017ing}). 
We remind the reader that $Z_{1}$ is not required to be boost invariant, but can be related to the boost invariant Lorentz scalar pressure $P$ via the grand canonical potential $\Omega$ (see Appendix \ref{app:gas} for more details)
\begin{equation}\label{eq:grandpot}
	\Omega
	=
	-
	\frac{Z_{1}}{\beta}e^{\beta\mu}
	\,,
	\quad
	P=\frac{Z_{1}}{V\beta}e^{\beta\mu}\,,
\end{equation}
where $\mu=\tilde{\mu}/\gamma$ is the chemical potential and $\tilde{\mu}$ is the rest-frame chemical potential and the grand potential is related to the pressure $P$ via $\Omega=-PV$.

For simplicity, we focus on a gas of massless particles, with hamiltonian 
\begin{equation}
H_{1}(p)=(\vec{p}^{\,2}c^2)^{1/2}=|\vec p\,|c\ ,
\end{equation}
and specify to the case of three spatial dimensions ($d=3$). The single particle partition function can be worked out to be
\begin{equation}\label{eq:part1}
    Z_{1}
    =~
    \frac{V}{h^{3}}\frac{4\pi}{\beta v}
    \int^{\infty}_{0} {\rm d}p\, p 
    \sinh\left(\beta v p\right)
    e^{-\beta pc}
   \\=~
   \frac{8\pi V\gamma^4}{h^3c^3\beta^3}
    \ ,
\end{equation}
which is the partition function of massless relativistic particles with Boltzmann statistics. This result is obvious when $v<c$, and in a relativistic theory we would only consider this case. But, in anticipation of the Carroll limit, we will also consider other cases. Notice first that, for $v>c$ with real velocity $v$, the integral diverges and is not well defined. 
There is a way to define it if we consider $\beta$ and $v$ both to lie in the complex plane. In that case we find that for the integral to converge we require $\text{Re}(\beta v) < \text{Re}(\beta c)$ which leads to
\begin{equation}\label{ineq}
    \text{Re}\,\beta~\text{Re}\,(v-c)
    -
    \text{Im}\,\beta~\text{Im}\,v<0
    \,.
\end{equation}
When this condition is satisfied, the resulting partition function is still given by \eqref{eq:part1} with complexified parameters for $\beta$ and $v$. This expression for the partition function is in fact also well-defined for $v>c$ with real $v$ and $\beta$, but then the inequality \eqref{ineq} is not satisfied and therefore this case cannot be interpreted as coming from a partition function. 

There are many solutions in the complex plane satisfying \eqref{ineq}, all related by analytic continuation. One particular solution that leads to a real partition function is obtained by taking $\beta $ real and $v$ purely imaginary with $c\neq 0$.
This can be realized by taking the chemical potentials $v^i$ purely imaginary and hence the partition function is a Fourier transform. A purely imaginary $v$ can no longer be interpreted as the average total velocity, but there is still a quantity with the dimension of a velocity, and we can therefore consider a Carroll regime where this velocity is much larger than the speed of light. This is one of the cases that we will work out below (spacelike case).

\subsubsection*{Timelike Carroll case}

We consider the option 
\begin{equation}
    v=0 \,,
    \qquad 
    \gamma\rightarrow 1\,,
\end{equation}
as $c\rightarrow 0$. 
Plugging \eqref{eq:part1} into \eqref{eq:grandpot} we then find
\begin{equation}\label{eq:leadingpressure2}
    P= \frac{8\pi  }{h^3c^{3}\beta^4}e^{\beta\mu}
    \,.
\end{equation}
Notice that this result diverges in the strict Carroll limit as $P\sim c^{-3}$. It turns out that the leading order pressure will not produce consistent thermodynamical relations with the energy-momentum tensor. Let us show this by redoing the limit and setting $v$ identically equal to zero, so we can take $v^i=c^2v_{(2)}^i+O(c^4)$. Let us consider this case for a Boltzmann gas of free massless relativistic particles. The pressure is \eqref{eq:grandpot} combined with \eqref{eq:part1}. This is a function of $T, \mu, v^i$. The first law states that
\begin{equation}
    {\rm d} P=s{\rm d}T+n{\rm d}\mu+\mathcal{P}_i{\rm d}v^i=\frac{\tilde{\mathcal{E}}+P}{T}{\rm d}T+nT{\rm d}\frac{\mu}{T}+\mathcal{P}_i {\rm d}v^i\,,
\end{equation}
where we used $\tilde{\mathcal{E}}+P=sT+n\mu$ with $\tilde{\mathcal{E}}$ the internal energy\footnote{The LAB frame energy density $\mathcal{E}$ depends on the extensive conserved quantities $s, n, \mathcal{P}_i$. The internal energy, or rest-frame energy, is the energy density $\tilde{\mathcal{E}}$ which depends on $s, n$ and $v^i$. For a boost invariant system the $v^i$ dependence can be absorbed in $s$ and $n$, so that the internal energy only depends on the rest-frame entropy and particle number densities. The internal and LAB frame energies are thus Legendre transforms of each other where $\mathcal{P}_i$-dependence is traded for $v^i$-dependence and vice versa, i.e. we have $\mathcal{E}=\tilde{\mathcal{E}}+\mathcal{P}_i v^i$.}. The momentum density is computed to be
\begin{equation}
    \mathcal{P}_i=\frac{4}{c^2}\gamma^2 P v^i\,.
\end{equation}
The LAB frame energy-momentum tensor is\footnote{There is also a $U(1)$ current $J^0=n$ and $J^i=nv^i$ but this will play no role in our discussion.}
\begin{equation}
    T^0{}_0 = -\mathcal{E}\,,\qquad T^i{}_0=-\left(\mathcal{E}+P\right)v^i\,,\qquad T^0{}_j=\mathcal{P}_j\,,\qquad T^i{}_j=P\delta^i_j+\mathcal{P}_j v^i\,,
\end{equation}
where $\mathcal{E}=\tilde{\mathcal{E}}+\mathcal{P}_i v^i=3P+\frac{4}{c^2}\gamma^2 P v^2$. 
This is the equation of state of a scale invariant system, thus $\tilde{\mathcal{E}}=3P$ which obeys the ideal gas law $P=k_B n T$ (see Appendix \ref{app:gas} for more details).
If we expand this around $c=0$ by setting 
$v^i=c^2v_{(2)}^i+O(c^4)$ and taking $T$ and $\mu$ to be $O(1)$ with the leading order terms again denoted by $T$ and $\mu$ we find that the pressure becomes the one given in \eqref{eq:leadingpressure2}, while the energy-momentum tensor becomes
\begin{equation}
    T^0{}_0 = -\tilde{\mathcal{E}}=-3P\,,\qquad T^i{}_0=0\,,\qquad T^0{}_j=4 P v_{(2)}^j\,,\qquad T^i{}_j=P\delta^i_j\,.
\end{equation}
Note that the pressure \eqref{eq:leadingpressure2} does not depend on $v_{(2)}^i$. This energy-momentum tensor is not thermodynamic since it cannot be obtained from a first law applied to $P$ in \eqref{eq:leadingpressure2}.

\subsubsection*{Spacelike Carroll case}

In this case we take the chemical potential $v$ to be purely imaginary. In the small $|c/v|$ limit, i.e. the Carroll regime, we have that $\gamma \to |c/v|$ and we find for the pressure
\begin{equation}\label{eq:leadingpressure}
    P= \frac{8\pi  c }{h^3\beta^4v^{4}}e^{\beta\mu}
    \,.
\end{equation}
This result can only be obtained for $v\neq 0$. The Carroll regime now yields an energy-momentum tensor of the form of a spacelike Carroll fluid as presented in \eqref{eq:EMTcaseI} with thermodynamical relations that are consistent with the leading order pressure in \eqref{eq:leadingpressure}. This fluid explicitly satisfies also $\mathcal{E}=-P$, as can be shown explicitly by using \eqref{eq:part1} to compute
\begin{equation}
    -\mathcal{E}
    =
    \frac{N}{V}\partial_{\beta }\log Z_{1}
    -
    \frac{N}{\beta V}
    v^{i}\partial_{v^{i}}\log Z_{1}
    =
    \frac{N}{\beta V}
    =
    P
    \,.
\end{equation}
For more details we refer the reader to Appendix \ref{app:gas}.

 It would be interesting to understand better the physical interpretation of having imaginary chemical potentials $v^i$. If this can be justified physically, our construction of the partition function gives a microscopic description of a system with an equation of state with $w=-1$. 

\section*{Acknowledgements}

It is a pleasure to thank the organizers and participants of the workshops on Carrollian and Non-Lorentzian Geometry in Vienna, Mons and Edinburgh, where some of these results were presented. In particular we thank Arjun Bagchi, Eric Bergshoeff, Laura Donnay, Daniel Grumiller, Marc Henneaux, Gerben Oling, Simon Pekar and Marios Petropoulos for useful discussions. JdB is supported by the European Research Council under the European Union's Seventh Framework Programme (FP7/2007-2013), ERC Grant agreement ADG 834878.
JH is supported by the Royal Society University Research Fellowship Renewal “Non-Lorentzian String Theory” (grant number URF\textbackslash R\textbackslash 221038). The work of NO is supported in part by VR project grant 2021-04013 and Villum Foundation Experiment project 00050317, ``Exploring the wonderland of Carrollian physics: Extreme gravity, spacetime horizons and supersonic fluids''. Nordita is supported in part by Nordforsk. 
WS is supported by the Icelandic Research Fund via the Grant of Excellence titled “Quantum Fields and Quantum Geometry” and by the University of Iceland Research Fund.

\appendix

\section{Quantum mechanical toy model}
\label{qmtoymodel}

In this appendix we consider a quantum mechanical toy model that is representative for the type of Hamiltonian
one obtains in case of the magnetic Carroll scalar field theory.

Consider two commuting harmonic oscillators $X$ and $Y$, i.e. $[X,X^{\dagger}]=[Y,Y^{\dagger}]=1$ and the operator
\be H=(X+Y^{\dagger})(X^{\dagger}+Y) \; . \
\ee 
What are the eigenstates of this operator? Notice that $\langle \psi | H | \psi \rangle\geq 0$ so eigenvalues
of $H$ will be non-negative. 
Notice that $H=e^{-X^{\dagger}Y^{\dagger}} XY e^{X^{\dagger}Y^{\dagger}}$ so we might as well
ask about the eigenstates of the operator $XY$ (though one might need to worry about normalizability).

Consider the states 
\be
|\psi\rangle_{E,k} = \sum_{l=0}^{\infty} E^l \frac{(X^{\dagger})^{k+l} (Y^{\dagger})^l}{(k+l)! l!} |0\rangle
\ee
and similariy 
\be
|\chi\rangle_{E,k} = \sum_{l=0}^{\infty} E^l \frac{(X^{\dagger})^{l} (Y^{\dagger})^{k+l}}{l! (k+l)!} |0\rangle\,,
\ee
where $|0\rangle$ is the groundstate of $XY$'.
These series also appear in modified Bessel functions. Because
\be
XY (X^{\dagger})^{k} (Y^{\dagger})^l |0\rangle = kl (X^{\dagger})^{k-1} (Y^{\dagger})^{l-1} |0\rangle\,,
\ee
we immediately see that both states are eigenstates of $XY$ with eigenvalue $E$. For $k=0$ the two states agree.
For $E=0$ only the first term survives leading to states of the form $(X^{\dagger})^{k}|0 \rangle $ and
$(Y^{\dagger})^{k}|0 \rangle $. We conclude that the eigenstates of $H$ with eigenvalue $E$ are of the form $e^{-X^{\dagger}Y^{\dagger}} |\psi\rangle_{E,k} $
and $ e^{-X^{\dagger}Y^{\dagger}} |\chi\rangle_{E,k}$. 

We can try to compute the overlap of these states, so let us consider
\be
Z={}_{E_1,k}\langle \psi| e^{-XY}  e^{-X^{\dagger}Y^{\dagger}} |\psi\rangle_{E_2,k}
\ee
Expanding the exponentials leads to 
\be
Z=\sum_{l,l',p,q} \langle 0 | (-1)^{p+q} E_1^l E_2^{l'} 
\frac{ X^{k+l+p} Y^{l+p} (X^{\dagger})^{k+l'+q} (Y^{\dagger})^{l'+q} }{ (k+l)! l! p! (k+l')! l'! q!}
|0\rangle
\ee
which evaluates to ($l+p=l'+q=m$) 
\be
\sum_{l,l',m} \frac{(-1)^{l+l'} E_1^l E_2^{l'} (k+m)! m! }{ (k+l)! l! (m-l)! (k+l')! l'! (m-l') }
\ee
The sum over $m$ always diverges. We can already see this when we compute the norm of 
$e^{-X^{\dagger} Y^{\dagger} }|0\rangle$ which is infinite. So at best these energy eigenstates are
delta-function normalizable. 

We can also do the sums over $l$ and $l'$ first which leads to 
\be
\sum_m \frac{(k+m)!}{k!^2 m!} {}_1 F_1 (-m,k+1,E_1) {}_1 F_1 (-m,k+1,E_2)\ ,
\ee
or equivalently in terms of Laguerre polynomials
\be
\sum_m \frac{m!}{(m+k)!} L^{(k)}_{m}(E_1) L^{(k)}_{m}(E_2)\ .
\ee
Happily, Laguerre polynomials form a set of orthogonal polynomials on $[0,\infty) $ with measure $x^k e^{-x}$,
so that
\be 
\int_0^{\infty} {\rm d}E E^k e^{-E} L^{(k)}_{m}(E) L^{(k)}_{n}(E)  = \frac{ (n+k)! }{n!} \delta_{n,m} \ ,
\ee
which implies in particular that
\be 
E_1^k e^{-E_1}\sum_m \frac{m!}{(m+k)!} L^{(k)}_{m}(E_1) L^{(k)}_{m}(E_2) = \delta(E_1-E_2)\ ,
\ee 
so that the states are indeed delta-function normalizable.

To summarize: the spectrum of the theory consists of states labeled by $E\geq 0$ and $k\in\mathbb Z$ with 
\be 
|E,k\rangle = E^{|k|/2} e^{-E/2} e^{-X^{\dagger} Y^{\dagger} } \sum_{l=0}^{\infty} E^l \frac{(X^{\dagger})^{k+l} (Y^{\dagger})^l}{(k+l)! l!} |0\rangle \ ,
\ee
for $k\geq 0$ and a similar expression with $X$ and $Y$ interchanged for $k<0$. The inner product of these
states is 
\be 
\langle E_1,k_1 | E_2,k_2 \rangle = \delta(E_1-E_2) \delta_{k_1,k_2} \ .
\ee 
As a check, we compute the overlap of the ground state with energy eigenstates giving
\be \langle 0 | E,k \rangle = e^{-E/2} \delta_{k,0}\ . 
\ee
Therefore, $|0\rangle = \int {\rm d}E e^{-E/2} |E,0\rangle$ and we easily check that the norm of this state is
$\int_0^{\infty} e^{-E} {\rm d}E =1$.
The states with negative energy do not appear (and are presumably not even delta-function normalizable) given
the general argument above.

\section{From Galilean to Carrollian theories and back \label{app:map}}

In this appendix we present a Lagrangian method which produces a magnetic Carroll theory
from a given Galilean theory (for a closely related 
construction of non-Lorentzian models from a seed Lagrangian see \cite{Bergshoeff:2022qkx}). 
For simplicity, we give the argument here for a scalar field but we expect the method to generalize to
other fields.

Consider a Lagrangian $\mathcal{L}$ for a real scalar field $\phi$ that is Lorentz invariant and where $\phi$ is a Lorentz scalar, i.e.
\begin{equation}
    \mathcal{L}=\mathcal{L}(\phi, \dot\phi, \partial_i\phi)\,,
\end{equation}
where $\mathcal{L}$ infinitesimally transforms as 
\begin{equation}\label{eq:Lscalar}
\delta_L\mathcal{L}=\xi^\mu_L\partial_\mu\mathcal{L}\,,
\end{equation}
with 
\begin{equation}\label{eq:xi}
    \xi^\mu_L\partial_\mu=b^i x^i\partial_t+t b^i\partial_i\,.
\end{equation}
We have set $c=1$. The Lagrangian is a Lorentz scalar and $\delta_L\phi=\xi^\mu_L\partial_\mu\phi$. We assume that $\mathcal{L}$ can be written as the sum of a Lagrangian that is Galilean invariant and one that is Carroll invariant, i.e. we assume that
\begin{equation}
    \mathcal{L}=\mathcal{L}_C(\dot\phi, \phi)+\mathcal{L}_G(\partial_i\phi, \phi)\,.
\end{equation}
In this split $\mathcal{L}_C(\dot\phi, \phi)$ is an electric Carroll theory and $\mathcal{L}_G(\partial_i\phi, \phi)$ a magnetic Galilean theory. This assumption does not apply to higher-derivative theories but maybe the argument can be generalised to those cases. We can write 
infinitesimally any Lorentz transformation \eqref{eq:xi} as the sum of a Galilean and a Carroll transformation,
\begin{equation}
    \xi^\mu_L\partial_\mu=b^i x^i\partial_t+t b^i\partial_i=\xi^\mu_C\partial_\mu+\xi^\mu_G\partial_\mu\,.
\end{equation}
Write \eqref{eq:Lscalar} as
\begin{equation}\label{eq:Lscalar2}
\left(\delta_C+\delta_G\right)\left(\mathcal{L}_C+\mathcal{L}_G\right)=\left(\xi^\mu_C+\xi^\mu_G\right)\partial_\mu\left(\mathcal{L}_C+\mathcal{L}_G\right)\,.
\end{equation}
The fact that $\mathcal{L}_C$ and $\mathcal{L}_G$ are Carroll and Galilean scalars means that they obey the property that
\begin{equation}
    \delta_C\mathcal{L}_C=\xi^\mu_C\partial_\mu\mathcal{L}_C\,,
\end{equation}
where we transform $\phi$ as a Carroll scalar, $\delta_C\phi=\xi^\mu_C\partial_\mu\phi$, and a similar statement applies to the Galilean Lagrangian. After some algebra we then find
\begin{equation}
    \delta_G\mathcal{L}_C+\delta_C\mathcal{L}_G=\xi^\mu_G\partial_\mu\mathcal{L}_C+\xi^\mu_C\partial_\mu\mathcal{L}_G\,.
\end{equation}
Isolating $\delta_C\mathcal{L}_G$ and using that $\mathcal{L}_G=\mathcal{L}_G(\phi,\partial_i\phi)$ and that $\mathcal{L}_C=\mathcal{L}_C(\phi,\dot\phi)$ and computing the variations and derivatives using the chain rule, we obtain 
\begin{equation}\label{eq:seedL}
    \delta_C\mathcal{L}_G=-\frac{\partial\mathcal{L}_C}{\partial\dot\phi}b^i\partial_i\phi+\xi^\mu_C\partial_\mu\mathcal{L}_G\,.
\end{equation}
It can be shown that $\frac{\partial\mathcal{L}_C}{\partial\dot\phi}$ is a Carroll scalar, 
\begin{equation}
    \delta_C\left(\frac{\partial\mathcal{L}_C}{\partial\dot\phi}\right)=\frac{\partial^2\mathcal{L}_C}{\partial\phi\partial\dot\phi}\delta_C\phi+\frac{\partial^2\mathcal{L}_C}{\partial\dot\phi\partial\dot\phi}\delta_C\dot\phi=\xi^\mu_C\partial_\mu\left(\frac{\partial\mathcal{L}_C}{\partial\dot\phi}\right)\,.
\end{equation}

Equation \eqref{eq:seedL} is the main observation from which the rest follows. One can construct a new Carroll theory by starting with $\mathcal{L}_G$ and adding to it a Lagrange multiplier term proportional to $\frac{\partial\mathcal{L}_C}{\partial\dot\phi}$. In other words, define
\begin{equation}
\label{newL}
    \tilde{\mathcal{L}}_C=\mathcal{L}_G+\chi \frac{\partial\mathcal{L}_C}{\partial\dot\phi}\,.
\end{equation}
This new Lagrangian will be a Carroll scalar if $\chi$ transform as
\begin{equation}
    \delta\chi=\xi_C^\mu\partial_\mu\chi+b^i\partial_i\phi\,.
\end{equation}
This works because $\frac{\partial\mathcal{L}_C}{\partial\dot\phi}$ is a Carroll scalar.

In particular, applying this to the case for which one chooses the Carroll action to be the one
of the electric theory and the standard Galilean action
\begin{equation}
    \mathcal{L}_C=\frac{1}{2}\dot\phi^2-V(\phi)\,,\qquad\mathcal{L}_G=-\frac{1}{2}\partial_i\phi\partial_i\phi\,,
\end{equation}
it follows that \eqref{newL} generates the magnetic Carroll theory.

This idea also works in the other direction, so that it is possible to create a new Galilean theory from a Carroll theory by writing
\begin{equation}
    \tilde{\mathcal{L}}_G=\mathcal{L}_C+\chi^i\frac{\partial\mathcal{L}_G}{\partial\partial_i\phi}\,.
\end{equation}

We expect the procedure above to be easily applicable to Maxwell actions. Furthermore, in the same spirit
probably GR can also be viewed as an appropriate sum of a Carrollian and Galilean gravity theory. However, since in this case there are no global symmetries we expect the details to be somewhat different. 
It would be interesting to examine this further as well as what happens with higher-derivative theories such
as Born-Infeld.

\section{Details on gasses}\label{app:gas}

In this appendix we review massless and massive relativistic Boltzmann gasses and verify that their resulting energy momentum tensors reproduce the expected perfect fluid energy-momentum tensors. The single particle partition function is given by \eqref{eq:Z1}:
\begin{equation}\begin{aligned}
	Z_{1}(T,V,v^{i})
	=
	\frac{V}{h^{d}}\int {\rm d}^{d}p\,e^{-{\beta} H_{1}(p)+{\beta} v^{i}p_{i}}	
	\ .
\end{aligned}\end{equation}
The canonical $N$ particle partition function for a Blotzmann gas of $N$ free particles is given by $Z=(Z_{1}^N)/N!$ and the grand canonical partition function $\mathcal{Z}$ can be written as
\begin{equation}
    \log \mathcal{Z}
    =
    e^{\beta\mu}Z_{1}
    \,,
\end{equation}
where $\mu$ is the chemical potential. This allows us to write the pressure and grand potential $\Omega$ as\footnote{Using ${\rm d}\Omega=- S {\rm d}T - P{\rm d}V - P_{i}{\rm d}v^{i} -N {\rm d}\mu$.}
\begin{equation}\label{eq:pres}
    P
    =
    -\frac{\Omega}{V}
    =
    \frac{1}{V\beta} \log\mathcal{Z}
    =
    \frac{1}{V\beta}e^{\beta\mu}Z_{1}
     \,.
\end{equation}
The particle number is given by
\begin{equation}
    N
    =
    -
    \left(
        \frac{\partial \Omega}{\partial\mu}
    \right)_{V,T,v^{i}}
    =
    e^{\beta\mu}Z_{1} 
    \,,
\end{equation}
which when inserting the pressure as given in \eqref{eq:pres} we readily recognize the ideal gas law:
\begin{equation}\label{eq:ideal_gas}
     P V
    =
    \frac{N}{\beta} 
\end{equation}
The energy density of the system follows from 
\begin{equation}\label{eq:energy}
    \tilde{\mathcal{E}}
    =
    - \frac{\partial}{\partial\beta}\log Z
    =
    N \langle H_{1}\rangle_{1}
    -
    N v^{i}\langle p_{i}\rangle_{1}
    \,,
\end{equation}
where the brackets indicate with subscript 1 indicate average with respect to the single particle partition function and $p_{i}$ is the momentum that appears in the integral. Furthermore we'll denote $-T^{0}{_0}=\mathcal{E}= N \langle H_{1}\rangle_{1}$ such that $\tilde{\mathcal{E}}+P_{i}v^{i}=\mathcal{E}$, where $T^{0}{_j}=P_{j}=N\langle p_{j}\rangle_{1}$ is generalized momentum. 
We can furthermore compute
\begin{equation}
    \langle p_{i}\rangle_{1}
    =
    \frac{1}{\beta}\frac{\partial}{\partial v^{i}}\log\left(
        Z_{1}
    \right)
    \,.
\end{equation}
Let us now compute the spatial stress tensor $T^{i}{_j}=N\langle p_{i}\frac{\partial H_{1}}{\partial p_{j}}\rangle_{1}$ and energy flux $T^{i}{_0}=N\langle H_{1}\frac{\partial H_{1}}{\partial p_{j}}\rangle_{1}$. We observe
\begin{equation}
    T^{i}{_j}
    =
    -\frac{N}{\beta}\langle
        p_{i}\frac{\partial}{\partial p_{j}}
    \rangle_{1}
    +
    N\langle p_{j}\rangle v^{i}
    =
    PV\delta^{i}{_j}
    +
    P_{j}v^{i}
    \,,
\end{equation}
\begin{equation}
    T^{i}{_0}
    =
    -\frac{N}{\beta}\langle
        H_{1}\frac{\partial}{\partial p_{i}}
    \rangle_{1}
    +
    N\langle 
        H_{1}
    \rangle_{1}v_{i}
    =
    PV\langle \frac{\partial H_{1}}{\partial p_{i}}\rangle_{1}
    +
    \mathcal{E} v_{i}
    \,,
\end{equation}
where $\langle \frac{\partial H_{1}}{\partial p_{i}}\rangle_{1}=-\frac{1}{\beta}\langle \frac{\partial}{\partial p_{i}}\rangle_{1}v^{i}$, where the first term vanishes as it is a total derivative.
We now collect all the entries of the energy momentum tensor and divide by volume $V$:
\begin{equation}
    T^{0}{_0}
    =
    -\mathcal{E}
    \,,
\end{equation}
\begin{equation}
    T^{i}{_0}
    =
    (P
    +
    \mathcal{E})v_{i}
    \,,
\end{equation}
\begin{equation}
    T^{0}{_j}
    =
    \mathcal{P}_{j}
    \,,
\end{equation}
\begin{equation}
    T^{i}{_j}
    =
    P\delta^{i}{_j}
    +
    \mathcal{P}_{j}v^{i}
    \,.
\end{equation}
Taking the trace we find
\begin{equation}
    T^{0}{_0}
    +
    T^{i}{_i}
    =
    dP-\tilde{\mathcal{E}}
    \,.
\end{equation}

As an example, consider a $d$-dimensional massless relativistic gas, which has the following single particle partition function:
\begin{equation}
    Z_{1}
    =
    2^{d}V\frac{\pi^{\frac{d-1}{2}}}{h^d}\frac{\gamma^{d+1}}{(c\beta)^{d}}
    \Gamma\left[\frac{d+1}{2}\right]
    \,.
\end{equation}
Using \eqref{eq:energy} we find
\begin{equation}
    \tilde{\mathcal{E}}=
    -N\frac{\partial}{\partial\beta}\log Z_{1}
    =
    \frac{d}{\beta}N
    =
    d P V
    \,,
    \quad\Leftrightarrow
    \quad
    \tilde{\mathcal{E}}
    -dP
    =
    0
    \,,
\end{equation}
where in the last equality we used the idea gas law \eqref{eq:ideal_gas}. This moreover implies that the energy momentum tensor is traceless.

\bibliographystyle{JHEP}
\bibliography{ref}

\providecommand{\href}[2]{#2}\begingroup\raggedright\begin{thebibliography}{100}

\bibitem{Levy1965}
J.-M. Levy-Leblond, {\it Une nouvelle limite non-relativiste du groupe de
  poincar\'e},  {\em Annales de l'institut Henri Poincar\'e (A) Physique
  th\'eorique} {\bf 3} (1965), no.~1 1--12.

\bibitem{sen1966analogue}
N.~Sen~Gupta, {\it On an analogue of the galilei group},  {\em Il Nuovo Cimento
  A (1965-1970)} {\bf 44} (1966) 512--517.

\bibitem{Bacry:1968zf}
H.~Bacry and J.~Levy-Leblond, {\it {Possible kinematics}},  {\em J.Math.Phys.}
  {\bf 9} (1968) 1605--1614.

\bibitem{Bergshoeff:2014jla}
E.~Bergshoeff, J.~Gomis and G.~Longhi, {\it {Dynamics of Carroll Particles}},
  {\em Class.Quant.Grav.} {\bf 31} (2014), no.~20 205009
  [\href{http://arXiv.org/abs/1405.2264}{{\tt 1405.2264}}].

\bibitem{Casalbuoni:2023bbh}
R.~Casalbuoni, D.~Dominici and J.~Gomis, {\it {Two interacting conformal
  Carroll particles}},  \href{http://arXiv.org/abs/2306.02614}{{\tt
  2306.02614}}.

\bibitem{Marsot:2021tvq}
L.~Marsot, {\it {Planar Carrollean dynamics, and the Carroll quantum
  equation}},  \href{http://arXiv.org/abs/2110.08489}{{\tt 2110.08489}}.

\bibitem{Figueroa:2023jpi}
J.~Figueroa-O'Farrill, A.~P\'erez and S.~Prohazka, {\it {Carroll/fracton
  particles and their correspondence}},  {\em JHEP} {\bf 06} (2023) 207
  [\href{http://arXiv.org/abs/2305.06730}{{\tt 2305.06730}}].

\bibitem{deBoer:2021jej}
J.~de~Boer, J.~Hartong, N.~A. Obers, W.~Sybesma and S.~Vandoren, {\it {Carroll
  Symmetry, Dark Energy and Inflation}},  {\em Front. in Phys.} {\bf 10} (2022)
  810405 [\href{http://arXiv.org/abs/2110.02319}{{\tt 2110.02319}}].

\bibitem{Bondi:1962px}
H.~Bondi, M.~van~der Burg and A.~Metzner, {\it {Gravitational waves in general
  relativity. 7. Waves from axisymmetric isolated systems}},  {\em
  Proc.Roy.Soc.Lond.} {\bf A269} (1962) 21--52.

\bibitem{Duval:2014uva}
C.~Duval, G.~Gibbons and P.~Horvathy, {\it {Conformal Carroll groups and BMS
  symmetry}},  {\em Class.Quant.Grav.} {\bf 31} (2014) 092001
  [\href{http://arXiv.org/abs/1402.5894}{{\tt 1402.5894}}].

\bibitem{Duval:2014lpa}
C.~Duval, G.~Gibbons and P.~Horvathy, {\it {Conformal Carroll groups}},  {\em
  J.Phys.} {\bf A47} (2014) 335204 [\href{http://arXiv.org/abs/1403.4213}{{\tt
  1403.4213}}].

\bibitem{Hartong:2015usd}
J.~Hartong, {\it {Holographic Reconstruction of 3D Flat Space-Time}},  {\em
  JHEP} {\bf 10} (2016) 104 [\href{http://arXiv.org/abs/1511.01387}{{\tt
  1511.01387}}].

\bibitem{Bagchi:2016bcd}
A.~Bagchi, R.~Basu, A.~Kakkar and A.~Mehra, {\it {Flat Holography: Aspects of
  the dual field theory}},  {\em JHEP} {\bf 12} (2016) 147
  [\href{http://arXiv.org/abs/1609.06203}{{\tt 1609.06203}}].

\bibitem{Donnay:2022aba}
L.~Donnay, A.~Fiorucci, Y.~Herfray and R.~Ruzziconi, {\it {Carrollian
  Perspective on Celestial Holography}},  {\em Phys. Rev. Lett.} {\bf 129}
  (2022), no.~7 071602 [\href{http://arXiv.org/abs/2202.04702}{{\tt
  2202.04702}}].

\bibitem{Donnay:2022wvx}
L.~Donnay, A.~Fiorucci, Y.~Herfray and R.~Ruzziconi, {\it {Bridging Carrollian
  and Celestial Holography}},  \href{http://arXiv.org/abs/2212.12553}{{\tt
  2212.12553}}.

\bibitem{Saha:2023hsl}
A.~Saha, {\it {Carrollian approach to 1 + 3D flat holography}},  {\em JHEP}
  {\bf 06} (2023) 051 [\href{http://arXiv.org/abs/2304.02696}{{\tt
  2304.02696}}].

\bibitem{Bagchi:2022emh}
A.~Bagchi, S.~Banerjee, R.~Basu and S.~Dutta, {\it {Scattering Amplitudes:
  Celestial and Carrollian}},  {\em Phys. Rev. Lett.} {\bf 128} (2022), no.~24
  241601 [\href{http://arXiv.org/abs/2202.08438}{{\tt 2202.08438}}].

\bibitem{Bagchi:2023fbj}
A.~Bagchi, P.~Dhivakar and S.~Dutta, {\it {AdS Witten diagrams to Carrollian
  correlators}},  {\em JHEP} {\bf 04} (2023) 135
  [\href{http://arXiv.org/abs/2303.07388}{{\tt 2303.07388}}].

\bibitem{Penna:2018gfx}
R.~F. Penna, {\it {Near-horizon Carroll symmetry and black hole Love numbers}},
   \href{http://arXiv.org/abs/1812.05643}{{\tt 1812.05643}}.

\bibitem{Donnay:2019jiz}
L.~Donnay and C.~Marteau, {\it {Carrollian Physics at the Black Hole Horizon}},
   {\em Class. Quant. Grav.} {\bf 36} (2019), no.~16 165002
  [\href{http://arXiv.org/abs/1903.09654}{{\tt 1903.09654}}].

\bibitem{Freidel:2022bai}
L.~Freidel and P.~Jai-akson, {\it {Carrollian hydrodynamics from symmetries}},
  {\em Class. Quant. Grav.} {\bf 40} (2023), no.~5 055009
  [\href{http://arXiv.org/abs/2209.03328}{{\tt 2209.03328}}].

\bibitem{Redondo-Yuste:2022czg}
J.~Redondo-Yuste and L.~Lehner, {\it {Non-linear black hole dynamics and
  Carrollian fluids}},  {\em JHEP} {\bf 02} (2023) 240
  [\href{http://arXiv.org/abs/2212.06175}{{\tt 2212.06175}}].

\bibitem{Hawking:1974rv}
S.~W. Hawking, {\it {Black hole explosions}},  {\em Nature} {\bf 248} (1974)
  30--31.

\bibitem{Henneaux:1979vn}
M.~Henneaux, {\it {Geometry of Zero Signature Space-times}},  {\em
  Bull.Soc.Math.Belg.} {\bf 31} (1979) 47--63.

\bibitem{Dautcourt:1997hb}
G.~Dautcourt, {\it {On the ultrarelativistic limit of general relativity}},
  {\em Acta Phys.Polon.} {\bf B29} (1998) 1047--1055
  [\href{http://arXiv.org/abs/gr-qc/9801093}{{\tt gr-qc/9801093}}].

\bibitem{Hansen:2021fxi}
D.~Hansen, N.~A. Obers, G.~Oling and B.~T. S\o{}gaard, {\it {Carroll Expansion
  of General Relativity}},  {\em SciPost Phys.} {\bf 13} (2022), no.~3 055
  [\href{http://arXiv.org/abs/2112.12684}{{\tt 2112.12684}}].

\bibitem{Henneaux:2021yzg}
M.~Henneaux and P.~Salgado-Rebolledo, {\it {Carroll contractions of
  Lorentz-invariant theories}},  \href{http://arXiv.org/abs/2109.06708}{{\tt
  2109.06708}}.

\bibitem{Figueroa-OFarrill:2022mcy}
J.~Figueroa-O'Farrill, E.~Have, S.~Prohazka and J.~Salzer, {\it {The gauging
  procedure and carrollian gravity}},  {\em JHEP} {\bf 09} (2022) 243
  [\href{http://arXiv.org/abs/2206.14178}{{\tt 2206.14178}}].

\bibitem{Campoleoni:2022ebj}
A.~Campoleoni, M.~Henneaux, S.~Pekar, A.~P\'erez and P.~Salgado-Rebolledo, {\it
  {Magnetic Carrollian gravity from the Carroll algebra}},  {\em JHEP} {\bf 09}
  (2022) 127 [\href{http://arXiv.org/abs/2207.14167}{{\tt 2207.14167}}].

\bibitem{Belinski:2017fas}
V.~Belinski and M.~Henneaux, {\em {The Cosmological Singularity}}.
\newblock Cambridge Monogr.Math.Phys. Cambridge Univ. Pr., Cambridge, 2017.

\bibitem{Duval:2014uoa}
C.~Duval, G.~Gibbons, P.~Horvathy and P.~Zhang, {\it {Carroll versus Newton and
  Galilei: two dual non-Einsteinian concepts of time}},  {\em
  Class.Quant.Grav.} {\bf 31} (2014) 085016
  [\href{http://arXiv.org/abs/1402.0657}{{\tt 1402.0657}}].

\bibitem{Nzotungicimpaye:2014wya}
J.~Nzotungicimpaye, {\it {Kinematical versus Dynamical Contractions of the de
  Sitter Lie algebras}},  {\em J. Phys. Comm.} {\bf 3} (2019), no.~10 105003
  [\href{http://arXiv.org/abs/1406.0972}{{\tt 1406.0972}}].

\bibitem{Hartong:2015xda}
J.~Hartong, {\it {Gauging the Carroll Algebra and Ultra-Relativistic Gravity}},
   {\em JHEP} {\bf 08} (2015) 069 [\href{http://arXiv.org/abs/1505.05011}{{\tt
  1505.05011}}].

\bibitem{Bekaert:2015xua}
X.~Bekaert and K.~Morand, {\it {Connections and dynamical trajectories in
  generalised Newton-Cartan gravity II. An ambient perspective}},  {\em J.
  Math. Phys.} {\bf 59} (2018), no.~7 072503
  [\href{http://arXiv.org/abs/1505.03739}{{\tt 1505.03739}}].

\bibitem{Bergshoeff:2017btm}
E.~Bergshoeff, J.~Gomis, B.~Rollier, J.~Rosseel and T.~ter Veldhuis, {\it
  {Carroll versus Galilei Gravity}},  {\em JHEP} {\bf 03} (2017) 165
  [\href{http://arXiv.org/abs/1701.06156}{{\tt 1701.06156}}].

\bibitem{Duval:2017els}
C.~Duval, G.~W. Gibbons, P.~A. Horvathy and P.~M. Zhang, {\it {Carroll symmetry
  of plane gravitational waves}},  {\em Class. Quant. Grav.} {\bf 34} (2017),
  no.~17 175003 [\href{http://arXiv.org/abs/1702.08284}{{\tt 1702.08284}}].

\bibitem{Ciambelli:2018ojf}
L.~Ciambelli and C.~Marteau, {\it {Carrollian conservation laws and Ricci-flat
  gravity}},  {\em Class. Quant. Grav.} {\bf 36} (2019), no.~8 085004
  [\href{http://arXiv.org/abs/1810.11037}{{\tt 1810.11037}}].

\bibitem{Morand:2018tke}
K.~Morand, {\it {Embedding Galilean and Carrollian geometries I. Gravitational
  waves}},  {\em J. Math. Phys.} {\bf 61} (2020), no.~8 082502
  [\href{http://arXiv.org/abs/1811.12681}{{\tt 1811.12681}}].

\bibitem{Bergshoeff:2019ctr}
E.~Bergshoeff, J.~M. Izquierdo, T.~Ort\'\i{}n and L.~Romano, {\it {Lie Algebra
  Expansions and Actions for Non-Relativistic Gravity}},  {\em JHEP} {\bf 08}
  (2019) 048 [\href{http://arXiv.org/abs/1904.08304}{{\tt 1904.08304}}].

\bibitem{Ravera:2019ize}
L.~Ravera, {\it {AdS Carroll Chern-Simons supergravity in 2 + 1 dimensions and
  its flat limit}},  {\em Phys. Lett. B} {\bf 795} (2019) 331--338
  [\href{http://arXiv.org/abs/1905.00766}{{\tt 1905.00766}}].

\bibitem{Gomis:2019nih}
J.~Gomis, A.~Kleinschmidt, J.~Palmkvist and P.~Salgado-Rebolledo, {\it
  {Newton-Hooke/Carrollian expansions of (A)dS and Chern-Simons gravity}},
  {\em JHEP} {\bf 02} (2020) 009 [\href{http://arXiv.org/abs/1912.07564}{{\tt
  1912.07564}}].

\bibitem{Ciambelli:2019lap}
L.~Ciambelli, R.~G. Leigh, C.~Marteau and P.~M. Petropoulos, {\it {Carroll
  Structures, Null Geometry and Conformal Isometries}},  {\em Phys. Rev. D}
  {\bf 100} (2019), no.~4 046010 [\href{http://arXiv.org/abs/1905.02221}{{\tt
  1905.02221}}].

\bibitem{Ballesteros:2019mxi}
A.~Ballesteros, G.~Gubitosi and F.~J. Herranz, {\it {Lorentzian Snyder
  spacetimes and their Galilei and Carroll limits from projective geometry}},
  {\em Class. Quant. Grav.} {\bf 37} (2020), no.~19 195021
  [\href{http://arXiv.org/abs/1912.12878}{{\tt 1912.12878}}].

\bibitem{Bergshoeff:2020xhv}
E.~Bergshoeff, J.~M. Izquierdo and L.~Romano, {\it {Carroll versus Galilei from
  a Brane Perspective}},  {\em JHEP} {\bf 10} (2020) 066
  [\href{http://arXiv.org/abs/2003.03062}{{\tt 2003.03062}}].

\bibitem{Niedermaier:2020jdy}
M.~Niedermaier, {\it {Nonstandard Action of Diffeomorphisms and
  Gravity\textquoteright{}s Anti-Newtonian Limit}},  {\em Symmetry} {\bf 12}
  (2020), no.~5 752.

\bibitem{Gomis:2020wxp}
J.~Gomis, D.~Hidalgo and P.~Salgado-Rebolledo, {\it {Non-relativistic and
  Carrollian limits of Jackiw-Teitelboim gravity}},  {\em JHEP} {\bf 05} (2021)
  162 [\href{http://arXiv.org/abs/2011.15053}{{\tt 2011.15053}}].

\bibitem{Grumiller:2020elf}
D.~Grumiller, J.~Hartong, S.~Prohazka and J.~Salzer, {\it {Limits of JT
  gravity}},  {\em JHEP} {\bf 02} (2021) 134
  [\href{http://arXiv.org/abs/2011.13870}{{\tt 2011.13870}}].

\bibitem{Concha:2021jnn}
P.~Concha, D.~Pe\~nafiel, L.~Ravera and E.~Rodr\'\i{}guez, {\it
  {Three-dimensional Maxwellian Carroll gravity theory and the cosmological
  constant}},  {\em Phys. Lett. B} {\bf 823} (2021) 136735
  [\href{http://arXiv.org/abs/2107.05716}{{\tt 2107.05716}}].

\bibitem{Guerrieri:2021cdz}
A.~Guerrieri and R.~F. Sobreiro, {\it {Carroll limit of four-dimensional
  gravity theories in the first order formalism}},  {\em Class. Quant. Grav.}
  {\bf 38} (2021), no.~24 245003 [\href{http://arXiv.org/abs/2107.10129}{{\tt
  2107.10129}}].

\bibitem{Perez:2021abf}
A.~P\'erez, {\it {Asymptotic symmetries in Carrollian theories of gravity}},
  \href{http://arXiv.org/abs/2110.15834}{{\tt 2110.15834}}.

\bibitem{Figueroa-OFarrill:2021sxz}
J.~Figueroa-O'Farrill, E.~Have, S.~Prohazka and J.~Salzer, {\it {Carrollian and
  celestial spaces at infinity}},  {\em JHEP} {\bf 09} (2022) 007
  [\href{http://arXiv.org/abs/2112.03319}{{\tt 2112.03319}}].

\bibitem{Herfray:2021qmp}
Y.~Herfray, {\it {Carrollian manifolds and null infinity: A view from Cartan
  geometry}},  \href{http://arXiv.org/abs/2112.09048}{{\tt 2112.09048}}.

\bibitem{Baiguera:2022lsw}
S.~Baiguera, G.~Oling, W.~Sybesma and B.~T. S\o{}gaard, {\it {Conformal Carroll
  scalars with boosts}},  {\em SciPost Phys.} {\bf 14} (2023), no.~4 086
  [\href{http://arXiv.org/abs/2207.03468}{{\tt 2207.03468}}].

\bibitem{Perez:2022jpr}
A.~P\'erez, {\it {Asymptotic symmetries in Carrollian theories of gravity with
  a negative cosmological constant}},  {\em JHEP} {\bf 09} (2022) 044
  [\href{http://arXiv.org/abs/2202.08768}{{\tt 2202.08768}}].

\bibitem{Fuentealba:2022gdx}
O.~Fuentealba, M.~Henneaux, P.~Salgado-Rebolledo and J.~Salzer, {\it
  {Asymptotic structure of Carrollian limits of Einstein-Yang-Mills theory in
  four spacetime dimensions}},  {\em Phys. Rev. D} {\bf 106} (2022), no.~10
  104047 [\href{http://arXiv.org/abs/2207.11359}{{\tt 2207.11359}}].

\bibitem{Campoleoni:2022wmf}
A.~Campoleoni, L.~Ciambelli, A.~Delfante, C.~Marteau, P.~M. Petropoulos and
  R.~Ruzziconi, {\it {Holographic Lorentz and Carroll frames}},  {\em JHEP}
  {\bf 12} (2022) 007 [\href{http://arXiv.org/abs/2208.07575}{{\tt
  2208.07575}}].

\bibitem{deBoer:2017ing}
J.~de~Boer, J.~Hartong, N.~A. Obers, W.~Sybesma and S.~Vandoren, {\it {Perfect
  Fluids}},  {\em SciPost Phys.} {\bf 5} (2018), no.~1 003
  [\href{http://arXiv.org/abs/1710.04708}{{\tt 1710.04708}}].

\bibitem{Ciambelli:2018xat}
L.~Ciambelli, C.~Marteau, A.~C. Petkou, P.~M. Petropoulos and K.~Siampos, {\it
  {Covariant Galilean versus Carrollian hydrodynamics from relativistic
  fluids}},  {\em Class. Quant. Grav.} {\bf 35} (2018), no.~16 165001
  [\href{http://arXiv.org/abs/1802.05286}{{\tt 1802.05286}}].

\bibitem{Ciambelli:2018wre}
L.~Ciambelli, C.~Marteau, A.~C. Petkou, P.~M. Petropoulos and K.~Siampos, {\it
  {Flat holography and Carrollian fluids}},  {\em JHEP} {\bf 07} (2018) 165
  [\href{http://arXiv.org/abs/1802.06809}{{\tt 1802.06809}}].

\bibitem{Banerjee:2023jpi}
K.~Banerjee, R.~Basu, B.~Krishnan, S.~Maulik, A.~Mehra and A.~Ray, {\it
  {One-Loop Quantum Effects in Carroll Scalars}},
  \href{http://arXiv.org/abs/2307.03901}{{\tt 2307.03901}}.

\bibitem{Figueroa-OFarrill:2023qty}
J.~Figueroa-O'Farrill, A.~P\'erez and S.~Prohazka, {\it {Quantum
  Carroll/fracton particles}},  \href{http://arXiv.org/abs/2307.05674}{{\tt
  2307.05674}}.

\bibitem{Bagchi:2019xfx}
A.~Bagchi, A.~Mehra and P.~Nandi, {\it {Field Theories with Conformal
  Carrollian Symmetry}},  {\em JHEP} {\bf 05} (2019) 108
  [\href{http://arXiv.org/abs/1901.10147}{{\tt 1901.10147}}].

\bibitem{Bagchi:2019clu}
A.~Bagchi, R.~Basu, A.~Mehra and P.~Nandi, {\it {Field Theories on Null
  Manifolds}},  {\em JHEP} {\bf 02} (2020) 141
  [\href{http://arXiv.org/abs/1912.09388}{{\tt 1912.09388}}].

\bibitem{Banerjee:2020qjj}
K.~Banerjee, R.~Basu, A.~Mehra, A.~Mohan and A.~Sharma, {\it {Interacting
  Conformal Carrollian Theories: Cues from Electrodynamics}},  {\em Phys. Rev.
  D} {\bf 103} (2021), no.~10 105001
  [\href{http://arXiv.org/abs/2008.02829}{{\tt 2008.02829}}].

\bibitem{Chen:2021xkw}
B.~Chen, R.~Liu and Y.-f. Zheng, {\it {On Higher-dimensional Carrollian and
  Galilean Conformal Field Theories}},  {\em SciPost Phys.} {\bf 14} (2023) 088
  [\href{http://arXiv.org/abs/2112.10514}{{\tt 2112.10514}}].

\bibitem{Rivera-Betancour:2022lkc}
D.~Rivera-Betancour and M.~Vilatte, {\it {Revisiting the Carrollian scalar
  field}},  {\em Phys. Rev. D} {\bf 106} (2022), no.~8 085004
  [\href{http://arXiv.org/abs/2207.01647}{{\tt 2207.01647}}].

\bibitem{Saha:2022gjw}
A.~Saha, {\it {Intrinsic approach to 1 + 1D Carrollian Conformal Field
  Theory}},  {\em JHEP} {\bf 12} (2022) 133
  [\href{http://arXiv.org/abs/2207.11684}{{\tt 2207.11684}}].

\bibitem{Chen:2023pqf}
B.~Chen, R.~Liu, H.~Sun and Y.-f. Zheng, {\it {Constructing Carrollian Field
  Theories from Null Reduction}},  \href{http://arXiv.org/abs/2301.06011}{{\tt
  2301.06011}}.

\bibitem{Hao:2022xhq}
P.-X. Hao, W.~Song, Z.~Xiao and X.~Xie, {\it {A BMS-invariant free fermion
  model}},  \href{http://arXiv.org/abs/2211.06927}{{\tt 2211.06927}}.

\bibitem{Banerjee:2022ocj}
A.~Banerjee, S.~Dutta and S.~Mondal, {\it {Carroll fermions in two
  dimensions}},  \href{http://arXiv.org/abs/2211.11639}{{\tt 2211.11639}}.

\bibitem{Bagchi:2022eui}
A.~Bagchi, A.~Banerjee, R.~Basu, M.~Islam and S.~Mondal, {\it {Magic fermions:
  Carroll and flat bands}},  {\em JHEP} {\bf 03} (2023) 227
  [\href{http://arXiv.org/abs/2211.11640}{{\tt 2211.11640}}].

\bibitem{Basu:2018dub}
R.~Basu and U.~N. Chowdhury, {\it {Dynamical structure of Carrollian
  Electrodynamics}},  {\em JHEP} {\bf 04} (2018) 111
  [\href{http://arXiv.org/abs/1802.09366}{{\tt 1802.09366}}].

\bibitem{Islam:2023rnc}
M.~Islam, {\it {Carrollian Yang-Mills theory}},  {\em JHEP} {\bf 05} (2023) 238
  [\href{http://arXiv.org/abs/2301.00953}{{\tt 2301.00953}}].

\bibitem{Barducci:2018thr}
A.~Barducci, R.~Casalbuoni and J.~Gomis, {\it {Vector SUSY models with Carroll
  or Galilei invariance}},  {\em Phys. Rev. D} {\bf 99} (2019), no.~4 045016
  [\href{http://arXiv.org/abs/1811.12672}{{\tt 1811.12672}}].

\bibitem{Bidussi:2021nmp}
L.~Bidussi, J.~Hartong, E.~Have, J.~Musaeus and S.~Prohazka, {\it {Fractons,
  dipole symmetries and curved spacetime}},  {\em SciPost Phys.} {\bf 12}
  (2022), no.~6 205 [\href{http://arXiv.org/abs/2111.03668}{{\tt 2111.03668}}].

\bibitem{Figueroa-OFarrill:2023vbj}
J.~Figueroa-O'Farrill, A.~P\'erez and S.~Prohazka, {\it {Carroll/fracton
  particles and their duality}},  \href{http://arXiv.org/abs/2305.06730}{{\tt
  2305.06730}}.

\bibitem{Hao:2021urq}
P.-x. Hao, W.~Song, X.~Xie and Y.~Zhong, {\it {BMS-invariant free scalar
  model}},  {\em Phys. Rev. D} {\bf 105} (2022), no.~12 125005
  [\href{http://arXiv.org/abs/2111.04701}{{\tt 2111.04701}}].

\bibitem{Bagchi:2019unf}
A.~Bagchi, A.~Saha and Zodinmawia, {\it {BMS Characters and Modular
  Invariance}},  {\em JHEP} {\bf 07} (2019) 138
  [\href{http://arXiv.org/abs/1902.07066}{{\tt 1902.07066}}].

\bibitem{Yu:2022bcp}
Z.-f. Yu and B.~Chen, {\it {Free field realization of the BMS Ising model}},
  \href{http://arXiv.org/abs/2211.06926}{{\tt 2211.06926}}.

\bibitem{Oblak:2015sea}
B.~Oblak, {\it {Characters of the BMS Group in Three Dimensions}},  {\em
  Commun. Math. Phys.} {\bf 340} (2015), no.~1 413--432
  [\href{http://arXiv.org/abs/1502.03108}{{\tt 1502.03108}}].

\bibitem{Barnich:2015mui}
G.~Barnich, H.~A. Gonzalez, A.~Maloney and B.~Oblak, {\it {One-loop partition
  function of three-dimensional flat gravity}},  {\em JHEP} {\bf 04} (2015) 178
  [\href{http://arXiv.org/abs/1502.06185}{{\tt 1502.06185}}].

\bibitem{Bergshoeff:2022eog}
E.~Bergshoeff, J.~Figueroa-O'Farrill and J.~Gomis, {\it {A non-lorentzian
  primer}},  \href{http://arXiv.org/abs/2206.12177}{{\tt 2206.12177}}.

\bibitem{VandenBleeken:2017rij}
D.~Van~den Bleeken, {\it {Torsional Newton--Cartan gravity from the large c
  expansion of general relativity}},  {\em Class. Quant. Grav.} {\bf 34}
  (2017), no.~18 185004 [\href{http://arXiv.org/abs/1703.03459}{{\tt
  1703.03459}}].

\bibitem{Hansen:2018ofj}
D.~Hansen, J.~Hartong and N.~A. Obers, {\it {Action Principle for Newtonian
  Gravity}},  {\em Phys. Rev. Lett.} {\bf 122} (2019), no.~6 061106
  [\href{http://arXiv.org/abs/1807.04765}{{\tt 1807.04765}}].

\bibitem{Hansen:2020pqs}
D.~Hansen, J.~Hartong and N.~A. Obers, {\it {Non-Relativistic Gravity and its
  Coupling to Matter}},  {\em JHEP} {\bf 06} (2020) 145
  [\href{http://arXiv.org/abs/2001.10277}{{\tt 2001.10277}}].

\bibitem{Hartong:2022lsy}
J.~Hartong, N.~A. Obers and G.~Oling, {\it {Review on Non-Relativistic
  Gravity}},  \href{http://arXiv.org/abs/2212.11309}{{\tt 2212.11309}}.

\bibitem{Figueroa-OFarrill:2020gpr}
J.~Figueroa-O'Farrill, {\it {On the intrinsic torsion of spacetime
  structures}},  \href{http://arXiv.org/abs/2009.01948}{{\tt 2009.01948}}.

\bibitem{Kasner:1921zz}
E.~Kasner, {\it {Geometrical theorems on Einstein's cosmological equations}},
  {\em Am. J. Math.} {\bf 43} (1921) 217--221.

\bibitem{Oling:2021}
G.~Oling and B.~T. S\o{}gaard, {\it {unpublished}}, .

\bibitem{Einstein:1935tc}
A.~Einstein and N.~Rosen, {\it {The Particle Problem in the General Theory of
  Relativity}},  {\em Phys. Rev.} {\bf 48} (1935) 73--77.

\bibitem{Grumilleretal}
F.~Ecker, D.~Grumiller, J.~Hartong, A.~P\'erez, S.~Prohazka and R.~Troncoso,
  {\it {Carroll Black Holes}},  {\em to appear}.

\bibitem{ArjanvD}
A.~van Denzen, {\it {Geodescics in the Carroll limit}, {Master Thesis, Utrecht
  University, 2022};
  \url{https://studenttheses.uu.nl/handle/20.500.12932/536}}, .

\bibitem{Poovuttikul:2019ckt}
N.~Poovuttikul and W.~Sybesma, {\it {First order non-Lorentzian fluids, entropy
  production and linear instabilities}},
  \href{http://arXiv.org/abs/1911.00010}{{\tt 1911.00010}}.

\bibitem{Petkou:2022bmz}
A.~C. Petkou, P.~M. Petropoulos, D.~R. Betancour and K.~Siampos, {\it
  {Relativistic fluids, hydrodynamic frames and their Galilean versus
  Carrollian avatars}},  {\em JHEP} {\bf 09} (2022) 162
  [\href{http://arXiv.org/abs/2205.09142}{{\tt 2205.09142}}].

\bibitem{Freidel:2022vjq}
L.~Freidel and P.~Jai-akson, {\it {Carrollian hydrodynamics and symplectic
  structure on stretched horizons}},
  \href{http://arXiv.org/abs/2211.06415}{{\tt 2211.06415}}.

\bibitem{Banerjee:2012iz}
N.~Banerjee, J.~Bhattacharya, S.~Bhattacharyya, S.~Jain, S.~Minwalla and
  T.~Sharma, {\it {Constraints on Fluid Dynamics from Equilibrium Partition
  Functions}},  {\em JHEP} {\bf 09} (2012) 046
  [\href{http://arXiv.org/abs/1203.3544}{{\tt 1203.3544}}].

\bibitem{Jensen:2012jh}
K.~Jensen, M.~Kaminski, P.~Kovtun, R.~Meyer, A.~Ritz and A.~Yarom, {\it
  {Towards hydrodynamics without an entropy current}},  {\em Phys. Rev. Lett.}
  {\bf 109} (2012) 101601 [\href{http://arXiv.org/abs/1203.3556}{{\tt
  1203.3556}}].

\bibitem{Jensen:2014ama}
K.~Jensen, {\it {Aspects of hot Galilean field theory}},  {\em JHEP} {\bf 04}
  (2015) 123 [\href{http://arXiv.org/abs/1411.7024}{{\tt 1411.7024}}].

\bibitem{deBoer:2020xlc}
J.~de~Boer, J.~Hartong, E.~Have, N.~A. Obers and W.~Sybesma, {\it {Non-Boost
  Invariant Fluid Dynamics}},  \href{http://arXiv.org/abs/2004.10759}{{\tt
  2004.10759}}.

\bibitem{Bergshoeff:2022qkx}
E.~A. Bergshoeff, J.~Gomis and A.~Kleinschmidt, {\it {Non-Lorentzian theories
  with and without constraints}},  {\em JHEP} {\bf 01} (2023) 167
  [\href{http://arXiv.org/abs/2210.14848}{{\tt 2210.14848}}].

\end{thebibliography}\endgroup

\end{document}